\documentclass[12pt]{article}

\usepackage{scicite}

\usepackage{times}

\newenvironment{sciabstract}{%
\begin{quote} \bf}
{\end{quote}}

\usepackage{amsmath, amssymb}
\usepackage{mathtools}
\usepackage{physics}
\usepackage{pdfpages}
\usepackage[backend=bibtex,citestyle=numeric-comp,bibstyle=ieee,sorting=none,minbibnames=5,maxbibnames=5,firstinits=true]{biblatex}
\title{Two-electron two-nucleus effective Hamiltonian and the spin diffusion barrier}

\author
{Gevin von Witte$^{1,2}$, Sebastian Kozerke$^{1}$, Matthias Ernst$^{2\ast}$\\
\\
\normalsize{$^{1}$Institute for Biomedical Engineering, University and ETH Zurich, 8092 Zurich, Switzerland}\\
\normalsize{$^{2}$Institute of Molecular Physical Science, ETH Zurich, 8093 Zurich, Switzerland}\\
\\
\normalsize{$^\ast$To whom correspondence should be addressed; E-mail:   maer@ethz.ch}
}

\date{}

\begin{document} 


\baselineskip24pt


\maketitle


\begin{sciabstract}
Dynamic nuclear polarization (DNP) involves a polarization transfer from unpaired electrons to hyperfine coupled nuclei and can increase the sensitivity of nuclear magnetic resonance (NMR) signals by several orders of magnitude. 
The hyperfine coupling is considered to suppress nuclear dipolar flip-flop transitions, hindering the transport of nuclear hyperpolarization into the bulk (''spin-diffusion barrier'').
Possible polarization-transfer pathways leading to DNP and subsequent spin diffusion between hypershifted nuclei in a two-electron two-nucleus four-spin system are investigated. 
The Schrieffer-Wolff transformation is applied to characterize transitions that are only possible as second-order effects.
An energy-conserving electron-nuclear four-spin flip-flop is identified, which combines an electron dipolar with a nuclear dipolar flip-flop process, describing spin diffusion close to electrons.
The relevance of this process is supported by two-compartment model fits of HypRes-on experimental data.  
This suggests that all nuclear spins can contribute to the hyperpolarization of the bulk and the concept of a spin-diffusion barrier has to be reconsidered for samples with significant electron and nuclear dipolar couplings.

\end{sciabstract}

\section*{Introduction} 

In nuclear magnetic resonance (NMR), the problem of interactions between electron and nuclear spins has been discussed since at least the 1940s \cite{bloembergen_interaction_1949,khutsishvili_spin_1966,horvitz_nuclear_1971,wolfe_direct_1973,wittmann_electron-driven_2018,tan_three-spin_2019,stern_direct_2021,chessari_role_2023,thankamony_dynamic_2017,kundu_dnp_2019,Wenckebach,pang_hypershifted_2024}. 
Dynamic nuclear polarization (DNP) relies on the large thermal polarization and fast relaxation of unpaired electrons for transfer to low thermal polarization baths, typically slow relaxing nuclear spins. 
Microwave (MW) irradiation transfers polarization from the unpaired electrons (often called radicals in DNP) to hyperfine-coupled nuclei.
In DNP, the polarization transfer efficiency is not only limited by the polarization transfer from electrons to nuclei but also by the subsequent transport of the nuclear hyperpolarization into the bulk. 
Typically, nuclear spins in the bulk of the sample can be observed with inductive detection following a radio-frequency (RF) excitation pulse.
While hyperfine-coupled spins show the most efficient polarization transfer, they are strongly frequency shifted rendering them often unobservable in NMR (quenched, hidden or hypershifted spins). Accordingly, dipolar nuclear spin flip-flop processes are non-energy conserving. 
Nuclear spin flip-flops, described macroscopically by a nuclear spin-diffusion rate constant, subsequently spread the transferred polarization (homogeneously) throughout the sample.
Throughout this work we will refer to the recently coined term hypershifted spins \cite{pang_hypershifted_2024} to relate to strongly hyperfine-coupled spins that are difficult to observe with RF pulses.

In 1949, Bloembergen proposed the concept of a 'spin diffusion barrier' \cite{bloembergen_interaction_1949}, describing spins which are strongly coupled to unpaired electrons and, therefore, frequency shifted (hypershifted), such that they do not contribute to spin diffusion towards the bulk. 
$T_{1,\mathrm{e}}$ relaxation of the electrons will lead to a broadening of the hyperfine-split lines and eventually, for fast $T_{1,\mathrm{e}}$ times, to a population-averaged pseudo-contact shift \cite{bloembergen_shift_1950,bertini_magnetic_2002,parigi_magnetic_2019,pell_paramagnetic_2019}  that can be substantial under DNP conditions, since the polarization of the electrons will be high at low temperatures and high fields.
Several experiments have demonstrated indirectly \cite{wolfe_direct_1973,tan_three-spin_2019,jain_dynamic_2021,stern_direct_2021,chessari_role_2023,venkatesh_deuterated_2023} or directly \cite{pang_hypershifted_2024} an effective contribution of spins assumed to be within the spin diffusion barrier to the DNP process. 
These studies may question the size and existence of a spin diffusion barrier.  
Theoretical works aimed to explain these through relaxation processes, i.e. paramagnetic (electronic) relaxation causing a nuclear-nuclear flip-flop \cite{Horvitz1971,Buishvili1975,Sabirov1979,Atsarkin1980,chessari_role_2023}. 
In addition, the broadening of the zero-quantum (ZQ, nuclear flip-flop) line by the electron has been proposed as another pathway to make the spin diffusion close to electrons more efficient \cite{wittmann_electron-driven_2018}.
These models yield a strongly suppressed (vanishing) spin diffusion rate constant for spins less than several \AA\, away from the electron and a spin diffusion rate constant always smaller or equal to the one in the bulk.
In contrast, simulations of quantum dots suggest a spin diffusion coefficient around the electron exceeding its bulk value \cite{Gong2011} attributed to electron-mediated nuclear flip-flops described as two virtual electron-nuclear flip flops \cite{liu_control_2007}. 
In a similar direction, spin diffusion close to pairs of P1 centers in diamond is discussed in terms of two virtual electron-electron-nuclear triple spin flips \cite{pagliero_optically_2020,pagliero_magnetic_2021}.

For materials with a large electron line width and limited electron dipolar coupling, MW irradiation at a given frequency results in a hole burned into the electron spectrum \cite{kundu_theoretical_2019}. 
The resulting polarization difference between the hole and the rest of the electrons unaffected by the MW can be used to perform cross-effect (CE) DNP.
The minimum model to understand CE DNP consists of two electrons and a nucleus \cite{Hovav2012}. If the frequency difference between the electrons $\Delta\omega_\mathrm{e}=\omega_{\mathrm{e},1}-\omega_{\mathrm{e},2}$ becomes equal to the frequency of the nuclear Larmor frequency $\omega_\mathrm{n}$ (CE condition: $\Delta\omega_\mathrm{e} \approx \pm \omega_\mathrm{n}$), MW irradiation results in an efficient polarization transfer.
Thus, the fundamental process to generate hyperpolarization is a three-spin flip-flop-flip with an electronic flip-flop and a nuclear flip.

In the last decades, condensed matter systems have developed into one of the prime approaches for quantum information processing thanks to advanced manufacturing technology and tunability \cite{wolfowicz_quantum_2021}.
Defect centers in crystals such as the NV-center in diamond, phosphorous (P) dopants in silicon or quantum dots consist of a single or multiple unbound electrons surrounded by nuclear spins of the host crystal. 
Electron spins offer faster gate times and easier read-out at the expense of a shorter qubit coherence time ($T_2$). 
The opposite is true for nuclear spins. 
This has inspired the use of hybrid electron-nuclear spin systems with nuclear spins for processing or as a long coherence time qubit memory, which is read out through an electron \cite{morton_solid-state_2008,pla_high-fidelity_2013,madzik_precision_2022,noiri_fast_2022,reiner_high-fidelity_2024}. 
In either case of a hybrid electron-nuclear spin system, the coherence times of the electron and nuclear spins are highly dependent on the interactions between the two spin types.
Even if only the electron spin is used for a specific application, the interaction with background nuclear spins, e.g., \textsuperscript{13}C or \textsuperscript{29}Si with 1.1 and 4.7\% natural abundance, strongly influence the electron's relaxation \cite{khaetskii_electron_2002}.
Hence, isotope control, i.e. host crystals containing only a reduced or vanishing amount of nuclear isotopes with a magnetic moment, represents an efficient strategy to prolong electron coherence times \cite{witzel_electron_2010,tyryshkin_electron_2012,saeedi_room-temperature_2013,muhonen_storing_2014,bourassa_entanglement_2020}. 
Similar relaxation dependencies between electrons and nuclei have been studied in NMR, electron paramagnetic resonance (EPR) and DNP. 
In NMR and DNP, nuclear relaxation by nearby electrons (paramagnetic relaxation) has been studied extensively \cite{pell_paramagnetic_2019}.
In addition, the shortening effects of nuclear fluctuations, e.g. nuclear flip-flops, often called nuclear spin diffusion, reorientation of chemical (methyl) groups or tunneling, on the electronic phase memory or coherence times have been investigated in EPR \cite{Mims1972,zecevic_dephasing_1998,soetbeer_regularized_2021,jeschke_nuclear_2023,eggeling_quantifying_2023}.

In this work, we study a four-spin model consisting of two electrons and two nuclei to describe possible spin-transfer processes near a defect center (electron). 
We apply a lowest order Schrieffer-Wolff transformation to the spin system to calculate an effective Hamiltonian describing the electron-nuclear spin dynamics (Sec.~1). 
It is shown that, among the matrix elements describing polarization transfers, there is an electron-nuclear four-spin flip-flop potentially mediating nuclear spin diffusion close to paramagnetic defects.
To study if electron-nuclear four-spin flip-flops could explain recent experimental evidence of spin diffusion between hypershifted and bulk nuclei, we simulate HypRes-on experimental data \cite{chessari_role_2023}. 
To this end, the previously introduced one-compartment model of hyperpolarization \cite{von_witte_modelling_2023,von_witte_relaxation_2024} is extended to two coupled compartments (Fig.~2 and Sec.~S3, Supplementary Material). Simulation results suggest similar scaling of DNP injection by triple spin flips and inter-compartment coupling with applied MW power in agreement with the hypothesis that electron-nuclear four-spin flip-flops in DNP cause spin transport from hypershifted to bulk nuclei.

\section{Two-electron two-nucleus spin system} \label{sec:EffectiveHamiltonian}

The Hamiltonian of a two-electron two-nucleus spin system in the laboratory frame assuming identical frequencies for the two nuclei ($\omega_{n,1}=\omega_{n,2}=\omega_\mathrm{n}$) and allowing for different electron frequencies (i.e., due to g-anisotropy or two different radicals; $\omega_{\mathrm{e},a}=\omega_\mathrm{e}+\Delta \omega_{\mathrm{e},a}$ and $\omega_{\mathrm{e},b}=\omega_\mathrm{e}+\Delta \omega_{\mathrm{e},b}$) is given by
\begin{align}\label{eq:Hamiltonian}
    \mathbf{H} = \omega_{\mathrm{e},a} S_a^z + \omega_{\mathrm{e},b} S_b^z + &\Vec{S_a} \mathbf{D}_\mathrm{ee} \Vec{S_b} +  \Vec{S_\epsilon} \mathbf{A}_{\epsilon i} \Vec{I_i} + ... \nonumber \\
    + \omega_\mathrm{n} (I_1^z+I_2^z) + &\Vec{I_1} \mathbf{d}_\mathrm{nn} \Vec{I_2} 
\end{align}
with the Einstein sum convention of double occurring indices. 
For the hyperfine coupling $\mathbf{A}$, electron dipolar coupling $\mathbf{D}_\mathrm{ee}$ and nuclear dipolar coupling $\mathbf{d}_\mathrm{nn}$, the following general form with a quantization along the $z$-axis is used

\begin{subequations}
\begin{align}
   \mathbf{A}_{\epsilon i} &= 
   \begin{pmatrix} 
   A_{\epsilon i}^{++} & A_{\epsilon i}^{+-} & A_{\epsilon i}^{+z} \\
   A_{\epsilon i}^{-+} & A_{\epsilon i}^{--} & A_{\epsilon i}^{-z} \\  
   A_{\epsilon i}^{z+} & A_{\epsilon i}^{z-} & A_{\epsilon i}^{zz}
   \end{pmatrix} 
   \\ \mathbf{D}_\mathrm{ee} &= 
   \begin{pmatrix} 
   D^{++} & D^{+-} & D^{+z} \\
   D^{-+} & D^{--} & D^{-z} \\  
   D^{z+} & D^{z-} & D^{zz}
   \end{pmatrix}
   \\  \mathbf{d}_\mathrm{nn} &= 
   \begin{pmatrix} 
   d^{++} & d^{+-} & d^{+z} \\
   d^{-+} & d^{--} & d^{-z} \\  
   d^{z+} & d^{z-} & d^{zz}
   \end{pmatrix}  ~~~~~~ .
\end{align}
\end{subequations}

Spin interactions are often written in the $xyz$ rather than in the $+-z$ basis as used in our notation. 
For the translation between the two, we find 
\begin{subequations} \label{eq:conversionFromXYZ}
    \begin{align}
        A^{z+} = A^{+z} = (A^{z-})^* = (A^{-z})^* &= \frac{A^{xz} + iA^{yz}}{2} \label{eq:Az+} \\
        A^{++} = (A^{--})^* &= \frac{A^{xx} - A^{yy} + 2iA_{xy} }{4} \\
        A^{+-} = A^{-+} &= \frac{A^{xx} + A^{yy}}{4} \\
        A^{zz} &= A^{zz} ~~~~ .
    \end{align}
\end{subequations}

The hyperfine coupling consists in general of a dipolar part and an isotropic Fermi-contact part with the latter taking a diagonal form in the $xyz$ basis ($\bar{a}\mathbb{I}_{xyz}$). 
The Fermi-contact term is important for s-like electron orbitals with short electron-nuclear distances or delocalized electrons, e.g., phosphorous dopants in silicon or quantum dots. 
The Fermi-contact term can exceed several hundred MHz while the dipolar hyperfine coupling for an \textsuperscript{1}H nuclear spin 3 or 5\,\AA\, away from the (point charge) electron is around 3 or 0.6\,MHz, respectively.
Since only the $A^{zz}$, $A^{+-}$ and $A^{-+}$ terms from Eq.~\eqref{eq:conversionFromXYZ} depend on the Fermi-contact hyperfine coupling, only these might exceed a few MHz in cases with the dipolar hyperfine coupling of a few MHz or less.
If the wave functions of the two electrons overlap, this results in an isotropic exchange interaction $\mathbf{J}_\mathrm{ee}$, similar to the Fermi-contact part.
Hence, an eventual $\mathbf{J}_\mathrm{ee}$ can be absorbed into $\mathbf{D}_\mathrm{ee}$.

\begin{figure*}[!htb]\centering
	\includegraphics[width=\linewidth]{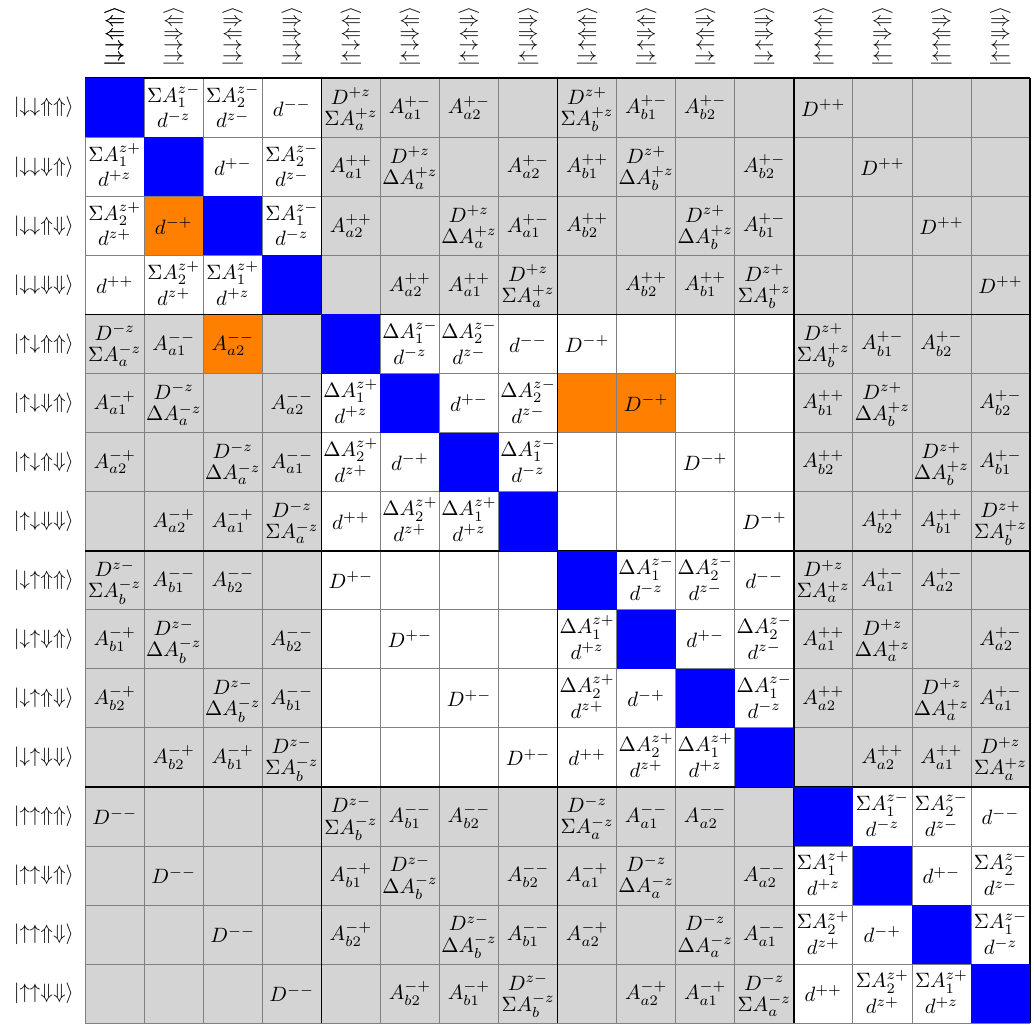}
	\caption{\textbf{Two-electron two-nucleus spin system involving electron and nuclear dipolar as well as hyperfine couplings.} 
    The diagonal matrix elements (colored in blue) form $\mathbf{H}_0$ in the following and are omitted for clarity with an example given by Eq.~\eqref{eq:energyLevel}. 
    Off-diagonal elements (without sign) compose $\mathbf{V}$. $\mathbf{V}$ consists of an inner part with quadratic blocks along the diagonal (colored in white) which conserve the total electron quantum number, while the outer part involves net electron flips (shaded in gray).
    More details are given in the main text.
    For elements highlighted in orange, the effective Hamiltonian elements will be discussed in detail in the main text.}
	\label{fig:Fig1}
\end{figure*}

Throughout this work, we assume a positive gyromagnetic ratio $\gamma$ ($<46$\,MHz/T) for nuclear spins, resulting in opposite preferred spin directions for electrons ($\gamma_e\approx-28$\,GHz/T) and nuclei.
We define $\omega_\mathrm{n}=|\omega_{n,L}|=|-\gamma_\mathrm{n} B_0|$ and $\omega_\mathrm{e}=|\omega_{\mathrm{e},L}|=|-\gamma_e B_0|$ such that both are positive frequencies.
The different sign of the gyromagnetic ratios of electrons and nuclei leads to opposite magnetic quantum numbers defining the ground state i.e. for electrons the spin-down state ($m_S=-1/2$), denoted by $\downarrow$, has lower energy than the spin-up state ($m_S=+1/2$), denoted by $\uparrow$.
For nuclei, this is inverted with $m_I=+1/2$ (spin-up, denoted by $\Uparrow$) lower in energy than $m_I=-1/2$ (spin-down, denoted by $\Downarrow$).
This notation with $\uparrow$, $\downarrow$, $\Uparrow$ and $\Downarrow$ is more common in physics compared to the $\alpha$ ($m_I=+1/2$) and $\beta$ ($m_I=-1/2$) notation in magnetic resonance (NMR and EPR) and offers in the current case the advantage that electron and nuclear spins are easy to distinguish.

The 16-by-16 matrix of the four-spin model of Eq.~\eqref{eq:Hamiltonian} is sketched in Fig.~2 and can be rewritten as $\mathbf{H} = \mathbf{H}_0 + \mathbf{V} = \mathbf{H}_0 + \mathbf{V}_\mathrm{inner} + \mathbf{V}_\mathrm{outer}$.
$\mathbf{H}_0$ is the diagonal part of the Hamiltonian containing the electron and nuclear Zeeman energy and the $zz$ elements of the hyperfine and dipolar couplings with an example for the diagonal energy levels given by
\begin{align}
        E_3&=E_{\downarrow\downarrow\Uparrow\Downarrow} = -\frac{1}{2}(\omega_\mathrm{e}+\Delta\omega_{\mathrm{e},a}) - \frac{1}{2}(\omega_\mathrm{e}+\Delta\omega_{\mathrm{e},b}) + D^{zz} \nonumber \\ 
        &- \left(\frac{1}{2}\omega_\mathrm{n}+A_{a1}^{zz}+A_{b1}^{zz}\right) + \left(\frac{1}{2}\omega_\mathrm{n}+A_{a2}^{zz}+A_{b2}^{zz}\right) - d^{zz} \label{eq:energyLevel}~~~~~.
\end{align}
The 16 energy levels can be grouped into four-spin-up and spin-down parallel electron states each as well as eight (two times four) antiparallel electron states as indicated by white shadings in Fig.~1.
These quadratic blocks around the diagonal form $\mathbf{V}_\mathrm{inner}$ and are characterized by conservation of the total electron quantum number ($m_S=m_{S_a}+m_{S_b}$).
The ten remaining 4-by-4 blocks form $\mathbf{V}_\mathrm{outer}$ and result in a change of the total electron quantum number $m_S$ (net electron flip).
Within the diagonal 4-by-4 blocks, the nuclear dipolar and hyperfine couplings can cause transitions between the nuclear spin states. 
All matrix elements outside the diagonal 4-by-4 blocks involve electron flips or flip-flops either through hyperfine or electron dipolar couplings.

For intermediate (tens to hundreds of mT depending on the other contributions to $\mathbf{H}$) to high magnetic fields, the electron Zeeman energy is much larger than all spin-spin interactions. 
In such a case, the parallel and antiparallel electron states are energetically well separated. 
Regarding the other contributions to $\mathbf{H}$: The electron frequency offsets can be hundreds of MHz at higher fields for defects/ radicals with significant g-factor anisotropy. 
Nuclear Larmor frequencies $\omega_\mathrm{n}$ can vary between a few MHz for low-$\gamma$ nuclei and intermediate fields of up to several hundred MHz at higher fields for high-$\gamma$ nuclei.
Hyperfine couplings can span from a few kHz for rather distant nuclei to hundreds of MHz for electrons localized on a specific atom although in this case most of the coupling would arise from the isotropic Fermi-contact part which only affects $A^{+-}$, $A^{-+}$ and $A^{zz}$ (see above and Eq.~\eqref{eq:conversionFromXYZ}).
Electron dipolar couplings can range into several MHz for close-by electrons.
Electron exchange couplings $\mathbf{J}_\mathrm{ee}$, eventually absorbed into $\mathbf{D}_\mathrm{ee}$, can range much higher and the same arguments as for the Fermi-contact part of the hyperfine coupling would apply.
Nuclear dipolar couplings range from hundreds of Hz for low-$\gamma$, low-abundance nuclei to several kHz for high-$\gamma$, high-abundance nuclei like \textsuperscript{1}H.

To simplify the notation in the following, we define 

\begin{subequations}\label{eq:rewriteHyperfine}
\begin{align}
    \Sigma A_i^{z,+/-/z} &= A_{ai}^{z,+/-/z} + A_{bi}^{z,+/-/z}\\ 
    \Delta A_i^{z,+/-/z} &= A_{ai}^{z,+/-/z} - A_{bi}^{z,+/-/z} \\
    \Sigma A_\epsilon^{+/-/z,z} &= A_{\epsilon 1}^{+/-/z,z} + A_{\epsilon 2}^{+/-/z,z}\\
    \Delta A_\epsilon^{+/-/z,z} &= A_{\epsilon 1}^{+/-/z,z} - A_{\epsilon 2}^{+/-/z,z}\\
    \Delta \omega_\mathrm{e} &= \Delta \omega_{\mathrm{e},a} - \Delta \omega_{\mathrm{e},b} ~~~~.
\end{align}
\end{subequations}
The commas in the superscript of Eq.~\eqref{eq:rewriteHyperfine} are only written here for clarity and will be omitted in the following, i.e. $A_{\epsilon i}^{z,z} = A_{\epsilon i}^{zz}$.

Nuclear spin transitions within the diagonal 4-by-4 blocks are suppressed by the separation between the energy levels, e.g.
\begin{subequations}
    \begin{align}
        E_2 - E_3 &= \Sigma A_1^{zz} - \Sigma A_2^{zz} = \Delta A_a^{zz} + \Delta A_b^{zz} \\ 
        E_4 - E_3 &=  \omega_\mathrm{n} + \frac{\Sigma A_1^{zz}}{2} + \frac{d^{zz}}{2} 
    \end{align}
\end{subequations}
unless the $\omega_\mathrm{n}$ matches the hyperfine couplings or the hyperfine couplings would be symmetric, causing an energy level degeneracy.
Outside of the diagonal 4-by-4 blocks, all elements of $\mathbf{V}_\mathrm{outer}$ are much smaller than $\omega_\mathrm{e}$.
For a large enough electron energy offset $\Delta\omega_\mathrm{e} = \Delta \omega_{\mathrm{e},a} - \Delta \omega_{\mathrm{e},b}$, electronic flip-flops by $D^{+-}$ and $D^{-+}$ are suppressed unless an energy level degeneracy would occur for a special combination of electron energy offsets and hyperfine couplings.
Therefore, for large enough magnetic fields, hyperfine couplings and electron energy offsets, the spin dynamics in the 4-spin system is suppressed to first order. 

The Schrieffer-Wolff transformation generates an effective Hamiltonian by perturbative diagonalization if an energy gap exists that is much bigger than the off-diagonal matrix elements \cite{bravyi_schriefferwolff_2011}.
The effective Hamiltonian is calculated by $\mathbf{H}^\mathrm{eff}=e^\mathbf{S} \mathbf{H} e^{-\mathbf{S}}$ where $\mathbf{S}$ is given in lowest order by $\mathbf{V} + [\mathbf{S},\mathbf{H}_0] = 0$ (off-diagonal $\mathbf{V}$ and diagonal $\mathbf{H}_0$) such that $\mathbf{H}^\mathrm{eff}=\mathbf{H}_0 + 1/2[\mathbf{S},\mathbf{V}] + \mathcal{O}(\mathbf{V}^3)$.
In this work, we apply two separate Schrieffer-Wolff transformations to $\mathbf{V}_\mathrm{inner}$ and $\mathbf{V}_\mathrm{outer}$ although this is identical to applying it to $\mathbf{V}$ with the current structure as discussed in Sec.~S1, Supplementary Material.
Applying the Schrieffer-Wolff transformation separately to $\mathbf{V}_\mathrm{inner}$ and $\mathbf{V}_\mathrm{outer}$ ensures that the off-diagonal perturbation is smaller than the energy gap of the diagonal as the $A_{\epsilon i}^{+-}$ elements can exceed $\omega_\mathrm{n}$ but not $\omega_\mathrm{e}$.
However, $\Delta\omega_\mathrm{e}$ might not always be larger than $D^{+-}$ causing a breakdown of the Schrieffer-Wolff-transformation.
Assuming that the Schrieffer-Wolff transformation can be applied, an example for the renormalized energies in the effective Hamiltonian is given by

\begin{align}
    E^\mathrm{eff}_{3} &= E_3 - \frac{1}{2} \left[\frac{4d^{-+}d^{+-}}{\Delta A_a^{zz} + \Delta A_b^{zz}} + \frac{(\Sigma A_1^{z-} + d^{-z})(\Sigma A_1^{z+} + d^{+z})}{\Sigma A_1^{zz} + d^{zz} + 2\omega_\mathrm{n}}\right.   \nonumber \\
    &- \frac{(\Sigma A_2^{z-} - d^{z-})(\Sigma A_2^{z+} - d^{z+})}{\Sigma A_2^{zz} - d^{zz} + 2 \omega_\mathrm{n}}  \left.\vphantom{\frac{1}{\Delta A_1^{zz}}}\right] + \mathcal{O}\left(\omega_\mathrm{e}^{-1}\right) \label{eq:energyLevelTransformed}
\end{align} 
The other energies can be calculated with a Mathematica notebook as shown in Sec.~S6, Supplementary Materials.

In the following, we will discuss several matrix elements of $\mathbf{H}^\mathrm{eff}$.
Specifically, we will look into different processes ranging from no electron flips (just their passive presence) over single electron flips (single-quantum transition) to  electron flip-flops (zero-quantum transition).

\subsection{Electron-mediated spin diffusion (EMSD)}
This process describes a nuclear flip-flop in the passive presence of the electrons, e.g., $\ket{\downarrow\downarrow\Downarrow\Uparrow} \rightarrow \ket{\downarrow\downarrow\Uparrow\Downarrow}$ connecting two states that are separated by an energy on the order of the nuclear Larmor frequency.

\begin{align}
    H^\mathrm{eff}_{3,2} &= \frac{1}{4}\sum_i\frac{(\Sigma A_1^{z+} - d^{+z})(\Sigma A_2^{z-} - d^{z-})}{\Sigma A_i^{zz} - d^{zz} + 2\omega_\mathrm{n}}  \label{eq:matrixElementEMSD}   \\
    &- \frac{1}{4}\sum_i\frac{(\Sigma A_1^{z+} + d^{+z})(\Sigma A_2^{z-} + d^{z-})}{\Sigma A_i^{zz} + d^{zz} + 2\omega_\mathrm{n}}  \nonumber \\
    &+\sum_{\epsilon\neq \kappa} \sum_{i\neq j}\left[\frac{2A_{\epsilon 1}^{++} A_{\epsilon2}^{--}}{A_{\kappa i}^{zz} - A_{\epsilon j}^{zz} - d^{zz} + D^{zz} - 2\omega_{\mathrm{e},\epsilon} + 2\omega _n} \right. \nonumber \\
    &+\frac{2A_{\epsilon 1}^{-+} A_{\epsilon 2}^{+-}}{A_{\epsilon i}^{zz} - A_{\kappa j}^{zz} - d^{zz} + D^{zz} - 2\omega_{\mathrm{e},\epsilon} - 2\omega _n} \left.\vphantom{\frac{1}{\Delta A_1^{zz}}}\right] \nonumber \\
    &\approx -\sum_{\epsilon} \frac{A_{\epsilon 1}^{++} A_{\epsilon2}^{--}+A_{\epsilon 1}^{-+} A_{\epsilon 2}^{+-}}{ \omega_{\mathrm{e},\epsilon}}\nonumber
\end{align}
The first two terms are mostly negligible as these cancel out for $\mathbf{d}\ll \omega_\mathrm{n},~\Sigma A_i^{+/-/z}$. 
The latter two terms describe electron-mediated spin diffusion, discussed as a limiting process in quantum dots \cite{liu_control_2007,gong_dynamics_2011} and quantum computing \cite{monir_impact_2024}.   
Considering the tensorial nature of the hyperfine coupling, only parts of the hyperfine coupling contribute to the nuclear-nuclear spin flip-flops instead of the full coupling.
Furthermore, two different pathways exist, either through $A_{b1}^{++} A_{b2}^{--}$ or $A_{b1}^{-+} A_{b2}^{+-}$.
Since electron-mediated spin diffusion scales as hyperfine coupling squared divided by electron Larmor frequency, its frequency is in the range of Hz.
Thus, it is a low frequency, broad non-resonant matrix element because the denominator is dominated by $\omega_\mathrm{e}$.
In contrast, all the following discussed matrix elements are (partially) resonant and only become relevant over a rather narrow frequency interval.
    
In DNP this term might transport polarization between hypershifted nuclei under all conditions.
The magnitude of the rate constant will depend, in a perturbation treatment \cite{ernst_chapter_1998} on the square of the coupling term in the Hamiltonian and the intensity of the zero-quantum line at frequency zero.

\subsection{Electron-nuclear flip-flip or flip-flop}

This process describes a joint one-nucleus-one-electron flip-flip (double-quantum, DQ) or flip-flop (zero-quantum, ZQ) process, e.g., $\ket{\downarrow\downarrow\Uparrow\Downarrow} \rightarrow \ket{\uparrow\downarrow\Uparrow\Uparrow}$ with energies separated on the order of the electron Larmor frequency.
\begin{align}
    H^\mathrm{eff}_{5,3} &= \frac{1}{2} \left[\frac{\left(\Delta A_a^{+z} - D^{+z}\right)\left(\Delta A_2^{z+} + d^{z+}\right)}{\Delta A_2^{zz} + d^{zz} - 2\omega_\mathrm{n}} \right. \label{eq:matrixElementSE} \\ 
    &- \frac{\left(\Sigma A_a^{+z} - D^{+z}\right)\left(\Sigma A_2^{z+} - d^{z+}\right)}{\Sigma A_2^{zz} - d^{zz} + 2\omega_\mathrm{n}} + \frac{2A_{a1}^{+-}d^{++}}{\Sigma A_a^{zz} - \Sigma A_b^{zz} - 4\omega_\mathrm{n}} \nonumber \\
    &+ \frac{2A_{b2}^{++}D^{+-}}{\Sigma A_a^{zz} - \Sigma A_b^{zz} + \Delta\omega_\mathrm{e}}  - \frac{2A_{a1}^{++}d^{-+}}{\Delta A_a^{zz} + \Delta A_b^{zz}} \left.\vphantom{\frac{\left(\Sigma A\right)}{\Delta A_1^{zz}}}\right] + \mathcal{O}\left(\omega_\mathrm{e}^{-1}\right) \nonumber
\end{align}

The probability of such a transition driven by the Hamiltonian of Eq. \eqref{eq:matrixElementSE} is negligible but under MW irradiation it becomes important for solid effect (SE) DNP.
In quantum information processing, this matrix element describes an electron-nucleus two qubit gate that can be used to initialize the nuclear qubits \cite{reiner_high-fidelity_2024}.

Since the electron nuclear flip-flop only requires a two spin system (one electron, one nucleus), we can simplify Eq.~\eqref{eq:matrixElementSE} to 
\begin{align}
     H^\mathrm{eff,1e1n}_{2,3} &= - \frac{A^{+z} A^{z+}}{4} \left(\frac{1}{A^{zz} - 2\omega_n} + \frac{1}{A^{zz} + 2\omega_n} \right) + \mathcal{O}\left(\omega_\mathrm{e}^{-1}\right)\nonumber \\ 
     &= - \frac{(A^{z+})^2}{2} \frac{A^{zz}}{(A^{zz})^2 - \left(2\omega_\mathrm{n}\right)^2} + \mathcal{O}\left(\omega_\mathrm{e}^{-1}\right) \label{eq:matrixElementSE_1e1n}
\end{align}
where we used $A^{z+} = A^{+z}$ from Eq.~\eqref{eq:Az+}.
This polarization transfer might be responsible for the observed near-unity polarization in optically pumped quantum dots \cite{millington-hotze_approaching_2024}.
 
If MW irradiation is applied to the electron-nuclear spin system in DNP, this transition would be called the solid effect (SE).
MW irradiation is tuned to $\omega_\mathrm{e}-\omega_\mathrm{n}$ (ZQ) or $\omega_\mathrm{e}+\omega_\mathrm{n}$ (DQ) to create a nuclear hyperpolarization \cite{Wenckebach,thankamony_dynamic_2017}. 
In Sec.~S5, solid effect (SE) and resonant mixing (RM) are derived in a one-electron-one-nucleus spin system with MW irradiation, underlining the ability of the used Schrieffer-Wolff approach to describe the known and unknown processes in electron-nuclear spin systems.

\subsection{Triple spin flips} \label{sec:tripleSpinFlip}

This process describes a joint electronic flip-flop and a nuclear flip, e.g., $\ket{\downarrow\uparrow\Uparrow\Uparrow} \rightarrow \ket{\uparrow\downarrow\Downarrow\Uparrow}$ of two states that are separated by energies on the order of the nuclear Larmor frequency.
\begin{align} 
    H^\mathrm{eff}_{6,9} &= -\frac{D^{+-}}{2}\left[\frac{\Delta A_1^{z-} + d^{-z}}{\Sigma A_a^{zz} - \Sigma A_b^{zz} + 2\Delta\omega_\mathrm{e}} - \frac{\Delta A_1^{z-} - d^{-z}}{\Delta A_a^{zz} - \Delta A_b^{zz} - 2\Delta\omega_\mathrm{e}} \right.  \nonumber \\
    &+ \frac{\Delta A_1^{z-} + d^{-z}}{\Delta A_1^{zz} + d^{zz} - 2\omega_\mathrm{n}} - \frac{\Delta A_1^{z-} - d^{-z}}{\Delta A_1^{zz} - d^{zz} + 2\omega_\mathrm{n}} \left.\vphantom{\frac{1}{\Delta A_1^{zz}}}\right] + \mathcal{O}\left(\omega_\mathrm{e}^{-1}\right) \label{eq:matrixElementTripleSpinFlip}
\end{align}
Triple spin flips with an electronic flip-flop and the flip of a hyperfine coupled nucleus (flip-flop-flip transition) are the basis for cross effect (CE) and thermal mixing (TM) DNP \cite{wenckebach_dynamic_2019,wenckebach_dynamic_2019-1,wenckebach_electron_2021}.
For the CE, DNP is efficient if $\Delta\omega_\mathrm{e}\simeq \pm \omega_\mathrm{n}$, creating an energy level degeneracy with the energy difference of the electrons available to flip a nuclear spin \cite{Wenckebach,Hovav2012,thankamony_dynamic_2017,kundu_dnp_2019}. 
Ignoring the hyperfine couplings in the denominator and all nuclear dipolar couplings gives in good approximation
\begin{align} 
    H^\mathrm{eff}_{6,10} \approx -\frac{D^{+-} \Delta A_1^{z-}}{2}\left[\frac{\omega_\mathrm{n}-\Delta\omega_\mathrm{e}}{\Delta\omega_\mathrm{e} \omega_\mathrm{n}}\right]
    \overset{\Delta\omega_\mathrm{e}=-\omega_\mathrm{n}}{=} -\frac{D^{+-} \Delta A_1^{z-}}{\omega_\mathrm{n}}
\end{align}
for the polarization transfer by triple spin flips if the matching condition is fulfilled, reproducing the triple spin flip result from Ref.~\cite{wenckebach_dynamic_2019}. 
To generate hyperpolarization, MW irradiation is required to generate a population imbalance between the two electrons involved in the CE process \cite{Hovav2012,thankamony_dynamic_2017,kundu_dnp_2019}. 
Such a Hamiltonian will not only drive heteronuclear polarization transfer but also homonuclear zero-quantum polarization transfer on the electrons which is mechanistically very similar to the MIRROR \cite{scholz_mirror_2008,scholz_mirror-cp_2009} experiment.

\subsection{Electron-nuclear four-spin flip-flops} 

This process describes a joint electron flip-flop and nuclear flip-flop, e.g., $\ket{\downarrow\uparrow\Uparrow\Downarrow} \rightarrow \ket{\uparrow\downarrow\Downarrow\Uparrow}$ where the energy difference of the two states is on the order of the difference frequency of the two electrons.
\begin{align}
    H^\mathrm{eff}_{6,10} &= -D^{+-}d^{-+}\left[\frac{1}{\Delta A_a^{zz} - \Delta A_b^{zz} + \Delta\omega_\mathrm{e}} \right. \label{eq:matrixElementFourSpin} \\
    &+ \frac{1}{\Delta A_a^{zz} - \Delta A_b^{zz} - \Delta\omega_\mathrm{e}} + \frac{2}{\Delta A_a^{zz} - \Delta A_b^{zz}} \left.\vphantom{\frac{1}{\Delta A_1^{zz}}}\right]  + \mathcal{O}\left(\omega_\mathrm{e}^{-1}\right) \nonumber \\
\end{align}


These terms have similarities to the cross-effect transition discussed above as they combine an electron flip-flop with a nuclear transition.
However, in this case the nuclear transition is a flip-flop mediated by the dipolar interaction (in total: flip-flop-flip-flop). 
These three terms scale as $D^{+-} d^{-+} \approx \mathcal{O}$(MHz $\cdot$ kHz) and become resonant if the nuclei have either identical hyperfine couplings $\Delta A_a^{zz} - \Delta A_b^{zz} \simeq 0$ or if the difference matches the electron frequency difference  $\Delta A_a^{zz} - \Delta A_b^{zz} \simeq \pm \Delta \omega_\mathrm{e}$.
The former case suggests that spin diffusion between spins with identical hyperfine coupling can be faster than bulk spin diffusion.
The latter case describes electron-nuclear four-spin flip-flops as an energy conserving process, independent of the interaction with MW photons or phonons causing nuclear spin and spectral diffusion close to the electron.

Similar to triple spin flips (cf. Sec.~1.3), this electron-nuclear four-spin flip-flop can drive homonculear zero-quantum polarization transfer on the electrons but in addition also on the nuclei and electron-nuclear zero- and double-quantum polarization transfer. 
However, the magnitude will be much smaller since the magnitude is determined by the product of the electron and the nuclear dipolar coupling while the cross-effect Hamiltonian contains the product of the electron dipolar coupling with the hyperfine coupling to the nuclei.

For a thermal electron polarization close to unity at liquid helium temperatures and few Tesla magnetic fields, electron-nuclear four-spin flip-flops are suppressed as few electron pairs with opposite polarization are available. 
However, this changes upon MW irradiation reducing the electron polarization (cf. Eq.~\eqref{eq:Torrey_model_simplified}) similar to the probability for triple spin flips to occur. 
        
Electron-nuclear four-spin flip-flops can be understood as a heteronuclear cross effect with the two nuclei having different resonance frequencies (in this case due to different hyperfine couplings).
A heteronuclear cross effect has been investiagated in Ref.~\cite{shimon_simultaneous_2015}.
A heteronuclear cross effect would explain the equilibration of polarization in samples containing more than one NMR-active nuclei (at least locally close to the electron) that has been usually described through a spin bath approach \cite{jahnig_spin-thermodynamic_2019,rodin_quantitative_2023} adapted from a classical description of thermal mixing. 

If dopant clusters with eventually multiple electrons shall be used for quantum information processing \cite{buch_spin_2013}, this term might limit the coherence and lifetime ($T_1$) of nuclear spin qubits.

\subsection{Discussion of the effective Hamiltonian}

In the above equations, we did not assume any particular symmetry of the spin interactions besides the existence of an intermediate to strong magnetic field, creating a quantization axis. 
Furthermore, we did not include any type of microwave (MW) irradiation unless explicitly stated.
Thus, all these processes occur in thermalized systems as often encountered in quantum information processing.
In DNP, for large enough electron frequency differences $\Delta\omega_\mathrm{e}$, MW irradiation at one of the electron frequencies causes (damped) Rabi oscillations  while the other electron remains unaffected.
Thus, MW irradiation creates a polarization difference between the two electrons available for transfer to nuclear spins either as (cross effect) triple spin flips or electron-nuclear four-spin flip-flops. 
The dependence of the triple spin flip rate and nuclear spin flip-flops close to the electron with MW power (electron saturation) is indirectly investigated in the next section. 

The above effective two-electron two-nucleus model is limited to processes involving two interactions at most as only the lowest order of the Schrieffer-Wolff transformation was used. 
Higher order Schrieffer-Wolff transformations \cite{bravyi_schriefferwolff_2011} could resolve this. 
Terms of the form $\mathbf{DAA}$ might show up, e.g., for four-spin cross effect \cite{shimon_simultaneous_2015}.
In the presented model, the transition matrix elements for such transitions are non-zero but a correct description is not possible as three interaction processes cannot be described with a lowest order Schrieffer-Wolff transformation.
Extension to higher order transformations in the lab frame might result in very long expressions.
Thus, effective Hamiltonians in the rotating frame can be used to simplify the result while retaining terms not scaling with $\mathcal{O}\left(\omega_\mathrm{e}^{-n}\right)$, $n\geq1$.

We tested the Schrieffer-Wolff transformation of the two-electron two-nucleus system in the electron rotating frame. 
For the electron-mediated spin diffusion (EMSD, Eq.~\eqref{eq:matrixElementEMSD}), triple spin flips (Eq.~\eqref{eq:matrixElementTripleSpinFlip}) and electron-nuclear four-spin flip-flops (Eq.~\eqref{eq:matrixElementFourSpin}), we found the same expressions as in the lab frame except from the truncated $\mathcal{O}\left(\omega_\mathrm{e}^{-1}\right)$ terms.
For the single electron flip processes, e.g. electron-nuclear flip-flip or flip-flops, the rotating frame transformation adds additional terms that were scaling in the lab frame with $\omega_{\mathrm{e},\epsilon}^{-1}$ and in the rotating frame will scale with $\Delta \omega_{\mathrm{e},\epsilon}^{-1}$.
Other terms scaling as $\mathcal{O}\left(\omega_\mathrm{e}^{-1}\right)$ in the lab frame are truncated.

Future studies might combine the effective Hamiltonian with relaxation or describe the coupling with phonons or (cavity) photons in a quantized approach, i.e. through creation and annihilation operators.
This might lead to the discovery of new quantum many-body effects causing hyperpolarization.

So far, we only discussed theoretically possible coherent polarization transfers.
While electron-mediated spin diffusion has been studied in quantum dots, electron-nuclear flip-flops and triple spin flips in DNP as described above, electron-nuclear four-spin flip-flops have not been studied experimentally. 
Since both DNP by triple spin flips and electron-nuclear four-spin flip-flops rely on an electronic flip-flops, both should scale similar with electron polarization saturation by MW irradiation.
By simulating the HypRes-on data from Ref.~\cite{chessari_role_2023} below, we show that the DNP injection (hyperpolarization rate constant) and the transport of hyperpolarization from the hypershifted to bulk spins both follow the saturation of the electron polarization with MW power, supporting the significance of electron-nuclear four-spin flip-flops in DNP.

\section{Electron saturation dependence of the spin transport between hypershifted and bulk nuclei} \label{sec:twoCompartmentModel}

In the following, we will apply a two-compartment model of hyperpolarization as sketched in Fig.~2 and discussed in detail in Sec.~S6, Supplementary Material, to the HypRes-on data from Ref.~\cite{chessari_role_2023}.
This approach enables us to quantify the increase in coupling between the hypershifted and bulk spins by microwave (MW) irradiation which is considered to describe the spin diffusion close to the electron.
The sample used in these experiments is TEMPOL in \textsuperscript{1}H glassy matrices in which DNP is commonly attributed to triple spin flips.
For triple spin flips, the polarization difference between the two electron spins leads to the nuclear hyperpolarization. 
If the coupling between the hypershifted and the bulk spin compartments shows the same dependence on the electron saturation by MW irradiation, this would be a strong indication that electron-nuclear four-spin flip-flops with the dependence on the electron spin polarization difference are the main process for nuclear flip-flops (spin diffusion) close to electrons. 

\begin{figure}[tb]\centering
	\includegraphics[width=0.6\linewidth]{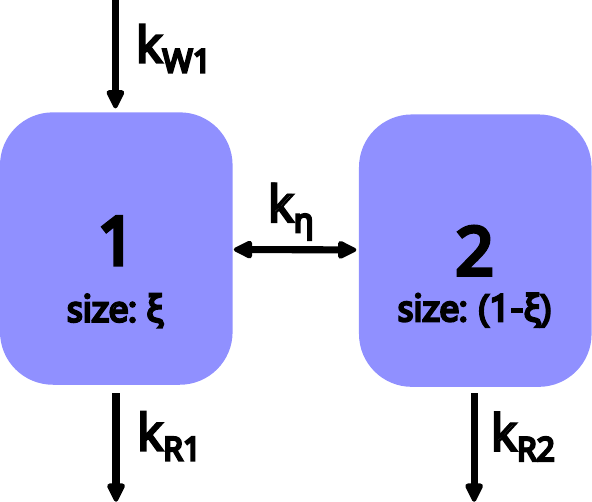}
	\caption{\textbf{Coupled two-compartment model of hyperpolarization.} 
    The injection of polarization into the first compartment is given by $k_{W1}$. 
    The coupling to an eventual second compartment is given by $k_\eta$. 
    The polarization in the two compartments decays with $k_{\mathrm{R1}}$ and $k_{\mathrm{R2}}$.
    $\xi$ describes the relative size of the first compartment and $(1-\xi)$ that of the second.}
	\label{fig:BasicIdea}
\end{figure}

In MW-on HypRes or HypRes-on experiments, the sample was first hyperpolarized before broadband saturation pulses were applied to saturate the bulk nuclear polarization. 
During the saturation, the MW was switched to the frequency of the other DNP lobe, reversing the sign of the DNP injection, and eventually its power was adjusted \cite{chessari_role_2023}. 
This creates two competing polarization dynamics: First, the positive polarization from the build-up is still stored in the hypershifted spins close to the electrons and diffuses into the bulk, with a time constant given by the inter-compartment coupling term. 
Second, the negative DNP process injects hyperpolarization with the opposite sign first into the unobservable (hypershifted) first compartment and then into the bulk spins. 
We note the similar polarization maximum during the HypRes-on experiment for all MW powers (cf. Fig.~S1), possibly suggesting a similar scaling of DNP injection and inter-compartment coupling (spin diffusion from hypershifted to bulk spins) with MW power. 

A system compromised of the RF ''invisible'' hypershifted spins and the bulk spins can be described by a two-compartment model. 
To model this, we extend the previously introduced one-compartment rate equation model \cite{von_witte_modelling_2023,von_witte_relaxation_2024} to a coupled two-compartment model: Fig.~2 sketches the basic idea of the model with DNP/ hyperpolarization injection (creation) only into the first compartment (relative size $\xi$), a coupling between the two compartments and a separate relaxation rate constant for each compartment. For simplicity and in analogy to the one-compartment model \cite{von_witte_modelling_2023}, we ignore a thermal equilibrium polarization as nuclear polarization enhancements exceeding 100 can be achieved in many materials, rendering the thermal polarization small compared to typical measurement uncertainties. 
The details of the model can be found in Sec.~S3, Supplementary Materials.

The parameters of the best fits to the HypRes-on data with the two-compartment model as described in Eqs.~(S18), Supplementary Material, are shown in Fig.~3.
The fitting is insensitive to the relaxation rates of the two compartments owing to rather low polarization levels and short experimental durations and, hence, the relaxation rate constants are set to zero. This leaves the DNP injection into the first compartment $k_{W1}$ and the inter-compartment coupling $k_\eta$ as the remaining fit parameters.
More details on the simulations including the fits to the experimental data can be found in Sec. S4, Supplementary Materials. 

\begin{figure}[!tb]\centering 
	\includegraphics[width=\linewidth]{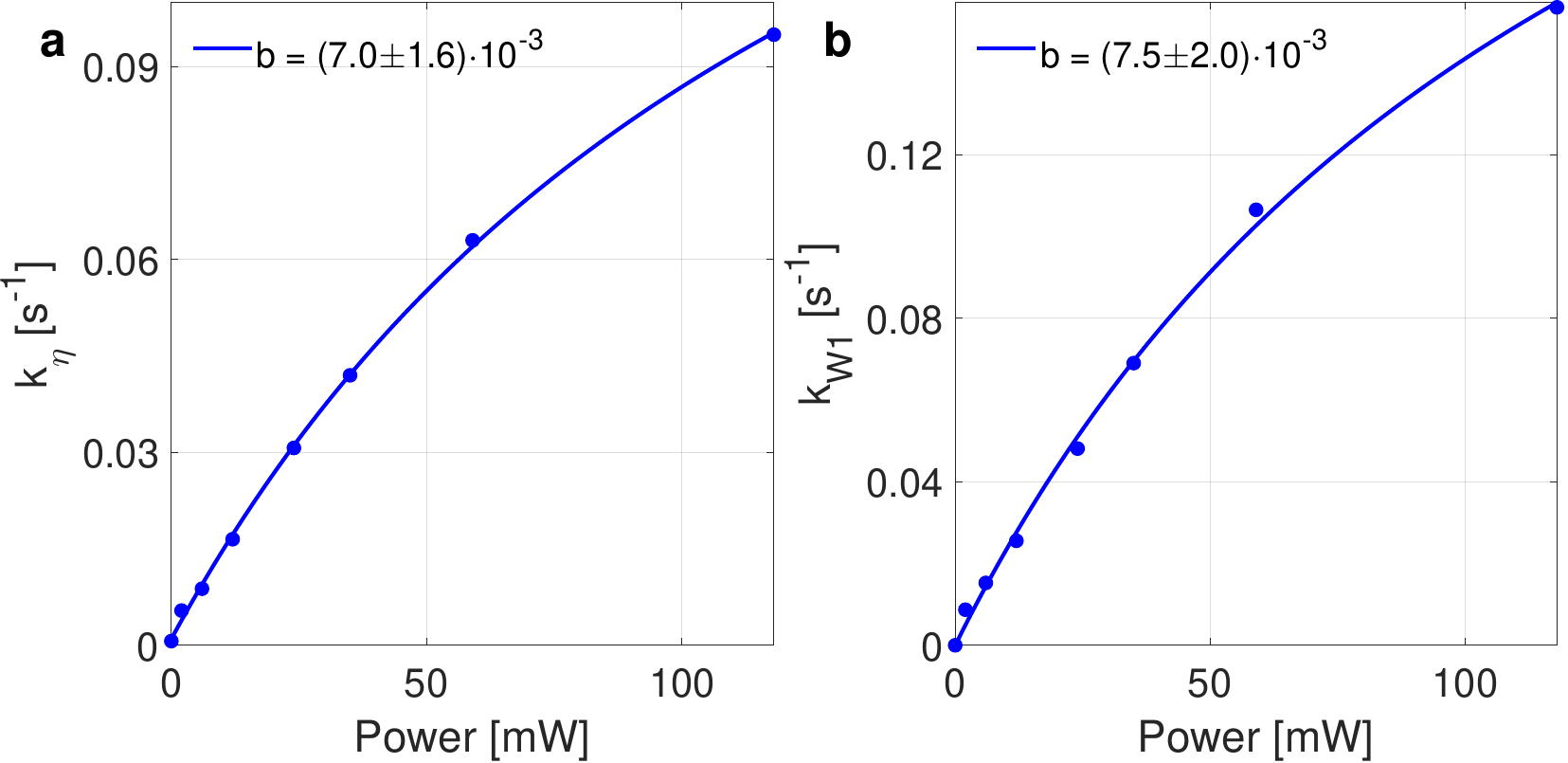}
	\caption{\textbf{Two-compartment modelling of HypRes-on data.} 
    Power dependence of the inter-compartment coupling constant $k_\eta$ \textbf{(a)} and injection parameter $k_{W1}$ \textbf{(b)} describing the experimental data from \cite{chessari_role_2023}. 
    The size of the first compartment was set to 7\% and the relaxation rate constants to zero. 
    More details about the simulations can be found in Sec.~S2, in particular Fig.~S1, Supplementary Material. 
    The coupling constant $k_\eta$ and DNP injection $k_{W1}$ are fitted with $a\left(1-\frac{1}{bx+1}\right)$ (compare Eq.~\eqref{eq:Torrey_model_simplified}) with $bx = \gamma_\mathrm{e}^2B_{1,\mathrm{MW}}^2T_{2,\mathrm{e}}T_{1,\mathrm{e}}$ being the saturation factor and an additional offset for the coupling constant attributed to the finite thermal coupling \cite{stern_direct_2021}, although the offset is fitted to be effectively zero.
    The similar scaling of DNP injection and transport from the hidden into the bulk spins suggest a similar origin with the DNP injection originating from triple spin flips.}
	\label{fig:Simulation_HypResOnParameters}
\end{figure} 

We fit the best fit parameters as shown in Fig.~3 with a model describing the saturation of the electrons by the MW irradiation based on 
\begin{align}
    1-\frac{P_\mathrm{{e},\infty}}{P_{0,\mathrm{e}}} = 1- \frac{1}{\gamma_\mathrm{e}^2B_{1,\mathrm{MW}}^2T_{2,\mathrm{e}}T_{1,\mathrm{e}}+1}  \label{eq:Torrey_model_simplified}
\end{align}
with $T_\mathrm{1,e}$, $T_\mathrm{2,e}$ being the electronic relaxation times and $B_{1,\mathrm{MW}}$ the MW $B_1$ field.
Since we do not know the relationship between applied MW power and $B_{1,\mathrm{MW}}$, we use the generalized saturation parameter $b$ (cf. caption of Fig.~3).
Eq.~\eqref{eq:Torrey_model_simplified} is derived from the Torrey model \cite{torrey_transient_1949} of damped Rabi oscillations which in this case is equivalent to the $z$-part of the time-independent steady-state solutions of the Bloch equations (cf. Sec.~S5, Supplementary Material).

The coupling constant $k_\eta$ and DNP injection $k_{W1}$ show a nearly identical saturation parameter in Fig.~3, suggesting a common origin.
We note that the coupling and injection parameters are for the highest MW power only around one half of their fitted maximum value, allowing for a much higher DNP injection into the bulk if higher MW powers would be available.
However, higher MW power at liquid helium temperatures likely would not result in higher steady-state polarization as the relaxation scales linear with the electron saturation although the build-up time could be shortened \cite{von_witte_relaxation_2024}.

\section{Discussion} \label{sec:Discussion}

The two-electron two-nucleus spin system discussed in Sec.~1 describes two different processes possibly leading to nuclear spin diffusion around electrons: (i) Electron-mediated spin diffusion (EMSD) is present under any conditions and is non-resonant but strongly suppressed by its $\mathcal{O}\left(\omega_\mathrm{e}^{-1}\right)$ scaling. 
(ii) Electron-nuclear four-spin flip-flops are energy conserving if the resonance is met and do not involve an immediate interaction with the lattice or MW field but require electrons with different spin directions available as provided during MW irradiation (cf. Eq.~\eqref{eq:Torrey_model_simplified} or Ref.~\cite{thankamony_dynamic_2017}). 
Thus, under MW irradiation, the polarization transfer would be expected to scale similar to DNP relying on triple spin flips (as for 50\,mM TEMPOL in \textsuperscript{1}H glassy matrices) as both rely on the saturation of one electron population by MW irradiation - consistent with our results described in Fig.~3.

Hence, our work suggests that spin diffusion close to electrons for \textsuperscript{1}H-rich electron environments is relatively fast as electron-nuclear four-spin flip-flops enable nuclear flip-flops even for nuclei with different energies (nuclear spectral diffusion) due to hyperfine couplings.
For these electron-nuclear four-spin flip-flops, the electron polarization in at least parts of the electron spectrum needs to be clearly below unity, e.g. MW irradiation or not to high thermal electron polarizations, to enable electronic flip-flops.
This suggests that a spin diffusion barrier does not exist and all spins can contribute to the transport of nuclear hyperpolarization towards the bulk for large enough electronic and nuclear dipolar couplings, e.g. in \textsuperscript{1}H glassy matrices. 
This is supported by selective deuteration experiments in \textsuperscript{1}H-rich electron environments \cite{venkatesh_deuterated_2023,chatterjee_role_2023}, relaxation \cite{wolfe_direct_1973}, Hyp-Res \cite{stern_direct_2021,chessari_role_2023} and three-spin solid effect experiments \cite{tan_three-spin_2019}.
Future work might give a quantitative estimate of the spin diffusion close to the electron for a specific material, include relaxation effects and involve higher order Schrieffer-Wolff transformations \cite{bravyi_schriefferwolff_2011}.\\

\section*{Materials and Methods}

The Schrieffer-Wolff transformations were computed with Mathematica (cf. Sec.~S6, Supplementary Materials).
HypRes-on fits were performed with in-house developed Matlab scripts.



\printbibliography

\section*{Acknowledgments}
We thank Quentin Stern and Sami Jannin for providing the HypRes-on data. 
GvW thanks Konstantin Tamarov, Tom Wenckebach, Gian-Marco Camenisch and Mat\'ias Ch\'avez for discussions.

ME acknowledges support by the Schweizerischer Nationalfonds zur Förderung der Wissenschaftlichen Forschung (grant no. 200020\_188988 and 200020\_219375).
Financial support of the Horizon 2020 FETFLAG MetaboliQs grant is gratefully acknowledged.

The authors declare that they have no competing interests.

GvW developed the models, performed simulations and prepared the original draft. 
SK and ME acquired funding, provided resources and supervision. 
All authors conceptualized the research, reviewed and edited the draft.

All data needed to evaluate the conclusions in the paper are present in the paper and/or the Supplementary Materials.

\clearpage

\section*{Supplementary Materials}

\setcounter{equation}{0}
\setcounter{figure}{0}
\setcounter{table}{0}
\setcounter{section}{0}
\setcounter{page}{1}
\makeatletter
\renewcommand{\theequation}{S\arabic{equation}}
\renewcommand{\thefigure}{S\arabic{figure}}
\renewcommand{\thetable}{S\arabic{table}}
\renewcommand{\thesection}{S\arabic{section}}


\section{Additivity of the first-order Schrieffer-Wolff transformation for quadratic block-diagonal and off-block-diagonal interactions} \label{sec:SW_addivity}

\textbf{Definition:}
Let $\mathcal{P}_0$ be a projector on the diagonal matrix elements of a given n-dimensional square matrix $\mathbf{V}$ \cite{xiao_perturbative_2022}
\begin{align}
   \mathcal{P}_0  \bullet \mathbf{V} = \sum_{j=1}^n \ket{\psi_j}\bra{\psi_j}\mathbf{V} \ket{\psi_j}\bra{\psi_j} ~~~~~~.
\end{align}
and
\begin{align}
   \mathcal{Q}_0  \bullet \mathbf{V} = \mathop{\sum_{j,k=1}}_{j\neq k}^n \ket{\psi_j}\bra{\psi_j} \mathbf{V} \ket{\psi_k}\bra{\psi_k} 
\end{align}
for the off-diagonal part. 
Let 
\begin{align}
    \mathcal{N}_i= \{(j,k) ~|~ a_i\leq j,k \leq b_i,~ 1 \leq a_i, b_i \leq n,~ a_i, b_i, j, k \in \mathbb{N} \}
\end{align} 
be an interval of real numbers with $\mathcal{N}_i \cap \mathcal{N}_j = \emptyset ~ \forall i\neq j$ and $\mathcal{N} = \bigcup_i \mathcal{N}_i$
Then we can define a projector on the block diagonal part of the matrix given by all $\mathcal{N}_i$
\begin{align}
   \mathcal{Q}_\mathrm{inner}  \bullet \mathbf{V} = \mathop{\sum_{(j,k) \in \mathcal{N}}}_{j\neq k} \ket{\psi_j}\bra{\psi_j} \mathbf{V} \ket{\psi_k}\bra{\psi_k} 
\end{align}
and the corresponding projector on the other off-diagonal elements
\begin{align}
   \mathcal{Q}_\mathrm{outer}  \bullet \mathbf{V} = \mathop{\sum_{(j,k) \notin \mathcal{N}}}_{j\neq k} \ket{\psi_j}\bra{\psi_j} \mathbf{V} \ket{\psi_k}\bra{\psi_k} 
\end{align}

\textbf{Definition:} 
The first-order effective Hamiltonian $\mathbf{H}^\mathrm{eff}$ from the Schrieffer-Wolff transformation of $\mathbf{H} = \mathbf{H}_0 + \mathbf{V}$ with with a diagonal matrix $\mathbf{H}_0$, i.e. $\mathbf{H}_0 = \mathcal{P}_0 \bullet \mathbf{H}_0$ and an off-diagonal $\mathbf{V}$, i.e. $\mathbf{V} = \mathcal{Q}_0 \bullet \mathbf{V}$, is given by
\begin{align}
    \mathbf{H}^\mathrm{eff} = \mathbf{H}_0 + \frac{1}{2}[\mathbf{S},\mathbf{V}] + \mathcal{O}(\mathbf{V}^3)
\end{align}
with $\mathbf{S}$ defined by
\begin{align}
    \mathbf{V}+[\mathbf{S},\mathbf{H}_0] = 0 ~~~~~~.
\end{align}

\textbf{Theorem:}
For $\mathbf{V}_\mathrm{inner} + \mathbf{V}_\mathrm{outer} \coloneqq \mathcal{Q}_\mathrm{inner}  \bullet \mathbf{V} + \mathcal{Q}_\mathrm{outer}  \bullet \mathbf{V} = \mathbf{V}$, the first-order Schrieffer-Wolff transformation is given by $\mathbf{H}^\mathrm{eff} = \mathbf{H}_0 + \frac{1}{2}[\mathbf{S},\mathbf{V}] + \mathcal{O}(\mathbf{V}^3) = \mathbf{H}_0 + \frac{1}{2}[\mathbf{S}_\mathrm{inner},\mathbf{V}_\mathrm{inner}] + \frac{1}{2}[\mathbf{S}_\mathrm{outer},\mathbf{V}_\mathrm{outer}] + \mathcal{O}(\mathbf{V}^3)$ with $\mathbf{S} = \mathbf{S}_\mathrm{inner} + \mathbf{S}_\mathrm{outer}$.

\vspace{5mm}

\textbf{Proof:}
$\mathbf{S}_\mathrm{inner}$ and $\mathbf{S}_\mathrm{outer}$ are given by $\mathbf{V}_\mathrm{inner} + [\mathbf{S}_\mathrm{inner}, \mathbf{H}_0] = 0$ and $\mathbf{V}_\mathrm{outer} + [\mathbf{S}_\mathrm{outer}, \mathbf{H}_0] = 0$, respectively.
Since $\mathcal{P}_0 \bullet \mathbf{H}_0 = \mathbf{H}_0$, $\mathcal{Q}_\mathrm{inner} \bullet [\mathbf{S}_\mathrm{inner}, \mathbf{H}_0] = [\mathbf{S}_\mathrm{inner}, \mathbf{H}_0]$ and $\mathcal{Q}_\mathrm{outer} \bullet [\mathbf{S}_\mathrm{inner}, \mathbf{H}_0] = 0$ as well as $\mathcal{Q}_\mathrm{inner} \bullet [\mathbf{S}_\mathrm{outer}, \mathbf{H}_0] = 0$ and $\mathcal{Q}_\mathrm{outer} \bullet [\mathbf{S}_\mathrm{outer}, \mathbf{H}_0] = [\mathbf{S}_\mathrm{inner},\mathbf{H}_0]$.
Then 
\begin{align}
    (\mathbf{S}_\mathrm{inner} \mathbf{V}_\mathrm{inner})_{jk} =  
    \begin{cases}
        \sum_l (\mathbf{S}_\mathrm{inner})_{jl}(\mathbf{V}_\mathrm{inner})_{lk} & j,k \in \mathcal{N}\\
        0 & j \lor k \notin \mathcal{N}
    \end{cases}
\end{align}
and 
\begin{align}
    (\mathbf{S}_\mathrm{outer} \mathbf{V}_\mathrm{outer})_{jk} =  
    \begin{cases}
        \sum_l (\mathbf{S}_\mathrm{outer})_{jl}(\mathbf{V}_\mathrm{outer})_{lk} & j,k \notin \mathcal{N}\\
        0 & j \lor k \in \mathcal{N}
    \end{cases}
\end{align}
which results in $\mathbf{S} \mathbf{V}=\mathbf{S}_\mathrm{inner} \mathbf{V}_\mathrm{inner} + \mathbf{S}_\mathrm{outer} \mathbf{V}_\mathrm{outer}$. 
As the same holds for $\mathbf{V} \mathbf{S}$, $[\mathbf{S}, \mathbf{V}] = [\mathbf{S}_\mathbf{inner}, \mathbf{V}_\mathbf{inner}] + [\mathbf{S}_\mathbf{outer}, \mathbf{V}_\mathbf{outer}]$.
\hfill $\Box$

\vspace{5mm}

We speculate that the above theorem could be extended to arbitrary order for the above construction of $\mathbf{V}$ as the Schrieffer-Wolff transformation is given by $\mathbf{H}^\mathrm{eff}=e^{\mathbf{S}}\mathbf{H}e^{-\mathbf{S}}$ with a double Schrieffer-Wolff transformation taking the form $\mathbf{H}^\mathrm{eff}=e^{\mathbf{S}_2}e^{\mathbf{S}_1}\mathbf{H}e^{-\mathbf{S}_1}e^{-\mathbf{S}_2}$ and for $[\mathbf{S}_1, \mathbf{S}_2]=0$, $e^{\mathbf{S}_2}e^{\mathbf{S}_1} = e^{\mathbf{S}}$ (Baker-Campbell-Hausdorff formula). 
A proof for this is beyond the scope of the current manuscript.

\clearpage

\section{Derivation of solid effect and resonant mixing DNP} \label{sec:SI_SE_RM}

We study a one-electron-one-nucleus spin system with MW irradiation to illustrate that the Schrieffer-Wolff transformation applied to this spin system can describe solid effect (SE) and resonant mixing (RM) DNP.
Electron and nuclear Zeeman terms as well as the hyperfine coupling are as above. 
We add MW irradiation with the electron Rabi frequency $\omega_{1s}$ along the $x$-direction of the lab frame coordinate system ($\mathbf{H}_\mathrm{MW}=\omega_{1s}S^x \cos{(\omega_\mathrm{MW}t+\varphi)}=\omega_{1s}/2(S^+ + S^-)\cos{(\omega_\mathrm{MW}t+\varphi)}$). 
To avoid the time dependence of the MW irradiation, we transform into the rotating frame of the MW. 
To lowest order, this is can understood as a formalized approach to eliminate spin-spin processes that result in net electron flips as these violate energy conservation \cite{Wenckebach} and only the MW irradiation itself can cause net electron flips.
In the rotating frame, the one-electron-one-nucleus Hamiltonian takes the form

\begin{align}
    \mathbf{H}^\mathrm{1e1n,MW} =
    \begin{pmatrix}
        E_1^\mathrm{1e1n,MW} & A^{z-}/2  & \omega_{1s}/2  & 0 \\
        A^{z+}/2 & E_2^\mathrm{1e1n,MW} & 0 & \omega_{1s}/2 \\
        \omega_{1s}/2 & 0 & E_3^\mathrm{1e1n,MW} & -A^{z-}/2 \\
        0 & \omega_{1s}/2 & -A^{z+}/2 & E_4^\mathrm{1e1n,MW}
    \end{pmatrix}
\end{align}
with $E_1^\mathrm{1e1n,MW} = -(\omega_\mathrm{e} -\omega_\mathrm{MW}) - \omega_\mathrm{n} + A^{zz}$ and the other energies following the same approach as for the two-electron two-nucleus system in the lab frame (cf. Eq.~(4)).
The 'MW' in the superscript indicates the MW rotating frame in the following.
Applying the Schrieffer-Wolff transformation as above for this smaller rotating frame system gives

\begin{align}
    H^\mathrm{eff,1e1n,MW}_{2,3} &= \frac{A^{z+}\omega_{1s}/2}{2} \label{eq:SEwithAzz}\\
    &\left[\frac{2\omega_\mathrm{e}-2\omega_\mathrm{MW}}{(A^{zz})^2 -(2\omega_\mathrm{e}-2\omega_\mathrm{MW})^2} + \frac{2\omega_\mathrm{n}}{(A^{zz})^2 - \left(2\omega_\mathrm{n}\right)^2}  \right] \nonumber
\end{align}
which simplifies for $A^{zz}\ll \omega_\mathrm{n},~\omega_\mathrm{e}-\omega_\mathrm{MW}$ to
\begin{align}
    H^\mathrm{eff,1e1n,MW}_{2,3} \approx - \frac{A^{z+}\omega_{1s}/2}{2} \frac{\omega_\mathrm{e}-\omega_\mathrm{MW}+\omega_\mathrm{n}}{\omega_\mathrm{n}(\omega_\mathrm{e}-\omega_\mathrm{MW})} \label{eq:SEwithoutAzz} 
\end{align}
MW irradiation at $\omega_\mathrm{e}-\omega_\mathrm{MW}=\omega_\mathrm{n}$ in this limit gives 
\begin{align}
    H^\mathrm{eff,1e1n,MW}_\mathrm{SE,DQ} = -\frac{A^{z+}\omega_{1s}/2}{2\omega_\mathrm{n}} \label{eq:SE}
\end{align}
which is the well known scaling of the SE matrix element \cite{Hovav2010,Wenckebach,thankamony_dynamic_2017,tan_three-spin_2019}. 
    
Assuming $ \varepsilon \coloneq \omega_\mathrm{e} - \omega_\mathrm{MW} \rightarrow 0$, causes $A^{zz} \simeq \omega_\mathrm{e}-\omega_\mathrm{MW} \ll \omega_\mathrm{n}$.
For Eq.~\eqref{eq:SEwithAzz} this results in 
\begin{align}
    H^\mathrm{eff,1e1n,MW}_{RM} &\approx \frac{A^{z+}\omega_{1s}/2}{2} \left[\frac{2\varepsilon}{(A^{zz})^2 - (2\varepsilon)^2} - \frac{1}{2\omega_\mathrm{n}}  \right] \label{eq:RM}
\end{align}
which gives half the transiton matrix amplitude of the solid effect (cf. Eq.~\eqref{eq:SE}) for $\varepsilon = 0$ and includes a resonance condition for $A^{zz} = \pm 2\varepsilon$.
We identify this strong electron-nuclear transition as the recently introduced resonant mixing DNP \cite{quan_resonant_2023}. 
Resonant mixing can lead to a dispersive enhancement of the nuclear polarization around the electron resonance (single quantum EPR transition, SQ\textsubscript{e}, cf. Eq.~\eqref{eq:RM}).
For more details on this including a discussion of experimental evidence, the reader is referred to Ref.~\cite{quan_resonant_2023}.

If higher order Schrieffer-Wolff transformation \cite{bravyi_schriefferwolff_2011} are applied, this might also describe the three-spin solid effect (electron flip with two nuclei flipping) if MW irradiation is included \cite{tan_three-spin_2019}. 
At even higher orders, the four-spin solid effect (SE with three nuclei flipping) could be described \cite{tan_observation_2022}.

\clearpage

\section{Coupled two-compartment model} \label{sec:SI_2compartment}

We start with a recap of the previously introduced single homogeneous compartment model\cite{von_witte_modelling_2023}:
The hyperpolarization build-up can be described through a first-order differential equation with a hyperpolarization injection rate constant $k_\mathrm{W}$ and a relaxation rate constant of the build-up $k_\mathrm{R}^\mathrm{bup}$ 
\begin{align}
	\frac{\text{d}P}{\text{d}t} = (A-P) k_\mathrm{W} - k_\mathrm{R}^\mathrm{bup} P \label{eq:ODE_bup_1compartment}
\end{align}
with $A$ describing the theoretical maximum of hyperpolarization achievable, i.e., the thermal electron polarization in DNP.
The solution of Eq.~\eqref{eq:ODE_bup_1compartment} is a mono-exponential curve which can be compared with the phenomenological description of the build-up curve by $P(t) = P_0 (1-e^{-t/\tau_\mathrm{bup}})$ to express the experimental parameters in terms of model parameters. 
Here, $P_0$ is the steady-state polarization and $\tau_{\mathrm{bup}}$ the build-up time.

\begin{subequations} \label{eq:1compartment_bup_solution}
	\begin{align}
		\tau_{\mathrm{bup}}^{-1} &= k_\mathrm{W}+k_\mathrm{R}^\mathrm{bup} \label{eq:1compartment_bup_tau} \\
		P_0 &= \frac{Ak_\mathrm{W}}{k_{W}+k_\mathrm{R}^\mathrm{bup}} = Ak_\mathrm{W}\tau_{\mathrm{bup}} \label{eq:1compartment_bup_P0}
	\end{align}
\end{subequations}
 
For the decay, $k_\mathrm{W}$ would be set to zero (MW off), leading to $\tau_\mathrm{decay}^{-1}=k_\mathrm{R}^\mathrm{decay}$. 

Extending the one-compartment model to two uncoupled compartments with separate injection and relaxation rates is straightforward. 
Such a situation might be realized for a material consisting of two phases with different compositions (radical concentration, NMR-active spin concentration) such that each compartment follows its own mono-exponential build-up (cf. Eqs.~\eqref{eq:ODE_bup_1compartment}, \eqref{eq:1compartment_bup_solution}).
Crucially, spin diffusion between the two compartments needs to be suppressed, e.g., through a resonance frequency difference rendering inter-compartment nuclear flip-flops energy non-conserving.
If the frequency difference between the two compartments is small compared to the NMR linewidth such that the two compartments cannot be clearly discriminated through different peaks, the total measured signal describes the total magnetization created in the two compartments.
In such a case, the resulting build-up would take a bi-exponential form

\begin{align}
	P &= P_{0,2}  \left[\alpha  \left(1-e^{-t/\tau_1}\right) + (1-\alpha)  \left(1-e^{-t/\tau_2}\right)\right] \nonumber \\
	&= P_{0,2}  \left[1 - \alpha  e^{-t/\tau_1} - (1-\alpha)  e^{-t/\tau_2}\right] \label{eq:DoubleExpBup_experiment}
\end{align}
with the relative weight of the two time constants $\alpha$.
Experimentally, four parameters are extracted from the build-up while in the theoretical model five parameters are required: two injection and relaxation rates each giving rise to the two steady-state polarizations and build-up times as well as the relative size of the compartments.
Hence, for two uncoupled compartments, it is difficult to extract information about the individual compartments based on the above compartment model ansatz.

Hence, we focus on a coupled two-compartment model below: Fig.~1 sketches the basic idea of the model with DNP injection only into the first compartment (relative size $\xi$), a coupling between the two compartments and a separate relaxation rate constant for each compartment.
The resulting coupled differential equation system takes the form

\begin{subequations} \label{eq:ODE_bup_2compartment_SI}
	\begin{align}
		\frac{\mathrm{d}P_1}{\mathrm{d}t} &= (A-P_1) k_{W1} - k_{\mathrm{R1}}  P_1 - \frac{k_\eta}{\xi} (P_1-P_2) \label{eq:ODE_bup_2compartment_1_SI} \\
		\frac{\mathrm{d}P_2}{\mathrm{d}t} &= - k_{\mathrm{R2}}  P_2 + \frac{k_\eta}{1-\xi} (P_1-P_2) \label{eq:ODE_bup_2compartment_2_SI} 
	\end{align} 
\end{subequations}
with $P_1$, $P_2$, $k_{\mathrm{R1}}$ and $k_{\mathrm{R1}}$ being the polarizations and relaxation rates of the two compartments, respectively. DNP injection is only possible into the first compartment through the term $(A-P_1)k_{W1}$ with $k_{W1}$ being the DNP injection parameter. 
$A$ describes the theoretical maximum polarization, e.g., the thermal electron polarization in DNP. 
The idea behind this term is discussed in more detail in \cite{von_witte_modelling_2023}. 
$k_\eta$ is the inter-compartment coupling parameter. 
$\xi$ defines the relative size of the two compartments.
We note that the model can viewed as a generalization of the two-compartment model in \cite{stern_direct_2021} for build-ups and decays ($k_{W1}=0$ and a non-zero starting polarization for decays) although it was intended to shed light on bi-exponential build-ups, e.g., as observed in silicon \cite{atkins_synthesis_2013,kwiatkowski_exploiting_2018}. 
For simplicity and in analogy to the one-compartment model \cite{von_witte_modelling_2023}, we ignored a thermal equilibrium polarization as enhancements of 100 over the thermal equilibrium can be achieved in many materials, rendering the thermal polarization small compared to typical measurement uncertainties.

The coupling parameter $k_\eta$ in Eqs.~\eqref{eq:ODE_bup_2compartment_SI} is modulated by the compartment size $\xi$. 
$k_\eta$ describes a magnetization exchange between the two compartments and a large magnetization added to a small system leads to a drastic change of the polarization of the compartment as the polarization is a normalized magnetization. 
The total polarization of the system would be described by $P=\xi P_1 + (1-\xi)P_2$.

The structure of the coupling $(P_1-P_2)$ rather describes a polarization gradient and, hence, can be interpreted as a polarization flux according to Fick's first law of diffusion. 
Nuclear spin polarization usually spreads in a diffusive way (Fick's second law of diffusion):   

\begin{align}
	\frac{\partial P}{\partial t} = \nabla \left[D(x)  \nabla P\right] = D \Delta P \label{eq:spinDiffusion}
\end{align}
where an isotropic spin diffusion coefficient $D$ is assumed in the last step. $\Delta$ is the Laplace operator. Since our two-compartment model (see Fig.~1 and Eqs.~\eqref{eq:ODE_bup_2compartment_SI}) is independent of spatial variables, the Laplace operator is not defined, leaving us with the above two-compartment model. 
The spatially-dependent case (Eq.~\eqref{eq:spinDiffusion}) including relaxation and DNP injection has been solved in a numerically efficient way by Pinon and co-workers \cite{Pinon2017} and successfully applied to understand several complex materials better \cite{Zhao2018,Viger-Gravel2018,Bjorgvinsdottir2018,Pinon2018}. 
The spin diffusion model by Pinon and co-workers is rather focused on understanding the microscopic details, whereas the two-coupled compartment model is very macroscopic but offering intuitive understanding. 
Hence, the discussion of this model should give some general understanding of relationships between injection rate, relaxation and coupling on the one side and experimentally measured parameters on the other.

For simplicity, we first solve the coupled differential equation system with both compartments having the same size ($\xi=0.5$) as this eliminates $\xi$s from the equations and with it simplifying the notation. 
Extension to the general case with arbitrary $\xi$ is straightforward. Thus, we start from

\begin{subequations}
	\begin{align}
		\frac{\mathrm{d}P_1}{\mathrm{d}t} &= (A-P_1) k_{W1} - k_{\mathrm{R1}}  P_1 - k_\eta (P_1-P_2) \label{eq:ODE_bup_2compartment_1_SI_noXi} \\
		\frac{\mathrm{d}P_2}{\mathrm{d}t} &= - k_{\mathrm{R2}}  P_2 + k_\eta (P_1-P_2) \label{eq:ODE_bup_2compartment_2_SI_noXi} 
	\end{align} 
\end{subequations}
with the same parameters as in the main part except for assuming $\xi=0.5$ such that the two compartments have equal sizes. 

We start by solving the differential equation system by rewriting the second equation. \eqref{eq:ODE_bup_2compartment_2_SI_noXi}, to 
\begin{align}
	P_1 = \frac{\mathrm{d}P_2}{\mathrm{d}t} + k_{\mathrm{R2}}  P_2 + k_\eta P_2
\end{align}
and inserting this into the first equation, \eqref{eq:ODE_bup_2compartment_1_SI_noXi} to have a differential equation for $P_2$. We make and exponential ansatz $e^{rt}$ to solve the homogenous equation which gives us a polynomial equation for $r$ which we can easily solve to 
\begin{align}
	r_{1/2} = -\frac{1}{2}\left(k_{\mathrm{R1}} + k_{\mathrm{R2}} + k_{W1} + 2k_\eta\right)\pm\frac{1}{2} \sqrt{\left(k_{\mathrm{R1}}-k_{\mathrm{R2}}+k_{W1}\right)^2+4k_\eta^2} \label{eq:ODE_2_compartment_time_constants_exact}
\end{align}
From the inhomogenous case we find $C_0=\frac{(k_{\mathrm{R2}}+k_\eta)Ak_{W1}}{(k_{W1}+k_{\mathrm{R1}} + k_\eta)  (k_{\mathrm{R2}} +k_\eta)-k_\eta^2}$. We now can insert our solution for $P_2$ into the second equation, \eqref{eq:ODE_bup_2compartment_2_SI_noXi}, to find our solution for $P_1$. Our total polarization, which we measure experimentally, is given by $P=(P_1+P_2)/2$ as we have to average over the two compartments, and, thus, we find
\begin{align}
	P = &\frac{c_1}{2}  \left(1+\frac{r_1+k_{W1}+k_{\mathrm{R1}}+k_\eta}{k_\eta}\right)  e^{r_1t} + \frac{c_2}{2}  \left(1+\frac{r_2+k_{W1}+k_{\mathrm{R1}}+k_\eta}{k_\eta}\right) e^{r_2t} + ...\nonumber \\
	& + \frac{1}{2}\frac{(k_{\mathrm{R2}}+2k_\eta)Ak_{W1}}{(k_{W1}+k_{\mathrm{R1}}+k_\eta)(k_{\mathrm{R2}}+k_\eta)-k_\eta^2}
\end{align}
with $c_1$ and $c_2$ being constants of integration. We can set these by comparing our solution with an experimentally used model, e.g., a bi-exponential build-up (cf. \eqref{eq:DoubleExpBup_experiment})
\begin{align}
	P &= P_{0,2}  \left[\alpha  \left(1-e^{-t/\tau_1}\right) + (1-\alpha)  \left(1-e^{-t/\tau_2}\right)\right] \nonumber \\
	&= P_{0,2}  \left[1 - \alpha  e^{-t/\tau_1} - (1-\alpha)  e^{-t/\tau_2}\right] \label{eq:DoubleExpBup_experiment_SI}
\end{align}
for which we dropped the "bup" subscript for convenience. 
We can immediately read off
\begin{subequations}
	\begin{align}
		P_{0,2} &= \frac{1}{2}\frac{(k_{\mathrm{R2}}+2k_\eta)Ak_{W1}}{(k_{W1}+k_{\mathrm{R1}}+k_\eta)(k_{\mathrm{R2}}+k_\eta)-k_\eta^2} \\
		\tau_1^{-1} &= -r_1 \\ 
		\tau_2^{-1} &= -r_2
	\end{align}
\end{subequations}
Furthermore, we can choose $c_1$ and $c_2$ such that it reproduces the $-\alpha$ and $-(1-\alpha)$ prefactors of the exponentials. Our fourth equation for the theory-experimental correspondence comes from 
\begin{align}
	\frac{\mathrm{d}P_1}{\mathrm{d}t}(0) = -r_1\alpha P_{0,2} = Ak_{W1}
\end{align}
as we assume both compartments to be completely unpolarized initially.
With these four equations at hand, we could write the four theory parameters in terms of the four experimental parameters. However, this provides little insight as we encounter some complicated complex-valued fourth-order polynomial equations stemming from the square roots in the time constants. Thus, we take a different approach and find the parameter correspondence through analysis of the coupled differential equation system in the steady-state as well as boundary conditions.

\begin{figure}[!hb]
	\centering
	\includegraphics[width=\linewidth]{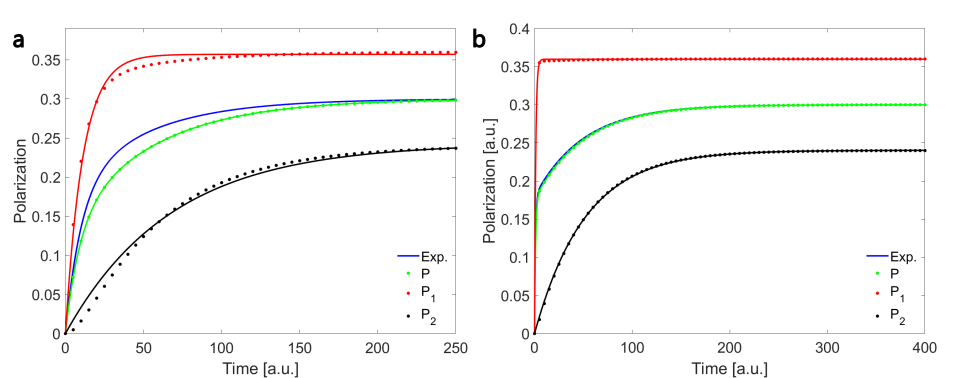}
	\caption{The differential equation model reproduces the experimental bi-exponential build-up for vastly different time constants between the two compartments but leads to imperfections for time constants on the same order of magnitude. (a) Simulation parameters: $P_{0,2}=0.3$, $\tau_1=1$, $\tau_2=50$, $\alpha=0.6$, $A=1$, $\xi=0.5$. Fitting the bi-exponential build-up resulting from numerical integration of the differential equation system gives $P_{0,2}=0.3$, $\alpha=0.4523$, $\tau_1=9.00$,  $\tau_2=55.41$. Note that the polarizations of the individual compartments cannot be fitted by mono-exponential models due to the interaction between the two. (b) Setting the fast build-up time to 1, leads to an excellent agreement between the theoretical and experimental model. The assumption that the two build-ups are at different time scales such that one is always at equilibrium on the time scale of the other, can be considered fulfilled in this case. The slow second compartment builds up exponentially as the initial delay to build up the polarization difference $P_1-P_2$ is negligible. Simulation parameters: $P_0=0.3$, $\tau_1=1$, $\tau_2=50$, $\alpha=0.6$, $A=1$, $\xi=0.5$}
	\label{fig:Bup_2compartment_TheoryExp}
\end{figure}

For large times our system is in a steady-state with the total polarization being $P_{0,2}$ and experimentally the two compartments having a polarization equal to $P_{1s} = 2\alpha P_{0,2}$ and $P_{2s} = 2(1-\alpha)P_{0,2}$. From equation \eqref{eq:ODE_bup_2compartment_2_SI_noXi} we find for large times
\begin{subequations}
	\begin{align}
		k_\eta(P_{1s}-P_{2s}) = k_{\mathrm{R2}} P_{2s} \Leftrightarrow k_{\mathrm{R2}} = k_\eta \frac{\alpha-(1-\alpha)}{1-\alpha} 
	\end{align}
\end{subequations}
which restrains $\alpha$ to be larger than $0.5$ to ensure that $k_{\mathrm{R2}} \ge 0$. 
Additionally, we notice that equation \eqref{eq:ODE_bup_2compartment_2_SI_noXi} is for large times identical to the one compartment rate equation discussed in \cite{von_witte_modelling_2023} $\left(\frac{\mathrm{d}P}{\mathrm{d}t}=(A-P) k_\mathrm{W}-k_\mathrm{R} P\right)$ upon substitution of $P_{1s}$ with $A$ and $k_\eta$ with $k_{W1}$. 
Thus, we find 
\begin{subequations}
	\begin{align}
		P_{2s} = k_\eta P_{1s}\tau_2 \Leftrightarrow \eta &= \frac{1-\alpha}{\alpha}\tau_2^{-1} \label{eq:eta_2compartment_bup} \\
		\Rightarrow k_{\mathrm{R2}} &= \tau_2^{-1} \left(2-\alpha^{-1}\right) \label{eq:R_2_2compartment_bup}
	\end{align}
\end{subequations}
Again we can use 
\begin{align}
	\frac{\mathrm{d}P_1}{\mathrm{d}t}(0) = \alpha P_{0,2}\tau_1^{-1} = Ak_{W1} \Leftrightarrow k_{W1} = \frac{P_{0,2}}{A}\alpha \tau_1^{-1} \label{eq:W_2compartment_bup}
\end{align}
For the relaxation rate of the first compartment we can use the steady-state condition of the overall system
\begin{subequations}
	\begin{align}
		(A-P_{1s})k_{W1} &= k_{\mathrm{R1}} P_{1s} + k_{\mathrm{R2}} P_{2s} \\
		k_{\mathrm{R1}} &= k_{W1} \left(\frac{A}{2\alpha P_{0,2}}-1\right) - \frac{1-\alpha}{\alpha}R_2 \label{eq:R_1_2compartment_bup}
	\end{align}
\end{subequations}
We note that inserting these parameters into the exact solution for the time constants of the coupled differential equation system, equation \eqref{eq:ODE_2_compartment_time_constants_exact}, does not reproduce the experimental time constants exactly. Such a discrepancy was to be expected as the differential equation system is not describing the experimentally occurring spin diffusion mediated build-up of polarization but rather describes a spin flux. A comparison between the experimental model and its corresponding theoretical build-up is shown in Fig.~\ref{fig:Bup_2compartment_TheoryExp}a. The discrepancy of the theoretical model is due to the structure of our coupling between the two compartments. We first need to build up a polarization difference between the two compartments before polarization can be transferred into the second compartment. This leads to a very slow initial build-up compared to the experimental model. Furthermore, this acts back on the first compartment as this is drained by the large polarization difference once it developed, leading to a slower build-up of polarization in the first compartment. 
Interestingly, once we fit the total polarization of our theoretical model, the faster time constant is smaller than in the experimental case but at the expense of a smaller balance parameter $\alpha$. This is a result of the internal interaction between the two compartments and fitting them jointly. Interestingly, the build-up of polarization within each compartment cannot be fitted well with a mono-exponential as shown in Fig.~\ref{fig:Bup_2compartment_TheoryExp}a whereas the total polarization can be fitted accurately with a bi-exponential as given by equation \eqref{eq:DoubleExpBup_experiment_SI}, underlining that the description of the exchange of polarization between the two compartments in our model is not trivial.   

We can find the same parameters for the experiment-theory correspondence with an alternative approach.
For this we solve our coupled differential equation (\eqref{eq:ODE_bup_2compartment_1_SI_noXi}-\eqref{eq:ODE_bup_2compartment_2_SI_noXi}) under the approximation that $P_1$ in the differential equation for $P_2$ (\eqref{eq:ODE_bup_2compartment_2_SI_noXi}) is time-independent and vice versa. This is similar to the Born-Oppenheimer approximation in quantum chemistry where the dynamics of electrons and nuclei is assumed to be at different time scales such that the electrons experience the nuclei as being at rest and vice versa. If the two time scales in the build-up are vastly different, this assumption would be valid as the polarization in the first compartment is already fully developed while the second compartment is still nearly unpolarized. This is shown and discussed in Fig.~\ref{fig:Bup_2compartment_TheoryExp}. In practice, this assumption will not fully hold as it is difficult to observe a component that builds-up orders of magnitude slower than the other. 

We start solving the differential equation system by first solving the second equation (\eqref{eq:ODE_bup_2compartment_2_SI_noXi}). This gives us 
\begin{align}
	P_2 = \frac{\eta}{k_{\mathrm{R2}} + k_\eta}  P_1 + c_2  e^{-(k_{\mathrm{R2}}+k_\eta)t}
\end{align}
We can insert this into the first equation (Eq.~\eqref{eq:ODE_bup_2compartment_1_SI_noXi}) and get 
\begin{align}
	P_1 = &\frac{Ak_{W1}}{k_{W1}+k_{\mathrm{R1}}+k_\eta-\frac{k_\eta^2}{k_{\mathrm{R2}}+k_\eta}} + c_1  e^{-\left(k_{W1}+k_{\mathrm{R1}}+k_\eta-\frac{k_\eta^2}{k_{\mathrm{R2}}+k_\eta}\right)t} + \nonumber \\ &\frac{k_\eta}{k_{W1}+k_{\mathrm{R1}}-k_{\mathrm{R2}}-\frac{k_\eta^2}{k_{\mathrm{R2}}+k_\eta}}  c_2  e^{-(k_{\mathrm{R2}}+k_\eta)t} 
\end{align}
The total polarization $P$ of the system, as measured experimentally, is given by the weighted sum of $P_1$ and $P_2$.
\begin{subequations}
	\begin{align}
		P &= \frac{P_1+P_2}{2} = \frac{1}{2}\left(1+\frac{k_\eta}{k_{\mathrm{R2}}+k_\eta}\right)  P_1 + \frac{c_2}{2}  e^{-(k_{\mathrm{R2}}+k_\eta)t} \nonumber \\
		&= \frac{1}{2}\left(1+\frac{k_\eta}{k_{\mathrm{R2}}+k_\eta}\right)  \left[\frac{Ak_{W1}}{k_{W1}+k_{\mathrm{R1}}+k_\eta-\frac{k_\eta^2}{k_{\mathrm{R2}}+k_\eta}} + \frac{c_1}{2}  e^{-\left(k_{W1}+k_{\mathrm{R1}}+k_\eta-\frac{k_\eta^2}{k_{\mathrm{R2}}+k_\eta}\right)t}\right] + ... \nonumber \\
		& ~~~~~ +\left[1+\frac{k_\eta}{k_{W1}+k_{\mathrm{R1}}-k_{\mathrm{R2}}-\frac{k_\eta^2}{k_{\mathrm{R2}}+k_\eta}}\right]  \frac{c_2}{2}  e^{-(k_{\mathrm{R2}}+k_\eta)t} \\
		&=  \frac{1}{2}\left(1+\frac{k_\eta}{k_{\mathrm{R2}}+k_\eta}\right)  \frac{Ak_{W1}}{k_{W1}+k_{\mathrm{R2}}+k_\eta-\frac{k_\eta^2}{k_{\mathrm{R2}}+k_\eta}} \left[1 + \frac{k_{W1}+k_{\mathrm{R1}}+k_\eta-\frac{k_\eta^2}{k_{\mathrm{R2}}+k_\eta}}{Ak_{W1}}  ... \right.\nonumber \\
		& ~~~~~ \left.  \left[c_1  e^{-\left(k_{W1}+k_{\mathrm{R1}}+k_\eta-\frac{k_\eta^2}{k_{\mathrm{R2}}+k_\eta}\right)t} +  \left(1+\frac{k_\eta}{k_{\mathrm{R2}}+k_\eta}\right)^{-1} ... \right. \right. \nonumber \\ 
		& ~~~~~ \left. \left.  \left(1+\frac{k_\eta}{k_{W1}+k_{\mathrm{R1}}-k_{\mathrm{R2}}-\frac{k_\eta^2}{k_{\mathrm{R2}}+k_\eta}}\right)  c_2  e^{-(k_{\mathrm{R2}}+k_\eta)t} \right] \right] \label{eq:DoubleExpBup_theory}
	\end{align}
\end{subequations}

To find the constants of integration, we rewrite equation \eqref{eq:DoubleExpBup_theory}
\begin{align}
	P = P_{0,2}  \left[1 - \frac{1}{Ak_{W1}\tau_1}  \left[c_1  e^{-t/\tau_1} + \frac{1}{1+k_\eta \tau_2}  \left(1+\frac{k_\eta}{\tau_1^{-1}-\tau_2^{-1}}\right)  c_2  e^{-t/\tau_2} \right] \right]
\end{align}
and can find as a boundary condition for large $t$
\begin{align}
	\frac{1}{Ak_{W1}\tau_1}  \left[\frac{1}{1+k_\eta \tau_2}  \left(1+\frac{k_\eta}{\tau_1^{-1}-\tau_2^{-1}}\right)  c_2 + c_1\right] = - (1-\alpha) - \alpha = -1
\end{align}
and choose 
\begin{subequations}
	\begin{align}
		\alpha &= \frac{c_1}{Ak_{W1}\tau_1} \\
		-(1-\alpha) &= \frac{c_2}{Ak_{W1}\tau_1}  \frac{1}{1+k_\eta \tau_2}  \left(1+\frac{k_\eta}{\tau_1^{-1}-\tau_2^{-1}}\right) \label{eq:DoubleExpBup_alpha}
	\end{align}
\end{subequations} 

If we compare \eqref{eq:DoubleExpBup_theory} with our bi-exponential, experimental model (Eq.~\eqref{eq:DoubleExpBup_experiment_SI}), we can read off equations for both time constants and the steady-state polarization $P_{0,2}$.
\begin{subequations}
	\begin{align}
		\tau_1^{-1} &= k_{W1}+k_{\mathrm{R1}}+k_\eta-\frac{k_\eta^2}{k_{\mathrm{R2}}+k_\eta} \label{eq:DoubleExpBup_tau1} \\
		\tau_2^{-1} &= k_{\mathrm{R2}}+k_\eta \label{eq:DoubleExpBup_tau2} \\
		P_{0,2} &= \frac{1}{2} \left(1+\frac{k_\eta}{k_{\mathrm{R2}}+k_\eta}\right)  \frac{Ak_{W1}}{k_{W1}+k_{\mathrm{R1}}+k_\eta-\frac{k_\eta^2}{k_{\mathrm{R2}}+k_\eta}} = \frac{1}{2}(1+k_\eta \tau_2)Ak_{W1}\tau_1 \label{eq:DoubleExpBup_P0}  \\
		\alpha &= (1+k_\eta \tau_2)^{-1}
	\end{align}
\end{subequations}
We can rewrite these four equations to
\begin{subequations}
	\begin{align}
		k_{\mathrm{R1}} &= \tau_1^{-1} - k_{W1} - k_\eta + k_\eta^2 \tau_2 \label{eq:DoubleTheoryBup_R1_initial}\\
		k_{\mathrm{R2}} &= \tau_2^{-1} - k_\eta \label{eq:DoubleTheoryBup_R2_initial} \\
		k_{W1} &= \frac{2 P_{0,2}}{(1+k_\eta \tau_2) A \tau_1} \label{eq:DoubleTheoryBup_W_initial}\\
		k_\eta &= \frac{1-\alpha}{\alpha}\tau_2^{-1}
	\end{align}
\end{subequations}
where we used the same arguments for $k_\eta$ as for equation \eqref{eq:eta_2compartment_bup}.
These four equations give the same differential equation parameters as our above approach with the boundary conditions and the steady-state analysis of the differential equation system. 

From here it is straightforward to include the size of the first compartment $\xi$ resulting in
\begin{subequations}
	\begin{align}
		\alpha &= \left(1+\frac{k_\eta}{\xi}\tau_2\right)^{-1} \\
		\tau_2^{-1} &= k_{\mathrm{R2}} + \frac{k_\eta}{1-\xi} \\
		\tau_1^{-1} &= k_{W1} + k_{\mathrm{R1}} + \frac{k_\eta}{\xi} - \frac{k_\eta^2}{\xi(1-\xi)} \left(k_{\mathrm{R2}}+\frac{k_\eta}{1-\xi}\right)^{-1} \nonumber \\
        &=  k_{W1} + k_{\mathrm{R1}} + \frac{k_\eta}{\xi} - \frac{k_\eta^2}{\xi(1-\xi)} \tau_2 \\
		P_{0,2} &= \frac{Ak_{W1}  \left[\xi+k_\eta\left(k_{\mathrm{R2}}+\frac{k_\eta}{1-\xi}\right)^{-1}\right]}{k_{W1} + k_{\mathrm{R1}} + \frac{k_\eta}{\xi} - \frac{k_\eta^2}{\xi(1-\xi)} \left(k_{\mathrm{R2}}+\frac{k_\eta}{1-\xi}\right)^{-1}} \nonumber \\
		&= Ak_{W1}\tau_1 \left[\xi + k_\eta \tau_2 \right] ~~~.
	\end{align}
\end{subequations}

For the decay, the injection term from equation \eqref{eq:ODE_bup_2compartment_1_SI_noXi} is eliminated. 
If we follow the above approach with different time scales of the compartments, the time constants are identical to the build-up case apart from the vanishing $k_{W1}$. 
Under the assumption of long build-up times before the decay such that both compartments reach their steady-state polarization, the initial polarization and coupling constant are the same as for the build-up. 
The exact solution for the decay case is given in \cite{stern_direct_2021} to understand the interaction between hypershifted spins with the RF-visible bulk.

\clearpage

\section{Two compartment HypRes-on fits} \label{sec:SI_twoCompartmentHypRes}

The simulations shown in Figs. 3 and S1 were performed with MATLAB. For the simulations of the HypRes-on (MW-on HypRes experiments) data from \cite{chessari_role_2023}, provided to us from Quentin Stern, we numerically integrated Eqs.~\eqref{eq:ODE_bup_2compartment_SI} (time slicing) and performed a grid search over 1 million iterations after pre-scanning the parameter ranges. 
The model is dominated by two competing processes: (i) Transport of polarization stored in the hypershifted spins with compartment size $\xi$ into the RF saturated bulk and (ii) DNP injection of the other DNP lobe with respect to the lobe used for the initial polarization build-up. 
For this, we set the polarization of the hidden compartment ($P_1$) to the estimated build-up polarization of 70\%. 
Since the experiments were performed at 7\;T and 1.2\;K, we set $A=-1$ (the minus sign stems from the choice of the DNP lobe). We assumed all relaxation being mediated by the electrons, hence, the relaxation of the second compartment $k_{\mathrm{R2}}$ was set to zero (in agreement with the data from \cite{stern_direct_2021}). 
The model parameters being varied in the grid search were initially the DNP injection parameter $k_{W1}$, relaxation rate $k_{\mathrm{R1}}$ and the size of the first hidden compartment $\xi$ as well as the inter-compartment coupling parameter $k_\eta$ (compare Eqs.~\eqref{eq:ODE_bup_2compartment_SI}). 
The model was found to be insensitive to any relaxation, likely due to the short experimental build-up duration of 25\,s. 
Thus, we set $k_{\mathrm{R1}} = 0$ for the simulations shown in this work. 
In a similar way, the first compartment size $\xi$ was initially used as a fit parameter but values were around 7\% and thus fixed to 7\% to simplify the final fits with finer parameter grids.
The parameter combinations with the lowest least square value are chosen and shown in Fig.~S1. 
The polarization of the second compartment at the beginning of the simulation (time zero) was set to zero, explaining the discrepancies for very short times. 
The best fit parameters are summarized in Fig.~3.

\begin{figure*}[!ht]\centering 
	\includegraphics[width=0.9\textwidth]{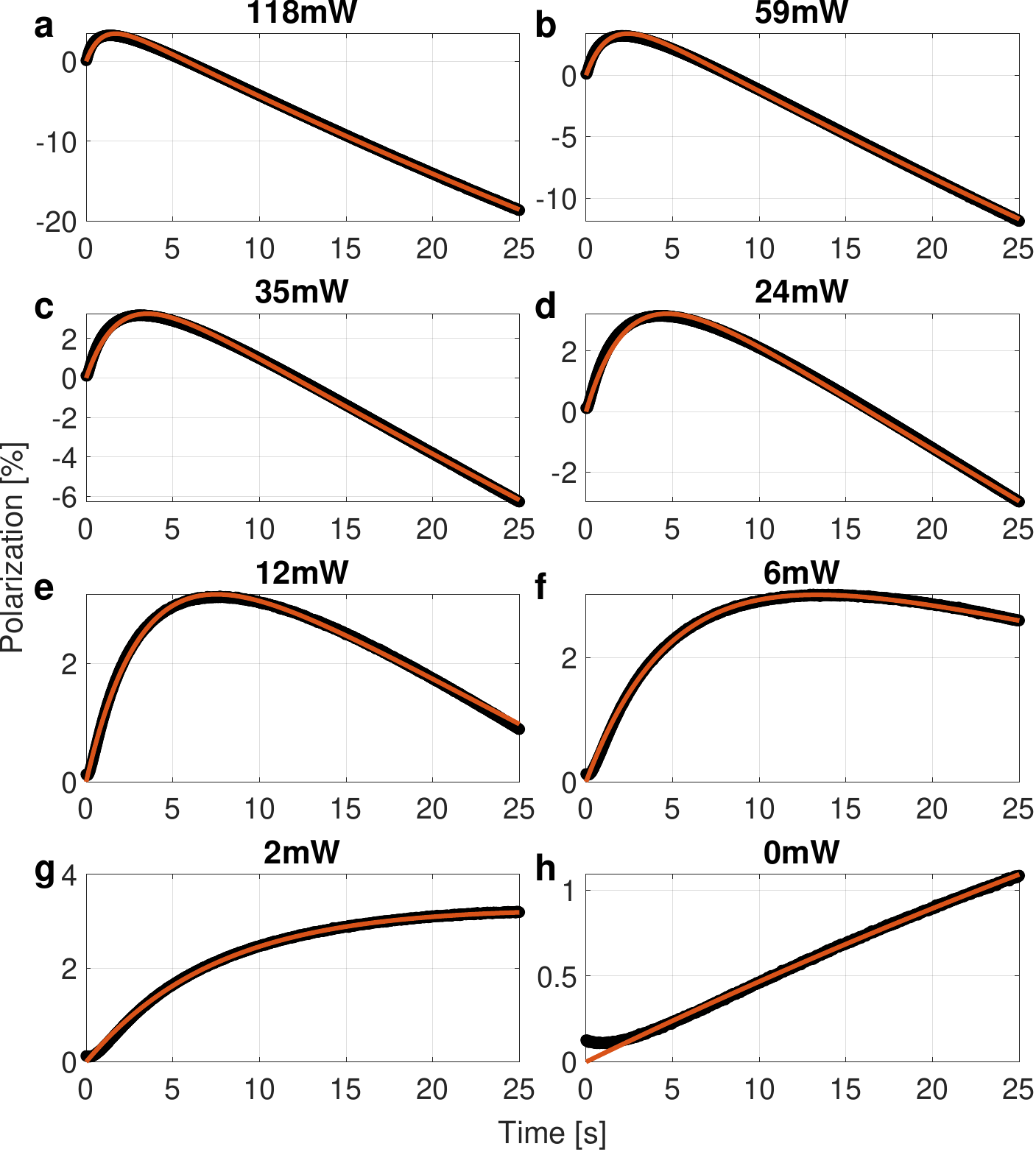}
	\caption{\textbf{Simulated and experimental HypRes-on data} Simulated HypRes-on in red and experimental data from \cite{chessari_role_2023} in black.}
	\label{fig:Simulation_HypResOn_Power}
\end{figure*} 

\clearpage

\section{Model for the electron saturation} \label{sec:SI_TorreyModel}

To fit the best model parameters as shown in Fig.~3, we adopted the Torrey model \cite{Torrey1949} of damped Rabi oscillations (spin Rabi oscillations in the presence of relaxation) to describe the reduced electron polarization under MW irradiation (partial saturation of the electrons).
For long time scales as in CW (continuous wave) DNP build-ups, only the time-independent terms need to be considered, although other terms might play a role in non-CW MW irradiation which potentially could result in higher electron recruitment. 
The time-independent part of the Torrey model describing the ratio between the electron polarization at infinite time under MW irradiation ($P_\mathrm{{e},\infty}$) to the thermal electron polarization of the system after rewriting from \cite{Torrey1949} takes the form
\begin{align}
	\frac{P_\mathrm{{e},\infty}}{P_{0,\mathrm{e}}} = \frac{\left(1-\frac{\omega_\mathrm{MW}}{\omega_{0,\mathrm{e}}}\right)^2\gamma_\mathrm{e}^2B_{1,\mathrm{MW}}^2T_{2,\mathrm{e}}^2+1}{\gamma_\mathrm{e}^2B_{1,\mathrm{MW}}^2T_{2,\mathrm{e}}\left(\left(1-\frac{\omega_\mathrm{MW}}{\omega_{0,\mathrm{e}}}\right)^2T_{2,\mathrm{e}}+T_{1,\mathrm{e}}\right)+1}  \label{eq:Torrey_model}
\end{align}
with the electron gyromagnetic ratio $\gamma_e$, its associated resonance frequency $\omega_{0,\mathrm{e}}=\gamma_e B_0$, relaxation times $T_{1,\mathrm{e}}$ and $T_{2,\mathrm{e}}$ as well as the MW frequency $\omega_\mathrm{MW}$ and the MW field strength $B_{1,\mathrm{MW}}$.
The resulting expression is identical to the steady-state solution of the z-magnetization of the Bloch equations for the electrons.
Electron spectral diffusion and electron line broadening are not explicitly included in the approach but might give rise to electron relaxation times differing from those observed in typical EPR measurements.

Since DNP employs (near-) resonant MW irradiation and $T_{2,\mathrm{e}}\ll T_{1,\mathrm{e}}$ at low temperatures and high electron concentrations, Eq.~\eqref{eq:Torrey_model} can be simplified to 
\begin{align}
    1-\frac{P_\mathrm{{e},\infty}}{P_{0,\mathrm{e}}} = 1- \frac{1}{\gamma_\mathrm{e}^2B_{1,\mathrm{MW}}^2T_{2,\mathrm{e}}T_{1,\mathrm{e}}+1}  
\end{align}

\clearpage

\section{Mathematica notebook of two-electron two-nucleus four-spin system}

The Schrieffer-Wolff transformations discussed in this work were computed with Mathamatica. 
The Mathematica notebook of the lab frame two-electron two-nucleus four-spin system is found below.
Mathematica notebooks for the rotating frame and the one-electron one-nucleus two spin system with MW irradiation were derived from the notebook below.



\section*{Supplementary material}

```
In[ ]:= << NMR11/StartUpNMR.m
       << mathematica/NMRmaer.m ;
       << Notation`;
```

# NMR with Mathematica
*Version 1.1*
*Marlies Brinksma*
*Email: mabi@solidmr.kun.nl*
© 1998, Marlies Brinksma, Matthias Ernst and Beat Meier, Department of Physical Chemistry, University of Nijmegen, The Netherlands

C:\Program Files\Wolfram Research\Mathematica\13.3\AddOns\Applications\NMR11\Spinsystem.m
  loaded...

NormalForm: $Pre is set to: CheckTimes

C:\Program Files\Wolfram Research\Mathematica\13.3\AddOns\Applications\NMR11\NormalForm.m
  loaded...

C:\Program Files\Wolfram Research\Mathematica\13.3\AddOns\Applications\NMR11\SOFunctions.m
  loaded...

⋯ SetDelayed: Tag SquareMatrixQ in SquareMatrixQ[mat_?MatrixQ] is Protected.

⋯ SetDelayed: Tag SquareMatrixQ in SquareMatrixQ[expr_] is Protected.

C:\Program Files\Wolfram Research\Mathematica\13.3\AddOns\Applications\NMR11\Matrix.m
  loaded...

C:\Program Files\Wolfram Research\Mathematica\13.3\AddOns\Applications\NMR11\QMFunctions.m
  loaded...

C:\Program Files\Wolfram Research\Mathematica\13.3\AddOns\Applications\NMR11\NMRFunctions.m
  loaded...

⋯ Get: Cannot open mathematica/NMRmaer.m.

```
In[ ]:= Symbolize[ω_{e2}] (*ω_e=-γ_eB_0>0 as γ_e<0 *)
       Symbolize[ω_{e1}] (*ω_e=-γ_eB_0>0 as γ_e<0 *)
       Symbolize[ω_n]  (*ω_n=-γ_nB_0<0 as γ_n>0 *)
       Symbolize[D^{++}]
       Symbolize[D^{+-}]
       Symbolize[D^{-+}]
       Symbolize[D^{--}]
       Symbolize[D^{z-}]
       Symbolize[D^{z+}]
       Symbolize[D^{+z}]
       Symbolize[D^{-z}]
       Symbolize[D^{zz}]

       Symbolize[d^{zz}]
       Symbolize[d^{++}]
       Symbolize[d^{+-}]
       Symbolize[d^{-+}]
       Symbolize[d^{--}]
       Symbolize[d^{z-}]
       Symbolize[d^{z+}]
```

```
Symbolize[d⁺ᶻ]
Symbolize[d⁻ᶻ]

Symbolize[A₁₃⁺⁺]
Symbolize[A₁₃⁺⁻]
Symbolize[A₁₃⁻⁺]
Symbolize[A₁₃⁻⁻]
Symbolize[A₁₃⁺ᶻ]
Symbolize[A₁₃⁻ᶻ]
Symbolize[A₁₃ᶻ⁺]
Symbolize[A₁₃ᶻ⁻]
Symbolize[A₁₃ᶻᶻ]

Symbolize[A₂₃⁺⁺]
Symbolize[A₂₃⁺⁻]
Symbolize[A₂₃⁻⁺]
Symbolize[A₂₃⁻⁻]
Symbolize[A₂₃⁺ᶻ]
Symbolize[A₂₃⁻ᶻ]
Symbolize[A₂₃ᶻ⁺]
Symbolize[A₂₃ᶻ⁻]
Symbolize[A₂₃ᶻᶻ]

Symbolize[A₁₄⁺⁺]
Symbolize[A₁₄⁺⁻]
Symbolize[A₁₄⁻⁺]
Symbolize[A₁₄⁻⁻]
Symbolize[A₁₄⁺ᶻ]
Symbolize[A₁₄⁻ᶻ]
Symbolize[A₁₄ᶻ⁺]
Symbolize[A₁₄ᶻ⁻]
Symbolize[A₁₄ᶻᶻ]

Symbolize[A₂₄⁺⁺]
Symbolize[A₂₄⁺⁻]
Symbolize[A₂₄⁻⁺]
Symbolize[A₂₄⁻⁻]
Symbolize[A₂₄⁺ᶻ]
Symbolize[A₂₄⁻ᶻ]
Symbolize[A₂₄ᶻ⁺]
Symbolize[A₂₄ᶻ⁻]
Symbolize[A₂₄ᶻᶻ]

Symbolize[A₁₃]
Symbolize[A₂₃]
Symbolize[A₁₄]
Symbolize[A₂₄]
```

```
Symbolize[D_12]
Symbolize[d_34]

Symbolize[S_1]
Symbolize[S_2]
Symbolize[I_3]
Symbolize[I_4]

Symbolize[S_1z]
Symbolize[S_2z]
Symbolize[I_3z]
Symbolize[I_4z]

Symbolize[H_0]
```

QMFunctions: Calculation done in the subspace: {}

```
In[ ]:= RegisterSpin[{1, 2, 3, 4}, {1 / 2, 1 / 2, 1 / 2, 1 / 2}]
```

Spinsystem: Spin 1 has been registered…

Spinsystem: Spin 2 has been registered…

Spinsystem: Spin 3 has been registered…

General: Further output of Spinsystem::Registered will be suppressed during this calculation.

```
In[ ]:= SetNucleus[{1, 2, 3, 4}, {"S", "S", "I", "I"}]
```
⌊ima⋯ ⌊imagir

```
In[ ]:= A_13 = {{A_13^{++}, A_13^{+-}, A_13^{+z}}, {A_13^{-+}, A_13^{--}, A_13^{-z}}, {A_13^{z+}, A_13^{z-}, A_13^{zz}}};
       A_14 = {{A_14^{++}, A_14^{+-}, A_14^{+z}}, {A_14^{-+}, A_14^{--}, A_14^{-z}}, {A_14^{z+}, A_14^{z-}, A_14^{zz}}};
       A_23 = {{A_23^{++}, A_23^{+-}, A_23^{+z}}, {A_23^{-+}, A_23^{--}, A_23^{-z}}, {A_23^{z+}, A_23^{z-}, A_23^{zz}}};
       A_24 = {{A_24^{++}, A_24^{+-}, A_24^{+z}}, {A_24^{-+}, A_24^{--}, A_24^{-z}}, {A_24^{z+}, A_24^{z-}, A_24^{zz}}};
       D_12 = {{D^{++}, D^{+-}, D^{+z}}, {D^{-+}, D^{--}, D^{-z}}, {D^{z+}, D^{z-}, D^{zz}}};
       d_34 = {{d^{++}, d^{+-}, d^{+z}}, {d^{-+}, d^{--}, d^{-z}}, {d^{z+}, d^{z-}, d^{zz}}};

In[ ]:= S_1z = {0, 0, S[1, z]};
       S_2z = {0, 0, S[2, z]};
       I_3z = {0, 0, S[3, z]};
       I_4z = {0, 0, S[4, z]};

In[ ]:= S_1 = {S[1, "+"], S[1, "-"], S[1, z]};
       S_2 = {S[2, "+"], S[2, "-"], S[2, z]};
       I_3 = {S[3, "+"], S[3, "-"], S[3, z]};
       I_4 = {S[4, "+"], S[4, "-"], S[4, z]};
```

*In[ ]:=* (*$\omega_e = -\gamma_e B_0 > 0$ as $\gamma_e < 0$ and $\omega_n = -\gamma_n B_0 < 0$ as $\gamma_n > 0$ throughout this work, requiring a - sign in front of the nuclear Zeeman energy term to have - $\omega_n$ as the ground state energy*)
Ham = $\omega_{e1}$ S[1, z] + $\omega_{e2}$ S[2, z] - $\omega_n$ (S[3, z] + S[4, z]) +
    $S_1.D_{12}.S_2 + I_3.d_{34}.I_4 + S_1.A_{13}.I_3 + S_1.A_{14}.I_4 + S_2.A_{23}.I_3 + S_2.A_{24}.I_4$

*Out[ ]=*
$\omega_{e1} S_{1,z} + (D^{--} S_{1,-} + D^{+-} S_{1,+} + D^{z-} S_{1,z}) S_{2,-} + (D^{-+} S_{1,-} + D^{++} S_{1,+} + D^{z+} S_{1,z}) S_{2,+} +$
$\omega_{e2} S_{2,z} + (D^{-z} S_{1,-} + D^{+z} S_{1,+} + D^{zz} S_{1,z}) S_{2,z} + (A_{13}^{--} S_{1,-} + A_{13}^{+-} S_{1,+} + A_{13}^{z-} S_{1,z}) I_{3,-} +$
$(A_{23}^{--} S_{2,-} + A_{23}^{+-} S_{2,+} + A_{23}^{z-} S_{2,z}) I_{3,-} + (A_{13}^{-+} S_{1,-} + A_{13}^{++} S_{1,+} + A_{13}^{z+} S_{1,z}) I_{3,+} +$
$(A_{23}^{-+} S_{2,-} + A_{23}^{++} S_{2,+} + A_{23}^{z+} S_{2,z}) I_{3,+} + (A_{13}^{-z} S_{1,-} + A_{13}^{+z} S_{1,+} + A_{13}^{zz} S_{1,z}) I_{3,z} +$
$(A_{23}^{-z} S_{2,-} + A_{23}^{+z} S_{2,+} + A_{23}^{zz} S_{2,z}) I_{3,z} + (A_{14}^{--} S_{1,-} + A_{14}^{+-} S_{1,+} + A_{14}^{z-} S_{1,z}) I_{4,-} +$
$(A_{24}^{--} S_{2,-} + A_{24}^{+-} S_{2,+} + A_{24}^{z-} S_{2,z}) I_{4,-} + (d^{--} I_{3,-} + d^{+-} I_{3,+} + d^{z-} I_{3,z}) I_{4,-} +$
$(A_{14}^{-+} S_{1,-} + A_{14}^{++} S_{1,+} + A_{14}^{z+} S_{1,z}) I_{4,+} + (A_{24}^{-+} S_{2,-} + A_{24}^{++} S_{2,+} + A_{24}^{z+} S_{2,z}) I_{4,+} +$
$(d^{-+} I_{3,-} + d^{++} I_{3,+} + d^{z+} I_{3,z}) I_{4,+} + (A_{14}^{-z} S_{1,-} + A_{14}^{+z} S_{1,+} + A_{14}^{zz} S_{1,z}) I_{4,z} +$
$(A_{24}^{-z} S_{2,-} + A_{24}^{+z} S_{2,+} + A_{24}^{zz} S_{2,z}) I_{4,z} + (d^{-z} I_{3,-} + d^{+z} I_{3,+} + d^{zz} I_{3,z}) I_{4,z} - \omega_n (I_{3,z} + I_{4,z})$

*In[ ]:=* $H_0$ = $\omega_{e1}$ S[1, z] + $\omega_{e2}$ S[2, z] - $\omega_n$ (S[3, z] + S[4, z]) + $S_{1z}.D_{12}.S_{2z}$ +
    $I_{3z}.d_{34}.I_{4z} + S_{1z}.A_{13}.I_{3z} + S_{1z}.A_{14}.I_{4z} + S_{2z}.A_{23}.I_{3z} + S_{2z}.A_{24}.I_{4z}$

*Out[ ]=*
$\omega_{e1} S_{1,z} + \omega_{e2} S_{2,z} + D^{zz} S_{1,z} S_{2,z} + A_{13}^{zz} S_{1,z} I_{3,z} +$
$A_{23}^{zz} S_{2,z} I_{3,z} + A_{14}^{zz} S_{1,z} I_{4,z} + A_{24}^{zz} S_{2,z} I_{4,z} + d^{zz} I_{3,z} I_{4,z} - \omega_n (I_{3,z} + I_{4,z})$

*In[ ]:=* V = Simplify[Ham - $H_0$]
         ⎣vereinfache

*Out[ ]=*
$D^{z-} S_{1,z} S_{2,-} + D^{z+} S_{1,z} S_{2,+} + A_{13}^{z-} S_{1,z} I_{3,-} + A_{23}^{--} S_{2,-} I_{3,-} + A_{23}^{+-} S_{2,+} I_{3,-} + A_{23}^{z-} S_{2,z} I_{3,-} +$
$A_{13}^{z+} S_{1,z} I_{3,+} + A_{23}^{-+} S_{2,-} I_{3,+} + A_{23}^{++} S_{2,+} I_{3,+} + A_{23}^{z+} S_{2,z} I_{3,+} + A_{23}^{-z} S_{2,-} I_{3,z} + A_{23}^{+z} S_{2,+} I_{3,z} +$
$A_{14}^{z-} S_{1,z} I_{4,-} + A_{24}^{--} S_{2,-} I_{4,-} + A_{24}^{+-} S_{2,+} I_{4,-} + A_{24}^{z-} S_{2,z} I_{4,-} + d^{--} I_{3,-} I_{4,-} + d^{+-} I_{3,+} I_{4,-} +$
$d^{z-} I_{3,z} I_{4,-} + A_{14}^{z+} S_{1,z} I_{4,+} + A_{24}^{-+} S_{2,-} I_{4,+} + A_{24}^{++} S_{2,+} I_{4,+} + A_{24}^{z+} S_{2,z} I_{4,+} + d^{-+} I_{3,-} I_{4,+} +$
$d^{++} I_{3,+} I_{4,+} + d^{z+} I_{3,z} I_{4,+} + A_{24}^{-z} S_{2,-} I_{4,z} + A_{24}^{+z} S_{2,+} I_{4,z} + d^{-z} I_{3,-} I_{4,z} + d^{+z} I_{3,+} I_{4,z} +$
$S_{1,-} (D^{--} S_{2,-} + D^{-+} S_{2,+} + D^{-z} S_{2,z} + A_{13}^{--} I_{3,-} + A_{13}^{-+} I_{3,+} + A_{13}^{-z} I_{3,z} + A_{14}^{--} I_{4,-} + A_{14}^{-+} I_{4,+} + A_{14}^{-z} I_{4,z}) +$
$S_{1,+} (D^{+-} S_{2,-} + D^{++} S_{2,+} + D^{+z} S_{2,z} + A_{13}^{+-} I_{3,-} + A_{13}^{++} I_{3,+} + A_{13}^{+z} I_{3,z} + A_{14}^{+-} I_{4,-} + A_{14}^{++} I_{4,+} + A_{14}^{+z} I_{4,z})$

*In[ ]:=* VMx = MatrixRepresentation[V, {1, 2, 3, 4}];

*In[ ]:=* H0Mx = MatrixRepresentation[$H_0$, {1, 2, 3, 4}];

```
In[ ]:= P = {
        {0, 0, 0, 0, 0, 0, 0, 0, 0, 0, 0, 0, 1, 0, 0, 0},
        {0, 0, 0, 0, 0, 0, 0, 0, 0, 0, 0, 0, 0, 0, 1, 0},
        {0, 0, 0, 0, 0, 0, 0, 0, 0, 0, 0, 0, 0, 1, 0, 0},
        {0, 0, 0, 0, 0, 0, 0, 0, 0, 0, 0, 0, 0, 0, 0, 1},
        {0, 0, 0, 0, 1, 0, 0, 0, 0, 0, 0, 0, 0, 0, 0, 0},
        {0, 0, 0, 0, 0, 0, 1, 0, 0, 0, 0, 0, 0, 0, 0, 0},
        {0, 0, 0, 0, 0, 1, 0, 0, 0, 0, 0, 0, 0, 0, 0, 0},
        {0, 0, 0, 0, 0, 0, 0, 1, 0, 0, 0, 0, 0, 0, 0, 0},
        {0, 0, 0, 0, 0, 0, 0, 0, 1, 0, 0, 0, 0, 0, 0, 0},
        {0, 0, 0, 0, 0, 0, 0, 0, 1, 0, 0, 0, 0, 0, 0, 0},
        {0, 0, 0, 0, 0, 0, 0, 0, 0, 0, 1, 0, 0, 0, 0, 0},
        {0, 0, 0, 0, 0, 0, 0, 0, 0, 0, 0, 1, 0, 0, 0, 0},
        {1, 0, 0, 0, 0, 0, 0, 0, 0, 0, 0, 0, 0, 0, 0, 0},
        {0, 0, 1, 0, 0, 0, 0, 0, 0, 0, 0, 0, 0, 0, 0, 0},
        {0, 1, 0, 0, 0, 0, 0, 0, 0, 0, 0, 0, 0, 0, 0, 0},
        {0, 0, 0, 1, 0, 0, 0, 0, 0, 0, 0, 0, 0, 0, 0, 0}};
Bs = {αααα, αααβ, ααβα, ααββ, αβαα, αβαβ,
      αββα, αβββ, βααα, βααβ, βαβα, βαββ, ββαα, ββαβ, βββα, ββββ};
P.Bs
```

Out[ ]=
{ββαα, βββα, ββαβ, ββββ, αβαα, αββα, αβαβ,
 αβββ, βααβ, βααα, βαβα, βαββ, αααα, ααβα, αααβ, ααββ}

```
In[ ]:= VM = P.VMx.Inverse[P]
```

Out[ ]=

$$\left\{\left\{0, -\frac{A_{13}^{z+}}{2} - \frac{A_{23}^{z+}}{2} + \frac{d^{+z}}{2}, -\frac{A_{14}^{z+}}{2} - \frac{A_{24}^{z+}}{2} + \frac{d^{z+}}{2}, d^{++}, \frac{A_{13}^{-z}}{2} + \frac{A_{14}^{-z}}{2} - \frac{D^{-z}}{2}, A_{13}^{-+}, A_{14}^{-+}, 0, A_{24}^{-+},\right.\right.$$
$$\left.\frac{A_{23}^{-z}}{2} + \frac{A_{24}^{-z}}{2} - \frac{D^{z-}}{2}, A_{23}^{-+}, 0, D^{--}, 0, 0, 0\right\}, \left\{-\frac{A_{13}^{z-}}{2} - \frac{A_{23}^{z-}}{2} + \frac{d^{z-}}{2}, 0, d^{-+}, -\frac{A_{14}^{z+}}{2} - \frac{A_{24}^{z+}}{2} - \frac{d^{z+}}{2},\right.$$
$$\left.A_{13}^{--}, -\frac{A_{13}^{-z}}{2} + \frac{A_{14}^{-z}}{2} - \frac{D^{-z}}{2}, 0, A_{14}^{-+}, 0, A_{23}^{--}, -\frac{A_{23}^{-z}}{2} + \frac{A_{24}^{-z}}{2} - \frac{D^{z-}}{2}, A_{24}^{-+}, 0, D^{--}, 0, 0\right\},$$
$$\left\{-\frac{A_{14}^{z-}}{2} - \frac{A_{24}^{z-}}{2} + \frac{d^{z-}}{2}, d^{+-}, 0, -\frac{A_{13}^{z+}}{2} - \frac{A_{23}^{z+}}{2} - \frac{d^{+z}}{2}, A_{14}^{--}, 0, \frac{A_{13}^{-z}}{2} - \frac{A_{14}^{-z}}{2} - \frac{D^{-z}}{2}, A_{13}^{-+},\right.$$
$$\left.\frac{A_{23}^{-z}}{2} - \frac{A_{24}^{-z}}{2} - \frac{D^{z-}}{2}, A_{24}^{--}, 0, A_{23}^{-+}, 0, 0, D^{--}, 0\right\}, \left\{d^{--}, -\frac{A_{14}^{z-}}{2} - \frac{A_{24}^{z-}}{2} - \frac{d^{z-}}{2}, -\frac{A_{13}^{z-}}{2} - \frac{A_{23}^{z-}}{2} - \frac{d^{-z}}{2},\right.$$
$$\left.0, 0, A_{14}^{--}, A_{13}^{--}, -\frac{A_{13}^{-z}}{2} - \frac{A_{14}^{-z}}{2} - \frac{D^{-z}}{2}, A_{23}^{--}, 0, A_{24}^{--}, -\frac{A_{23}^{-z}}{2} - \frac{A_{24}^{-z}}{2} - \frac{D^{z-}}{2}, 0, 0, 0, D^{--}\right\},$$
$$\left\{\frac{A_{13}^{+z}}{2} + \frac{A_{14}^{+z}}{2} - \frac{D^{+z}}{2}, A_{13}^{++}, A_{14}^{++}, 0, 0, \frac{A_{13}^{z+}}{2} - \frac{A_{23}^{z+}}{2} + \frac{d^{+z}}{2}, \frac{A_{14}^{z+}}{2} - \frac{A_{24}^{z+}}{2} + \frac{d^{z+}}{2}, d^{++}, 0, D^{+-}, 0,\right.$$
$$\left.0, \frac{A_{23}^{-z}}{2} + \frac{A_{24}^{-z}}{2} + \frac{D^{z-}}{2}, A_{23}^{-+}, A_{24}^{-+}, 0\right\}, \left\{A_{13}^{+-}, -\frac{A_{13}^{+z}}{2} + \frac{A_{14}^{+z}}{2} - \frac{D^{+z}}{2}, 0, A_{14}^{++}, \frac{A_{13}^{z-}}{2} - \frac{A_{23}^{z-}}{2} + \frac{d^{-z}}{2},\right.$$
$$\left.0, d^{-+}, \frac{A_{14}^{z+}}{2} - \frac{A_{24}^{z+}}{2} - \frac{d^{z+}}{2}, 0, 0, D^{+-}, 0, A_{23}^{--}, -\frac{A_{23}^{-z}}{2} + \frac{A_{24}^{-z}}{2} + \frac{D^{z-}}{2}, 0, A_{24}^{-+}\right\},$$
$$\left\{A_{14}^{+-}, 0, \frac{A_{13}^{+z}}{2} - \frac{A_{14}^{+z}}{2} - \frac{D^{+z}}{2}, A_{13}^{++}, \frac{A_{14}^{z-}}{2} - \frac{A_{24}^{z-}}{2} + \frac{d^{z-}}{2}, d^{+-}, 0, \frac{A_{13}^{z+}}{2} - \frac{A_{23}^{z+}}{2} - \frac{d^{+z}}{2}, D^{+-},\right.$$
$$\left.0, 0, 0, A_{24}^{--}, 0, \frac{A_{23}^{-z}}{2} - \frac{A_{24}^{-z}}{2} + \frac{D^{z-}}{2}, A_{23}^{-+}\right\}, \left\{0, A_{14}^{+-}, A_{13}^{+-}, -\frac{A_{13}^{+z}}{2} - \frac{A_{14}^{+z}}{2} - \frac{D^{+z}}{2}, d^{--},\right.$$

$$\frac{A_{14}^{z-}}{2} - \frac{A_{24}^{z-}}{2} - \frac{d^{z-}}{2}, \frac{A_{13}^{z-}}{2} - \frac{A_{23}^{z-}}{2} - \frac{d^{-z}}{2}, 0, 0, 0, 0, D^{+-}, 0, A_{24}^{--}, A_{23}^{--}, -\frac{A_{23}^{-z}}{2} - \frac{A_{24}^{-z}}{2} + \frac{D^{z-}}{2}\Big\},$$

$$\Big\{A_{24}^{+-}, 0, \frac{A_{23}^{+z}}{2} - \frac{A_{24}^{+z}}{2} - \frac{D^{z+}}{2}, A_{23}^{++}, 0, 0, D^{-+}, 0, 0, -\frac{A_{14}^{z-}}{2} + \frac{A_{24}^{z-}}{2} + \frac{d^{z-}}{2}, d^{+-}, -\frac{A_{13}^{z+}}{2} + \frac{A_{23}^{z+}}{2} - \frac{d^{+z}}{2},$$

$$A_{14}^{--}, 0, \frac{A_{13}^{-z}}{2} - \frac{A_{14}^{-z}}{2} + \frac{D^{-z}}{2}, A_{13}^{-+}\Big\}, \Big\{\frac{A_{23}^{+z}}{2} + \frac{A_{24}^{+z}}{2} - \frac{D^{z+}}{2}, A_{23}^{++}, A_{24}^{++}, 0, D^{-+}, 0, 0, 0,$$

$$-\frac{A_{14}^{z+}}{2} + \frac{A_{24}^{z+}}{2} + \frac{d^{z+}}{2}, 0, -\frac{A_{13}^{z+}}{2} + \frac{A_{23}^{z+}}{2} + \frac{d^{+z}}{2}, d^{++}, \frac{A_{13}^{-z}}{2} + \frac{A_{14}^{-z}}{2} + \frac{D^{-z}}{2}, A_{13}^{-+}, A_{14}^{-+}, 0\Big\},$$

$$\Big\{A_{23}^{+-}, -\frac{A_{23}^{+z}}{2} + \frac{A_{24}^{+z}}{2} - \frac{D^{z+}}{2}, 0, A_{24}^{++}, 0, D^{-+}, 0, 0, d^{+-}, -\frac{A_{13}^{z-}}{2} + \frac{A_{23}^{z-}}{2} + \frac{d^{-z}}{2}, 0, -\frac{A_{14}^{z+}}{2} + \frac{A_{24}^{z+}}{2} - \frac{d^{z+}}{2},$$

$$A_{13}^{--}, -\frac{A_{13}^{-z}}{2} + \frac{A_{14}^{-z}}{2} + \frac{D^{-z}}{2}, 0, A_{14}^{-+}\Big\}, \Big\{0, A_{24}^{+-}, A_{23}^{+-}, -\frac{A_{23}^{+z}}{2} - \frac{A_{24}^{+z}}{2} - \frac{D^{z+}}{2}, 0, 0, 0, D^{-+},$$

$$-\frac{A_{13}^{z-}}{2} + \frac{A_{23}^{z-}}{2} - \frac{d^{z-}}{2}, d^{--}, -\frac{A_{14}^{z-}}{2} + \frac{A_{24}^{z-}}{2} - \frac{d^{z-}}{2}, 0, 0, A_{14}^{--}, A_{13}^{--}, -\frac{A_{13}^{-z}}{2} - \frac{A_{14}^{-z}}{2} + \frac{D^{-z}}{2}\Big\},$$

$$\Big\{D^{++}, 0, 0, 0, \frac{A_{23}^{+z}}{2} + \frac{A_{24}^{+z}}{2} + \frac{D^{z+}}{2}, A_{23}^{++}, A_{24}^{++}, 0, A_{14}^{++}, \frac{A_{13}^{+z}}{2} + \frac{A_{14}^{+z}}{2} + \frac{D^{+z}}{2}, A_{13}^{++}, 0, 0,$$

$$\frac{A_{13}^{z+}}{2} + \frac{A_{23}^{z+}}{2} + \frac{d^{z+}}{2}, \frac{A_{14}^{z+}}{2} + \frac{A_{24}^{z+}}{2} + \frac{d^{z+}}{2}, d^{++}\Big\}, \Big\{0, D^{++}, 0, 0, A_{23}^{+-}, -\frac{A_{23}^{+z}}{2} + \frac{A_{24}^{+z}}{2} + \frac{D^{z+}}{2}, 0,$$

$$A_{24}^{++}, 0, A_{13}^{+-}, -\frac{A_{13}^{+z}}{2} + \frac{A_{14}^{+z}}{2} + \frac{D^{+z}}{2}, A_{14}^{++}, \frac{A_{13}^{z-}}{2} + \frac{A_{23}^{z-}}{2} + \frac{d^{-z}}{2}, 0, d^{-+}, \frac{A_{14}^{z+}}{2} + \frac{A_{24}^{z+}}{2} - \frac{d^{z+}}{2}\Big\},$$

$$\Big\{0, 0, D^{++}, 0, A_{24}^{+-}, 0, \frac{A_{23}^{+z}}{2} - \frac{A_{24}^{+z}}{2} + \frac{D^{z+}}{2}, A_{23}^{++}, \frac{A_{13}^{+z}}{2} - \frac{A_{14}^{+z}}{2} + \frac{D^{+z}}{2}, A_{14}^{+-}, 0, A_{13}^{++},$$

$$\frac{A_{14}^{z-}}{2} + \frac{A_{24}^{z-}}{2} + \frac{d^{z-}}{2}, d^{+-}, 0, \frac{A_{13}^{z+}}{2} + \frac{A_{23}^{z+}}{2} - \frac{d^{+z}}{2}\Big\}, \Big\{0, 0, 0, D^{++}, 0, A_{24}^{+-}, A_{23}^{+-}, -\frac{A_{23}^{+z}}{2} - \frac{A_{24}^{+z}}{2} + \frac{D^{z+}}{2},$$

$$A_{13}^{+-}, 0, A_{14}^{+-}, -\frac{A_{13}^{+z}}{2} - \frac{A_{14}^{+z}}{2} + \frac{D^{+z}}{2}, d^{--}, \frac{A_{14}^{z-}}{2} + \frac{A_{24}^{z-}}{2} - \frac{d^{z-}}{2}, \frac{A_{13}^{z-}}{2} + \frac{A_{23}^{z-}}{2} - \frac{d^{-z}}{2}, 0\Big\}\Big\}$$

*In[ ]:=* `H0M = P.H0Mx.Inverse[P]`
                                     inverse Matrix

Out[ ]=

$$\left\{\left\{-\frac{A_{13}^{zz}}{4}-\frac{A_{14}^{zz}}{4}-\frac{A_{23}^{zz}}{4}-\frac{A_{24}^{zz}}{4}+\frac{d^{zz}}{4}+\frac{D^{zz}}{4}-\frac{\omega_{e1}}{2}-\frac{\omega_{e2}}{2}-\omega_n, 0, 0, 0, 0, 0, 0, 0, 0, 0, 0, 0, 0, 0, 0, 0\right\},\right.$$

$$\left\{0, \frac{A_{13}^{zz}}{4}-\frac{A_{14}^{zz}}{4}+\frac{A_{23}^{zz}}{4}-\frac{A_{24}^{zz}}{4}-\frac{d^{zz}}{4}+\frac{D^{zz}}{4}-\frac{\omega_{e1}}{2}-\frac{\omega_{e2}}{2}, 0, 0, 0, 0, 0, 0, 0, 0, 0, 0, 0, 0, 0, 0\right\},$$

$$\left\{0, 0, -\frac{A_{13}^{zz}}{4}+\frac{A_{14}^{zz}}{4}-\frac{A_{23}^{zz}}{4}+\frac{A_{24}^{zz}}{4}-\frac{d^{zz}}{4}+\frac{D^{zz}}{4}-\frac{\omega_{e1}}{2}-\frac{\omega_{e2}}{2}, 0, 0, 0, 0, 0, 0, 0, 0, 0, 0, 0, 0, 0\right\},$$

$$\left\{0, 0, 0, \frac{A_{13}^{zz}}{4}+\frac{A_{14}^{zz}}{4}+\frac{A_{23}^{zz}}{4}+\frac{A_{24}^{zz}}{4}+\frac{d^{zz}}{4}+\frac{D^{zz}}{4}-\frac{\omega_{e1}}{2}-\frac{\omega_{e2}}{2}+\omega_n, 0, 0, 0, 0, 0, 0, 0, 0, 0, 0, 0, 0\right\},$$

$$\left\{0, 0, 0, 0, \frac{A_{13}^{zz}}{4}+\frac{A_{14}^{zz}}{4}-\frac{A_{23}^{zz}}{4}-\frac{A_{24}^{zz}}{4}+\frac{d^{zz}}{4}-\frac{D^{zz}}{4}+\frac{\omega_{e1}}{2}-\frac{\omega_{e2}}{2}-\omega_n, 0, 0, 0, 0, 0, 0, 0, 0, 0, 0, 0\right\},$$

$$\left\{0, 0, 0, 0, 0, -\frac{A_{13}^{zz}}{4}+\frac{A_{14}^{zz}}{4}+\frac{A_{23}^{zz}}{4}-\frac{A_{24}^{zz}}{4}-\frac{d^{zz}}{4}-\frac{D^{zz}}{4}+\frac{\omega_{e1}}{2}-\frac{\omega_{e2}}{2}, 0, 0, 0, 0, 0, 0, 0, 0, 0, 0\right\},$$

$$\left\{0, 0, 0, 0, 0, 0, \frac{A_{13}^{zz}}{4}-\frac{A_{14}^{zz}}{4}-\frac{A_{23}^{zz}}{4}+\frac{A_{24}^{zz}}{4}-\frac{d^{zz}}{4}-\frac{D^{zz}}{4}+\frac{\omega_{e1}}{2}-\frac{\omega_{e2}}{2}, 0, 0, 0, 0, 0, 0, 0, 0, 0\right\},$$

$$\left\{0, 0, 0, 0, 0, 0, 0, -\frac{A_{13}^{zz}}{4}-\frac{A_{14}^{zz}}{4}+\frac{A_{23}^{zz}}{4}+\frac{A_{24}^{zz}}{4}+\frac{d^{zz}}{4}-\frac{D^{zz}}{4}+\frac{\omega_{e1}}{2}-\frac{\omega_{e2}}{2}+\omega_n, 0, 0, 0, 0, 0, 0, 0, 0\right\},$$

$$\left\{0, 0, 0, 0, 0, 0, 0, 0, -\frac{A_{13}^{zz}}{4}+\frac{A_{14}^{zz}}{4}+\frac{A_{23}^{zz}}{4}-\frac{A_{24}^{zz}}{4}-\frac{d^{zz}}{4}-\frac{D^{zz}}{4}-\frac{\omega_{e1}}{2}+\frac{\omega_{e2}}{2}, 0, 0, 0, 0, 0, 0, 0\right\},$$

$$\left\{0, 0, 0, 0, 0, 0, 0, 0, 0, -\frac{A_{13}^{zz}}{4}-\frac{A_{14}^{zz}}{4}+\frac{A_{23}^{zz}}{4}+\frac{A_{24}^{zz}}{4}+\frac{d^{zz}}{4}-\frac{D^{zz}}{4}-\frac{\omega_{e1}}{2}+\frac{\omega_{e2}}{2}-\omega_n, 0, 0, 0, 0, 0, 0\right\},$$

$$\left\{0, 0, 0, 0, 0, 0, 0, 0, 0, 0, \frac{A_{13}^{zz}}{4}-\frac{A_{14}^{zz}}{4}-\frac{A_{23}^{zz}}{4}+\frac{A_{24}^{zz}}{4}-\frac{d^{zz}}{4}-\frac{D^{zz}}{4}-\frac{\omega_{e1}}{2}+\frac{\omega_{e2}}{2}, 0, 0, 0, 0, 0\right\},$$

$$\left\{0, 0, 0, 0, 0, 0, 0, 0, 0, 0, 0, \frac{A_{13}^{zz}}{4}+\frac{A_{14}^{zz}}{4}-\frac{A_{23}^{zz}}{4}-\frac{A_{24}^{zz}}{4}+\frac{d^{zz}}{4}-\frac{D^{zz}}{4}-\frac{\omega_{e1}}{2}+\frac{\omega_{e2}}{2}+\omega_n, 0, 0, 0, 0\right\},$$

$$\left\{0, 0, 0, 0, 0, 0, 0, 0, 0, 0, 0, 0, \frac{A_{13}^{zz}}{4}+\frac{A_{14}^{zz}}{4}+\frac{A_{23}^{zz}}{4}+\frac{A_{24}^{zz}}{4}+\frac{d^{zz}}{4}+\frac{D^{zz}}{4}+\frac{\omega_{e1}}{2}+\frac{\omega_{e2}}{2}-\omega_n, 0, 0, 0\right\},$$

$$\left\{0, 0, 0, 0, 0, 0, 0, 0, 0, 0, 0, 0, 0, -\frac{A_{13}^{zz}}{4}+\frac{A_{14}^{zz}}{4}-\frac{A_{23}^{zz}}{4}+\frac{A_{24}^{zz}}{4}-\frac{d^{zz}}{4}+\frac{D^{zz}}{4}+\frac{\omega_{e1}}{2}+\frac{\omega_{e2}}{2}, 0, 0\right\},$$

$$\left\{0, 0, 0, 0, 0, 0, 0, 0, 0, 0, 0, 0, 0, 0, \frac{A_{13}^{zz}}{4}-\frac{A_{14}^{zz}}{4}+\frac{A_{23}^{zz}}{4}-\frac{A_{24}^{zz}}{4}-\frac{d^{zz}}{4}+\frac{D^{zz}}{4}+\frac{\omega_{e1}}{2}+\frac{\omega_{e2}}{2}, 0\right\},$$

$$\left.\left\{0, 0, 0, 0, 0, 0, 0, 0, 0, 0, 0, 0, 0, 0, 0, -\frac{A_{13}^{zz}}{4}-\frac{A_{14}^{zz}}{4}-\frac{A_{23}^{zz}}{4}-\frac{A_{24}^{zz}}{4}+\frac{d^{zz}}{4}+\frac{D^{zz}}{4}+\frac{\omega_{e1}}{2}+\frac{\omega_{e2}}{2}+\omega_n\right\}\right\}$$

```mathematica
In[ ]:= SM = Table[SMe[i, k], {i, 1, 16}, {k, 1, 16}]
```

Out[ ]=

{{SMe[1, 1], SMe[1, 2], SMe[1, 3], SMe[1, 4], SMe[1, 5],
  SMe[1, 6], SMe[1, 7], SMe[1, 8], SMe[1, 9], SMe[1, 10], SMe[1, 11],
  SMe[1, 12], SMe[1, 13], SMe[1, 14], SMe[1, 15], SMe[1, 16]},
 {SMe[2, 1], SMe[2, 2], SMe[2, 3], SMe[2, 4], SMe[2, 5], SMe[2, 6],
  SMe[2, 7], SMe[2, 8], SMe[2, 9], SMe[2, 10], SMe[2, 11],
  SMe[2, 12], SMe[2, 13], SMe[2, 14], SMe[2, 15], SMe[2, 16]},
 {SMe[3, 1], SMe[3, 2], SMe[3, 3], SMe[3, 4], SMe[3, 5], SMe[3, 6],
  SMe[3, 7], SMe[3, 8], SMe[3, 9], SMe[3, 10], SMe[3, 11],
  SMe[3, 12], SMe[3, 13], SMe[3, 14], SMe[3, 15], SMe[3, 16]},
 {SMe[4, 1], SMe[4, 2], SMe[4, 3], SMe[4, 4], SMe[4, 5], SMe[4, 6],
  SMe[4, 7], SMe[4, 8], SMe[4, 9], SMe[4, 10], SMe[4, 11],
  SMe[4, 12], SMe[4, 13], SMe[4, 14], SMe[4, 15], SMe[4, 16]},
 {SMe[5, 1], SMe[5, 2], SMe[5, 3], SMe[5, 4], SMe[5, 5], SMe[5, 6],
  SMe[5, 7], SMe[5, 8], SMe[5, 9], SMe[5, 10], SMe[5, 11],
  SMe[5, 12], SMe[5, 13], SMe[5, 14], SMe[5, 15], SMe[5, 16]},
 {SMe[6, 1], SMe[6, 2], SMe[6, 3], SMe[6, 4], SMe[6, 5], SMe[6, 6], SMe[6, 7],
  SMe[6, 8], SMe[6, 9], SMe[6, 10], SMe[6, 11], SMe[6, 12], SMe[6, 13],
  SMe[6, 14], SMe[6, 15], SMe[6, 16]}, {SMe[7, 1], SMe[7, 2], SMe[7, 3],
  SMe[7, 4], SMe[7, 5], SMe[7, 6], SMe[7, 7], SMe[7, 8], SMe[7, 9], SMe[7, 10],
  SMe[7, 11], SMe[7, 12], SMe[7, 13], SMe[7, 14], SMe[7, 15], SMe[7, 16]},
 {SMe[8, 1], SMe[8, 2], SMe[8, 3], SMe[8, 4], SMe[8, 5], SMe[8, 6], SMe[8, 7],
  SMe[8, 8], SMe[8, 9], SMe[8, 10], SMe[8, 11], SMe[8, 12], SMe[8, 13],
  SMe[8, 14], SMe[8, 15], SMe[8, 16]}, {SMe[9, 1], SMe[9, 2], SMe[9, 3],
  SMe[9, 4], SMe[9, 5], SMe[9, 6], SMe[9, 7], SMe[9, 8], SMe[9, 9], SMe[9, 10],
  SMe[9, 11], SMe[9, 12], SMe[9, 13], SMe[9, 14], SMe[9, 15], SMe[9, 16]},
 {SMe[10, 1], SMe[10, 2], SMe[10, 3], SMe[10, 4], SMe[10, 5], SMe[10, 6],
  SMe[10, 7], SMe[10, 8], SMe[10, 9], SMe[10, 10], SMe[10, 11],
  SMe[10, 12], SMe[10, 13], SMe[10, 14], SMe[10, 15], SMe[10, 16]},
 {SMe[11, 1], SMe[11, 2], SMe[11, 3], SMe[11, 4], SMe[11, 5], SMe[11, 6],
  SMe[11, 7], SMe[11, 8], SMe[11, 9], SMe[11, 10], SMe[11, 11],
  SMe[11, 12], SMe[11, 13], SMe[11, 14], SMe[11, 15], SMe[11, 16]},
 {SMe[12, 1], SMe[12, 2], SMe[12, 3], SMe[12, 4], SMe[12, 5], SMe[12, 6],
  SMe[12, 7], SMe[12, 8], SMe[12, 9], SMe[12, 10], SMe[12, 11],
  SMe[12, 12], SMe[12, 13], SMe[12, 14], SMe[12, 15], SMe[12, 16]},
 {SMe[13, 1], SMe[13, 2], SMe[13, 3], SMe[13, 4], SMe[13, 5], SMe[13, 6],
  SMe[13, 7], SMe[13, 8], SMe[13, 9], SMe[13, 10], SMe[13, 11],
  SMe[13, 12], SMe[13, 13], SMe[13, 14], SMe[13, 15], SMe[13, 16]},
 {SMe[14, 1], SMe[14, 2], SMe[14, 3], SMe[14, 4], SMe[14, 5], SMe[14, 6],
  SMe[14, 7], SMe[14, 8], SMe[14, 9], SMe[14, 10], SMe[14, 11],
  SMe[14, 12], SMe[14, 13], SMe[14, 14], SMe[14, 15], SMe[14, 16]},
 {SMe[15, 1], SMe[15, 2], SMe[15, 3], SMe[15, 4], SMe[15, 5], SMe[15, 6],
  SMe[15, 7], SMe[15, 8], SMe[15, 9], SMe[15, 10], SMe[15, 11],
  SMe[15, 12], SMe[15, 13], SMe[15, 14], SMe[15, 15], SMe[15, 16]},
 {SMe[16, 1], SMe[16, 2], SMe[16, 3], SMe[16, 4], SMe[16, 5], SMe[16, 6],
  SMe[16, 7], SMe[16, 8], SMe[16, 9], SMe[16, 10], SMe[16, 11],
  SMe[16, 12], SMe[16, 13], SMe[16, 14], SMe[16, 15], SMe[16, 16]}}

*In[ ]:=* **solution = Solve[VM + SM.H0M - H0M.SM == 0 , Flatten[SM]]**

⋯ Solve: Equations may not give solutions for all "solve" variables.

*Out[ ]=*

$$\left\{\left\{\text{SMe}[1, 2] \to -\frac{-A_{13}^{z+} - A_{23}^{z+} + d^{+z}}{A_{13}^{zz} + A_{23}^{zz} - d^{zz} + 2\omega_n}, \text{SMe}[1, 3] \to -\frac{-A_{14}^{z+} - A_{24}^{z+} + d^{z+}}{A_{14}^{zz} + A_{24}^{zz} - d^{zz} + 2\omega_n}, \right.\right.$$

$$\text{SMe}[1, 4] \to -\frac{2 d^{++}}{A_{13}^{zz} + A_{14}^{zz} + A_{23}^{zz} + A_{24}^{zz} + 4\omega_n}, \text{SMe}[1, 5] \to -\frac{A_{13}^{-z} + A_{14}^{-z} - D^{-z}}{A_{13}^{zz} + A_{14}^{zz} - D^{zz} + 2\omega_{e1}},$$

$$\text{SMe}[1, 6] \to -\frac{2 A_{13}^{-+}}{A_{14}^{zz} + A_{23}^{zz} - d^{zz} - D^{zz} + 2\omega_{e1} + 2\omega_n}, \text{SMe}[1, 7] \to -\frac{2 A_{14}^{-+}}{A_{13}^{zz} + A_{24}^{zz} - d^{zz} - D^{zz} + 2\omega_{e1} + 2\omega_n},$$

$$\text{SMe}[1, 8] \to 0, \text{SMe}[1, 9] \to -\frac{2 A_{24}^{-+}}{A_{14}^{zz} + A_{23}^{zz} - d^{zz} - D^{zz} + 2\omega_{e2} + 2\omega_n},$$

$$\text{SMe}[1, 10] \to -\frac{A_{23}^{-z} + A_{24}^{-z} - D^{z-}}{A_{23}^{zz} + A_{24}^{zz} - D^{zz} + 2\omega_{e2}}, \text{SMe}[1, 11] \to -\frac{2 A_{23}^{-+}}{A_{13}^{zz} + A_{24}^{zz} - d^{zz} - D^{zz} + 2\omega_{e2} + 2\omega_n},$$

$$\text{SMe}[1, 12] \to 0, \text{SMe}[1, 13] \to -\frac{2 D^{--}}{A_{13}^{zz} + A_{14}^{zz} + A_{23}^{zz} + A_{24}^{zz} + 2\omega_{e1} + 2\omega_{e2}}, \text{SMe}[1, 14] \to 0,$$

$$\text{SMe}[1, 15] \to 0, \text{SMe}[1, 16] \to 0, \text{SMe}[2, 1] \to -\frac{A_{13}^{z-} + A_{23}^{z-} - d^{-z}}{A_{13}^{zz} + A_{23}^{zz} - d^{zz} + 2\omega_n},$$

$$\text{SMe}[2, 3] \to \frac{2 d^{-+}}{A_{13}^{zz} - A_{14}^{zz} + A_{23}^{zz} - A_{24}^{zz}}, \text{SMe}[2, 4] \to -\frac{-A_{14}^{z+} - A_{24}^{z+} - d^{z+}}{A_{14}^{zz} + A_{24}^{zz} + d^{zz} + 2\omega_n},$$

$$\text{SMe}[2, 5] \to -\frac{2 A_{13}^{--}}{A_{14}^{zz} - A_{23}^{zz} + d^{zz} - D^{zz} + 2\omega_{e1} - 2\omega_n}, \text{SMe}[2, 6] \to -\frac{A_{13}^{-z} - A_{14}^{-z} + D^{-z}}{A_{13}^{zz} - A_{14}^{zz} + D^{zz} - 2\omega_{e1}},$$

$$\text{SMe}[2, 7] \to 0, \text{SMe}[2, 8] \to \frac{2 A_{14}^{-+}}{A_{13}^{zz} - A_{24}^{zz} - d^{zz} + D^{zz} - 2\omega_{e1} - 2\omega_n},$$

$$\text{SMe}[2, 9] \to 0, \text{SMe}[2, 10] \to \frac{2 A_{23}^{--}}{A_{13}^{zz} - A_{24}^{zz} - d^{zz} + D^{zz} - 2\omega_{e2} + 2\omega_n},$$

$$\text{SMe}[2, 11] \to -\frac{A_{23}^{-z} - A_{24}^{-z} + D^{z-}}{A_{23}^{zz} - A_{24}^{zz} + D^{zz} - 2\omega_{e2}}, \text{SMe}[2, 12] \to -\frac{2 A_{24}^{-+}}{A_{14}^{zz} - A_{23}^{zz} + d^{zz} - D^{zz} + 2\omega_{e2} + 2\omega_n},$$

$$\text{SMe}[2, 13] \to 0, \text{SMe}[2, 14] \to \frac{2 D^{--}}{A_{13}^{zz} - A_{14}^{zz} + A_{23}^{zz} - A_{24}^{zz} - 2\omega_{e1} - 2\omega_{e2}},$$

$$\text{SMe}[2, 15] \to 0, \text{SMe}[2, 16] \to 0, \text{SMe}[3, 1] \to -\frac{A_{14}^{z-} + A_{24}^{z-} - d^{z-}}{A_{14}^{zz} + A_{24}^{zz} - d^{zz} + 2\omega_n},$$

$$\text{SMe}[3, 2] \to -\frac{2 d^{+-}}{A_{13}^{zz} - A_{14}^{zz} + A_{23}^{zz} - A_{24}^{zz}}, \text{SMe}[3, 4] \to -\frac{-A_{13}^{z+} - A_{23}^{z+} - d^{+z}}{A_{13}^{zz} + A_{23}^{zz} + d^{zz} + 2\omega_n},$$

$$\text{SMe}[3, 5] \to -\frac{2 A_{14}^{--}}{A_{13}^{zz} - A_{24}^{zz} + d^{zz} - D^{zz} + 2\omega_{e1} - 2\omega_n}, \text{SMe}[3, 6] \to 0,$$

$$\text{SMe}[3, 7] \to -\frac{A_{13}^{-z} - A_{14}^{-z} - D^{-z}}{A_{13}^{zz} - A_{14}^{zz} - D^{zz} + 2\omega_{e1}}, \text{SMe}[3, 8] \to \frac{2 A_{13}^{-+}}{A_{14}^{zz} - A_{23}^{zz} - d^{zz} + D^{zz} - 2\omega_{e1} - 2\omega_n},$$

$$\text{SMe}[3, 9] \to -\frac{A_{23}^{-z} - A_{24}^{-z} - D^{z-}}{A_{23}^{zz} - A_{24}^{zz} - D^{zz} + 2\omega_{e2}}, \text{SMe}[3, 10] \to \frac{2 A_{24}^{--}}{A_{14}^{zz} - A_{23}^{zz} - d^{zz} + D^{zz} - 2\omega_{e2} + 2\omega_n},$$

$$\text{SMe}[3, 11] \to 0, \text{SMe}[3, 12] \to -\frac{2 A_{23}^{-+}}{A_{13}^{zz} - A_{24}^{zz} + d^{zz} - D^{zz} + 2\omega_{e2} + 2\omega_n}, \text{SMe}[3, 13] \to 0,$$

$$\text{SMe}[3, 14] \to 0, \text{SMe}[3, 15] \to -\frac{2 D^{--}}{A_{13}^{zz} - A_{14}^{zz} + A_{23}^{zz} - A_{24}^{zz} + 2\omega_{e1} + 2\omega_{e2}},$$

$$\text{SMe}[3, 16] \to 0, \text{SMe}[4, 1] \to \frac{2 d^{--}}{A_{13}^{zz} + A_{14}^{zz} + A_{23}^{zz} + A_{24}^{zz} + 4\omega_n},$$

$$\text{SMe}[4, 2] \to -\frac{A_{14}^{z-} + A_{24}^{z-} + d^{z-}}{A_{14}^{zz} + A_{24}^{zz} + d^{zz} + 2\omega_n}, \text{SMe}[4, 3] \to -\frac{A_{13}^{z-} + A_{23}^{z-} + d^{-z}}{A_{13}^{zz} + A_{23}^{zz} + d^{zz} + 2\omega_n}, \text{SMe}[4, 5] \to 0,$$

$$\text{SMe}[4, 6] \to \frac{2 A_{14}^{--}}{A_{13}^{zz} + A_{24}^{zz} + d^{zz} + D^{zz} - 2\omega_{e1} + 2\omega_n}, \text{SMe}[4, 7] \to \frac{2 A_{13}^{--}}{A_{14}^{zz} + A_{23}^{zz} + d^{zz} + D^{zz} - 2\omega_{e1} + 2\omega_n},$$

$$\text{SMe}[4, 8] \to -\frac{A_{13}^{-z} + A_{14}^{-z} + D^{-z}}{A_{13}^{zz} + A_{14}^{zz} + D^{zz} - 2\omega_{e1}}, \text{SMe}[4, 9] \to \frac{2 A_{23}^{--}}{A_{13}^{zz} + A_{24}^{zz} + d^{zz} + D^{zz} - 2\omega_{e2} + 2\omega_n},$$

$$\text{SMe}[4, 10] \to 0, \text{SMe}[4, 11] \to \frac{2 A_{24}^{--}}{A_{14}^{zz} + A_{23}^{zz} + d^{zz} + D^{zz} - 2\omega_{e2} + 2\omega_n},$$

$$\text{SMe}[4, 12] \to -\frac{A_{23}^{-z} + A_{24}^{-z} + D^{z-}}{A_{23}^{zz} + A_{24}^{zz} + D^{zz} - 2\omega_{e2}}, \text{SMe}[4, 13] \to 0, \text{SMe}[4, 14] \to 0,$$

$$\text{SMe}[4, 15] \to 0, \text{SMe}[4, 16] \to \frac{2 D^{--}}{A_{13}^{zz} + A_{14}^{zz} + A_{23}^{zz} + A_{24}^{zz} - 2\omega_{e1} - 2\omega_{e2}},$$

$$\text{SMe}[5, 1] \to -\frac{-A_{13}^{+z} - A_{14}^{+z} + D^{+z}}{A_{13}^{zz} + A_{14}^{zz} - D^{zz} + 2\omega_{e1}}, \text{SMe}[5, 2] \to \frac{2 A_{13}^{++}}{A_{14}^{zz} - A_{23}^{zz} + d^{zz} - D^{zz} + 2\omega_{e1} - 2\omega_n},$$

$$\text{SMe}[5, 3] \to \frac{2 A_{14}^{++}}{A_{13}^{zz} - A_{24}^{zz} + d^{zz} - D^{zz} + 2\omega_{e1} - 2\omega_n}, \text{SMe}[5, 4] \to 0, \text{SMe}[5, 6] \to -\frac{-A_{13}^{z+} + A_{23}^{z+} - d^{+z}}{A_{13}^{zz} - A_{23}^{zz} + d^{zz} - 2\omega_n},$$

$$\text{SMe}[5, 7] \to -\frac{-A_{14}^{z+} + A_{24}^{z+} - d^{z+}}{A_{14}^{zz} - A_{24}^{zz} + d^{zz} - 2\omega_n}, \text{SMe}[5, 8] \to \frac{2 d^{++}}{A_{13}^{zz} + A_{14}^{zz} - A_{23}^{zz} - A_{24}^{zz} - 4\omega_n}, \text{SMe}[5, 9] \to 0,$$

$$\text{SMe}[5, 10] \to \frac{2 D^{+-}}{A_{13}^{zz} + A_{14}^{zz} - A_{23}^{zz} - A_{24}^{zz} + 2\omega_{e1} - 2\omega_{e2}}, \text{SMe}[5, 11] \to 0, \text{SMe}[5, 12] \to 0,$$

$$\text{SMe}[5, 13] \to -\frac{A_{23}^{-z} + A_{24}^{-z} + D^{z-}}{A_{23}^{zz} + A_{24}^{zz} + D^{zz} + 2\omega_{e2}}, \text{SMe}[5, 14] \to \frac{2 A_{23}^{-+}}{A_{13}^{zz} - A_{24}^{zz} + d^{zz} - D^{zz} - 2\omega_{e2} - 2\omega_n},$$

$$\text{SMe}[5, 15] \to \frac{2 A_{24}^{-+}}{A_{14}^{zz} - A_{23}^{zz} + d^{zz} - D^{zz} - 2\omega_{e2} - 2\omega_n}, \text{SMe}[5, 16] \to 0,$$

$$\text{SMe}[6, 1] \to \frac{2 A_{13}^{+-}}{A_{14}^{zz} + A_{23}^{zz} - d^{zz} - D^{zz} + 2\omega_{e1} + 2\omega_n}, \text{SMe}[6, 2] \to -\frac{-A_{13}^{+z} + A_{14}^{+z} - D^{+z}}{A_{13}^{zz} - A_{14}^{zz} + D^{zz} - 2\omega_{e1}},$$

$$\text{SMe}[6, 3] \to 0, \text{SMe}[6, 4] \to -\frac{2 A_{14}^{++}}{A_{13}^{zz} + A_{24}^{zz} + d^{zz} + D^{zz} - 2\omega_{e1} + 2\omega_n},$$

$$\text{SMe}[6, 5] \to -\frac{A_{13}^{z-} - A_{23}^{z-} + d^{-z}}{A_{13}^{zz} - A_{23}^{zz} + d^{zz} - 2\omega_n}, \text{SMe}[6, 7] \to -\frac{2 d^{-+}}{A_{13}^{zz} - A_{14}^{zz} - A_{23}^{zz} + A_{24}^{zz}},$$

$$\text{SMe}[6, 8] \to -\frac{-A_{14}^{z+} + A_{24}^{z+} + d^{z+}}{A_{14}^{zz} - A_{24}^{zz} - d^{zz} - 2\omega_n}, \text{SMe}[6, 9] \to 0, \text{SMe}[6, 10] \to 0,$$

$$\text{SMe}[6, 11] \to -\frac{2 D^{+-}}{A_{13}^{zz} - A_{14}^{zz} - A_{23}^{zz} + A_{24}^{zz} - 2\omega_{e1} + 2\omega_{e2}}, \text{SMe}[6, 12] \to 0,$$

$$\text{SMe}[6, 13] \to -\frac{2 A_{23}^{--}}{A_{13}^{zz} + A_{24}^{zz} + d^{zz} + D^{zz} + 2\omega_{e2} - 2\omega_n}, \text{SMe}[6, 14] \to -\frac{A_{23}^{-z} - A_{24}^{-z} - D^{z-}}{A_{23}^{zz} - A_{24}^{zz} - D^{zz} - 2\omega_{e2}},$$

$$\text{SMe}[6, 15] \to 0, \text{SMe}[6, 16] \to \frac{2 A_{24}^{-+}}{A_{14}^{zz} + A_{23}^{zz} - d^{zz} - D^{zz} - 2\omega_{e2} - 2\omega_n},$$

$$\text{SMe}[7, 1] \to \frac{2 A_{14}^{+-}}{A_{13}^{zz} + A_{24}^{zz} - d^{zz} - D^{zz} + 2\omega_{e1} + 2\omega_n}, \text{SMe}[7, 2] \to 0,$$

$$\text{SMe}[7, 3] \to -\frac{-A_{13}^{+z} + A_{14}^{+z} + D^{+z}}{A_{13}^{zz} - A_{14}^{zz} - D^{zz} + 2\omega_{e1}}, \text{SMe}[7, 4] \to -\frac{2 A_{13}^{++}}{A_{14}^{zz} + A_{23}^{zz} + d^{zz} + D^{zz} - 2\omega_{e1} + 2\omega_n},$$

$$\text{SMe}[7, 5] \to -\frac{A_{14}^{z-} - A_{24}^{z-} + d^{z-}}{A_{14}^{zz} - A_{24}^{zz} + d^{zz} - 2\omega_n}, \text{SMe}[7, 6] \to \frac{2 d^{+-}}{A_{13}^{zz} - A_{14}^{zz} - A_{23}^{zz} + A_{24}^{zz}},$$

$$\text{SMe}[7, 8] \to -\frac{-A_{13}^{z+} + A_{23}^{z+} + d^{+z}}{A_{13}^{zz} - A_{23}^{zz} - d^{zz} - 2\omega_n}, \text{SMe}[7, 9] \to \frac{2 D^{+-}}{A_{13}^{zz} - A_{14}^{zz} - A_{23}^{zz} + A_{24}^{zz} + 2\omega_{e1} - 2\omega_{e2}},$$

$$\text{SMe}[7, 10] \to 0, \text{SMe}[7, 11] \to 0, \text{SMe}[7, 12] \to 0,$$

$$\text{SMe}[7, 13] \to -\frac{2 A_{24}^{--}}{A_{14}^{zz} + A_{23}^{zz} + d^{zz} + D^{zz} + 2\omega_{e2} - 2\omega_n}, \text{SMe}[7, 14] \to 0,$$

$$\text{SMe}[7, 15] \to -\frac{A_{23}^{-z} - A_{24}^{-z} + D^{z-}}{A_{23}^{zz} - A_{24}^{zz} + D^{zz} + 2\omega_{e2}}, \text{SMe}[7, 16] \to \frac{2 A_{23}^{-+}}{A_{13}^{zz} + A_{24}^{zz} - d^{zz} - D^{zz} - 2\omega_{e2} - 2\omega_n},$$

$$\text{SMe}[8, 1] \to 0, \text{SMe}[8, 2] \to -\frac{2 A_{14}^{+-}}{A_{13}^{zz} - A_{24}^{zz} - d^{zz} + D^{zz} - 2\omega_{e1} - 2\omega_n},$$

$$\text{SMe}[8, 3] \to -\frac{2 A_{13}^{+-}}{A_{14}^{zz} - A_{23}^{zz} + d^{zz} + D^{zz} - 2\omega_{e1} - 2\omega_n}, \text{SMe}[8, 4] \to -\frac{-A_{13}^{+z} - A_{14}^{+z} - D^{+z}}{A_{13}^{zz} + A_{14}^{zz} + D^{zz} - 2\omega_{e1}},$$

$$\text{SMe}[8, 5] \to -\frac{2 d^{--}}{A_{13}^{zz} + A_{14}^{zz} - A_{23}^{zz} - A_{24}^{zz} - 4\omega_n}, \text{SMe}[8, 6] \to -\frac{A_{14}^{z-} - A_{24}^{z-} - d^{z-}}{A_{14}^{zz} - A_{24}^{zz} - d^{zz} - 2\omega_n},$$

$$\text{SMe}[8, 7] \to -\frac{A_{13}^{z-} - A_{23}^{z-} - d^{-z}}{A_{13}^{zz} - A_{23}^{zz} - d^{zz} - 2\omega_n}, \text{SMe}[8, 9] \to 0, \text{SMe}[8, 10] \to 0,$$

$$\text{SMe}[8, 11] \to 0, \text{SMe}[8, 12] \to -\frac{2 D^{+-}}{A_{13}^{zz} + A_{14}^{zz} - A_{23}^{zz} - A_{24}^{zz} - 2\omega_{e1} + 2\omega_{e2}},$$

$$\text{SMe}[8, 13] \to 0, \text{SMe}[8, 14] \to -\frac{2 A_{24}^{--}}{A_{14}^{zz} - A_{23}^{zz} - d^{zz} + D^{zz} + 2\omega_{e2} - 2\omega_n},$$

$$\text{SMe}[8, 15] \to -\frac{2 A_{23}^{--}}{A_{13}^{zz} - A_{24}^{zz} - d^{zz} + D^{zz} + 2\omega_{e2} - 2\omega_n}, \text{SMe}[8, 16] \to -\frac{A_{23}^{-z} + A_{24}^{-z} - D^{z-}}{A_{23}^{zz} + A_{24}^{zz} - D^{zz} - 2\omega_{e2}},$$

$$\text{SMe}[9, 1] \to \frac{2 A_{24}^{+-}}{A_{14}^{zz} + A_{23}^{zz} - d^{zz} - D^{zz} + 2\omega_{e2} + 2\omega_n}, \text{SMe}[9, 2] \to 0,$$

$$\text{SMe}[9, 3] \to -\frac{-A_{23}^{+z} + A_{24}^{+z} + D^{z+}}{A_{23}^{zz} - A_{24}^{zz} - D^{zz} + 2\omega_{e2}}, \text{SMe}[9, 4] \to -\frac{2 A_{23}^{++}}{A_{13}^{zz} + A_{24}^{zz} + d^{zz} + D^{zz} - 2\omega_{e2} + 2\omega_n},$$

$$\text{SMe}[9, 5] \to 0, \text{SMe}[9, 6] \to 0, \text{SMe}[9, 7] \to -\frac{2 D^{-+}}{A_{13}^{zz} - A_{14}^{zz} - A_{23}^{zz} + A_{24}^{zz} + 2\omega_{e1} - 2\omega_{e2}},$$

$$\text{SMe}[9, 8] \to 0, \text{SMe}[9, 10] \to -\frac{A_{14}^{z-} - A_{24}^{z-} - d^{z-}}{A_{14}^{zz} - A_{24}^{zz} - d^{zz} + 2\omega_n}, \text{SMe}[9, 11] \to -\frac{2 d^{+-}}{A_{13}^{zz} - A_{14}^{zz} - A_{23}^{zz} + A_{24}^{zz}},$$

$$\text{SMe}[9, 12] \to -\frac{-A_{13}^{z+} + A_{23}^{z+} - d^{+z}}{A_{13}^{zz} - A_{23}^{zz} + d^{zz} + 2\omega_n}, \text{SMe}[9, 13] \to -\frac{2 A_{14}^{--}}{A_{13}^{zz} + A_{24}^{zz} + d^{zz} + D^{zz} + 2\omega_{e1} - 2\omega_n},$$

$$\text{SMe}[9, 14] \to 0, \text{SMe}[9, 15] \to -\frac{A_{13}^{-z} - A_{14}^{-z} + D^{-z}}{A_{13}^{zz} - A_{14}^{zz} + D^{zz} + 2\omega_{e1}},$$

$$\text{SMe}[9, 16] \to \frac{2 A_{13}^{-+}}{A_{14}^{zz} + A_{23}^{zz} - d^{zz} - D^{zz} - 2\omega_{e1} - 2\omega_n}, \text{SMe}[10, 1] \to -\frac{-A_{23}^{+z} - A_{24}^{+z} + D^{z+}}{A_{23}^{zz} + A_{24}^{zz} - D^{zz} + 2\omega_{e2}},$$

$$\text{SMe}[10, 2] \to -\frac{2 A_{23}^{++}}{A_{13}^{zz} - A_{24}^{zz} - d^{zz} + D^{zz} - 2\omega_{e2} + 2\omega_n}, \; \text{SMe}[10, 3] \to -\frac{2 A_{24}^{++}}{A_{14}^{zz} - A_{23}^{zz} - d^{zz} + D^{zz} - 2\omega_{e2} + 2\omega_n},$$

$$\text{SMe}[10, 4] \to 0, \; \text{SMe}[10, 5] \to -\frac{2 D^{-+}}{A_{13}^{zz} + A_{14}^{zz} - A_{23}^{zz} - A_{24}^{zz} + 2\omega_{e1} - 2\omega_{e2}}, \; \text{SMe}[10, 6] \to 0,$$

$$\text{SMe}[10, 7] \to 0, \; \text{SMe}[10, 8] \to 0, \; \text{SMe}[10, 9] \to -\frac{-A_{14}^{z+} + A_{24}^{z+} + d^{z+}}{A_{14}^{zz} - A_{24}^{zz} - d^{zz} + 2\omega_n},$$

$$\text{SMe}[10, 11] \to -\frac{-A_{13}^{z+} + A_{23}^{z+} + d^{+z}}{A_{13}^{zz} - A_{23}^{zz} - d^{zz} + 2\omega_n}, \; \text{SMe}[10, 12] \to -\frac{2 d^{++}}{A_{13}^{zz} + A_{14}^{zz} - A_{23}^{zz} - A_{24}^{zz} + 4\omega_n},$$

$$\text{SMe}[10, 13] \to -\frac{A_{13}^{-z} + A_{14}^{-z} + D^{-z}}{A_{13}^{zz} + A_{14}^{zz} + D^{zz} + 2\omega_{e1}}, \; \text{SMe}[10, 14] \to -\frac{2 A_{13}^{-+}}{A_{14}^{zz} - A_{23}^{zz} - d^{zz} + D^{zz} + 2\omega_{e1} + 2\omega_n},$$

$$\text{SMe}[10, 15] \to -\frac{2 A_{14}^{-+}}{A_{13}^{zz} - A_{24}^{zz} - d^{zz} + D^{zz} + 2\omega_{e1} + 2\omega_n}, \; \text{SMe}[10, 16] \to 0,$$

$$\text{SMe}[11, 1] \to \frac{2 A_{23}^{+-}}{A_{13}^{zz} + A_{24}^{zz} - d^{zz} - D^{zz} + 2\omega_{e2} + 2\omega_n}, \; \text{SMe}[11, 2] \to -\frac{-A_{23}^{+z} + A_{24}^{+z} - D^{z+}}{A_{23}^{zz} - A_{24}^{zz} + D^{zz} - 2\omega_{e2}},$$

$$\text{SMe}[11, 3] \to 0, \; \text{SMe}[11, 4] \to -\frac{2 A_{24}^{++}}{A_{14}^{zz} + A_{23}^{zz} + d^{zz} + D^{zz} - 2\omega_{e2} + 2\omega_n},$$

$$\text{SMe}[11, 5] \to 0, \; \text{SMe}[11, 6] \to \frac{2 D^{-+}}{A_{13}^{zz} - A_{14}^{zz} - A_{23}^{zz} + A_{24}^{zz} - 2\omega_{e1} + 2\omega_{e2}},$$

$$\text{SMe}[11, 7] \to 0, \; \text{SMe}[11, 8] \to 0, \; \text{SMe}[11, 9] \to \frac{2 d^{-+}}{A_{13}^{zz} - A_{14}^{zz} - A_{23}^{zz} + A_{24}^{zz}},$$

$$\text{SMe}[11, 10] \to -\frac{A_{13}^{z-} - A_{23}^{z-} - d^{-z}}{A_{13}^{zz} - A_{23}^{zz} - d^{zz} + 2\omega_n}, \; \text{SMe}[11, 12] \to -\frac{-A_{14}^{z+} + A_{24}^{z+} - d^{z+}}{A_{14}^{zz} - A_{24}^{zz} + d^{zz} + 2\omega_n},$$

$$\text{SMe}[11, 13] \to -\frac{2 A_{13}^{--}}{A_{14}^{zz} + A_{23}^{zz} + d^{zz} + D^{zz} + 2\omega_{e1} - 2\omega_n}, \; \text{SMe}[11, 14] \to -\frac{A_{13}^{-z} - A_{14}^{-z} - D^{-z}}{A_{13}^{zz} - A_{14}^{zz} - D^{zz} - 2\omega_{e1}},$$

$$\text{SMe}[11, 15] \to 0, \; \text{SMe}[11, 16] \to \frac{2 A_{14}^{-+}}{A_{13}^{zz} + A_{24}^{zz} - d^{zz} - D^{zz} - 2\omega_{e1} - 2\omega_n}, \; \text{SMe}[12, 1] \to 0,$$

$$\text{SMe}[12, 2] \to \frac{2 A_{24}^{+-}}{A_{14}^{zz} - A_{23}^{zz} + d^{zz} - D^{zz} + 2\omega_{e2} + 2\omega_n}, \; \text{SMe}[12, 3] \to \frac{2 A_{23}^{+-}}{A_{13}^{zz} - A_{24}^{zz} + d^{zz} - D^{zz} + 2\omega_{e2} + 2\omega_n},$$

$$\text{SMe}[12, 4] \to -\frac{-A_{23}^{+z} - A_{24}^{+z} - D^{z+}}{A_{23}^{zz} + A_{24}^{zz} + D^{zz} - 2\omega_{e2}}, \; \text{SMe}[12, 5] \to 0, \; \text{SMe}[12, 6] \to 0, \; \text{SMe}[12, 7] \to 0,$$

$$\text{SMe}[12, 8] \to \frac{2 D^{-+}}{A_{13}^{zz} + A_{14}^{zz} - A_{23}^{zz} - A_{24}^{zz} - 2\omega_{e1} + 2\omega_{e2}}, \; \text{SMe}[12, 9] \to -\frac{A_{13}^{z-} - A_{23}^{z-} + d^{-z}}{A_{13}^{zz} - A_{23}^{zz} + d^{zz} + 2\omega_n},$$

$$\text{SMe}[12, 10] \to \frac{2 d^{--}}{A_{13}^{zz} + A_{14}^{zz} - A_{23}^{zz} - A_{24}^{zz} + 4\omega_n}, \; \text{SMe}[12, 11] \to -\frac{A_{14}^{z-} - A_{24}^{z-} + d^{z-}}{A_{14}^{zz} - A_{24}^{zz} + d^{zz} + 2\omega_n},$$

$$\text{SMe}[12, 13] \to 0, \; \text{SMe}[12, 14] \to \frac{2 A_{14}^{--}}{A_{13}^{zz} - A_{24}^{zz} + d^{zz} - D^{zz} - 2\omega_{e1} + 2\omega_n},$$

$$\text{SMe}[12, 15] \to \frac{2 A_{13}^{--}}{A_{14}^{zz} - A_{23}^{zz} + d^{zz} - D^{zz} - 2\omega_{e1} + 2\omega_n}, \; \text{SMe}[12, 16] \to -\frac{A_{13}^{-z} + A_{14}^{-z} - D^{-z}}{A_{13}^{zz} + A_{14}^{zz} - D^{zz} - 2\omega_{e1}},$$

$$\text{SMe}[13, 1] \to \frac{2 D^{++}}{A_{13}^{zz} + A_{14}^{zz} + A_{23}^{zz} + A_{24}^{zz} + 2\omega_{e1} + 2\omega_{e2}}, \; \text{SMe}[13, 2] \to 0,$$

$$\text{SMe}[13, 3] \to 0, \; \text{SMe}[13, 4] \to 0, \; \text{SMe}[13, 5] \to -\frac{-A_{23}^{+z} - A_{24}^{+z} - D^{z+}}{A_{23}^{zz} + A_{24}^{zz} + D^{zz} + 2\omega_{e2}},$$

$$\text{SMe}[13, 6] \to \frac{2 A_{23}^{++}}{A_{13}^{zz} + A_{24}^{zz} + d^{zz} + D^{zz} + 2\omega_{e2} - 2\omega_n}, \quad \text{SMe}[13, 7] \to \frac{2 A_{24}^{++}}{A_{14}^{zz} + A_{23}^{zz} + d^{zz} + D^{zz} + 2\omega_{e2} - 2\omega_n},$$

$$\text{SMe}[13, 8] \to 0, \quad \text{SMe}[13, 9] \to \frac{2 A_{14}^{++}}{A_{13}^{zz} + A_{24}^{zz} + d^{zz} + D^{zz} + 2\omega_{e1} - 2\omega_n},$$

$$\text{SMe}[13, 10] \to -\frac{-A_{13}^{+z} - A_{14}^{+z} - D^{+z}}{A_{13}^{zz} + A_{14}^{zz} + D^{zz} + 2\omega_{e1}}, \quad \text{SMe}[13, 11] \to \frac{2 A_{13}^{++}}{A_{14}^{zz} + A_{23}^{zz} + d^{zz} + D^{zz} + 2\omega_{e1} - 2\omega_n},$$

$$\text{SMe}[13, 12] \to 0, \quad \text{SMe}[13, 14] \to -\frac{-A_{13}^{z+} - A_{23}^{z+} - d^{+z}}{A_{13}^{zz} + A_{23}^{zz} + d^{zz} - 2\omega_n}, \quad \text{SMe}[13, 15] \to -\frac{-A_{14}^{z+} - A_{24}^{z+} - d^{z+}}{A_{14}^{zz} + A_{24}^{zz} + d^{zz} - 2\omega_n},$$

$$\text{SMe}[13, 16] \to \frac{2 d^{++}}{A_{13}^{zz} + A_{14}^{zz} + A_{23}^{zz} + A_{24}^{zz} - 4\omega_n}, \quad \text{SMe}[14, 1] \to 0,$$

$$\text{SMe}[14, 2] \to -\frac{2 D^{++}}{A_{13}^{zz} - A_{14}^{zz} + A_{23}^{zz} - A_{24}^{zz} - 2\omega_{e1} - 2\omega_{e2}}, \quad \text{SMe}[14, 3] \to 0, \quad \text{SMe}[14, 4] \to 0,$$

$$\text{SMe}[14, 5] \to -\frac{2 A_{23}^{+-}}{A_{13}^{zz} - A_{24}^{zz} + d^{zz} - D^{zz} - 2\omega_{e2} - 2\omega_n}, \quad \text{SMe}[14, 6] \to -\frac{-A_{23}^{+z} + A_{24}^{+z} + D^{z+}}{A_{23}^{zz} - A_{24}^{zz} - D^{zz} - 2\omega_{e2}},$$

$$\text{SMe}[14, 7] \to 0, \quad \text{SMe}[14, 8] \to \frac{2 A_{24}^{++}}{A_{14}^{zz} - A_{23}^{zz} - d^{zz} + D^{zz} + 2\omega_{e2} - 2\omega_n}, \quad \text{SMe}[14, 9] \to 0,$$

$$\text{SMe}[14, 10] \to \frac{2 A_{13}^{+-}}{A_{14}^{zz} - A_{23}^{zz} - d^{zz} + D^{zz} + 2\omega_{e1} + 2\omega_n}, \quad \text{SMe}[14, 11] \to -\frac{-A_{13}^{+z} + A_{14}^{+z} + D^{+z}}{A_{13}^{zz} - A_{14}^{zz} - D^{zz} - 2\omega_{e1}},$$

$$\text{SMe}[14, 12] \to -\frac{2 A_{14}^{++}}{A_{13}^{zz} - A_{24}^{zz} + d^{zz} - D^{zz} - 2\omega_{e1} + 2\omega_n}, \quad \text{SMe}[14, 13] \to -\frac{A_{13}^{z-} + A_{23}^{z-} + d^{-z}}{A_{13}^{zz} + A_{23}^{zz} + d^{zz} - 2\omega_n},$$

$$\text{SMe}[14, 15] \to -\frac{2 d^{-+}}{A_{13}^{zz} - A_{14}^{zz} + A_{23}^{zz} - A_{24}^{zz}}, \quad \text{SMe}[14, 16] \to -\frac{-A_{14}^{z+} - A_{24}^{z+} + d^{z+}}{A_{14}^{zz} + A_{24}^{zz} - d^{zz} - 2\omega_n},$$

$$\text{SMe}[15, 1] \to 0, \quad \text{SMe}[15, 2] \to 0, \quad \text{SMe}[15, 3] \to \frac{2 D^{++}}{A_{13}^{zz} - A_{14}^{zz} + A_{23}^{zz} - A_{24}^{zz} + 2\omega_{e1} + 2\omega_{e2}},$$

$$\text{SMe}[15, 4] \to 0, \quad \text{SMe}[15, 5] \to -\frac{2 A_{24}^{+-}}{A_{14}^{zz} - A_{23}^{zz} + d^{zz} - D^{zz} - 2\omega_{e2} - 2\omega_n}, \quad \text{SMe}[15, 6] \to 0,$$

$$\text{SMe}[15, 7] \to -\frac{-A_{23}^{+z} + A_{24}^{+z} - D^{z+}}{A_{23}^{zz} - A_{24}^{zz} + D^{zz} + 2\omega_{e2}}, \quad \text{SMe}[15, 8] \to \frac{2 A_{23}^{++}}{A_{13}^{zz} - A_{24}^{zz} - d^{zz} + D^{zz} + 2\omega_{e2} - 2\omega_n},$$

$$\text{SMe}[15, 9] \to -\frac{-A_{13}^{+z} + A_{14}^{+z} - D^{+z}}{A_{13}^{zz} - A_{14}^{zz} + D^{zz} + 2\omega_{e1}}, \quad \text{SMe}[15, 10] \to \frac{2 A_{14}^{+-}}{A_{13}^{zz} - A_{24}^{zz} - d^{zz} + D^{zz} + 2\omega_{e1} + 2\omega_n},$$

$$\text{SMe}[15, 11] \to 0, \quad \text{SMe}[15, 12] \to -\frac{2 A_{13}^{++}}{A_{14}^{zz} - A_{23}^{zz} + d^{zz} - D^{zz} - 2\omega_{e1} + 2\omega_n},$$

$$\text{SMe}[15, 13] \to -\frac{A_{14}^{z-} + A_{24}^{z-} + d^{z-}}{A_{14}^{zz} + A_{24}^{zz} + d^{zz} - 2\omega_n}, \quad \text{SMe}[15, 14] \to \frac{2 d^{+-}}{A_{13}^{zz} - A_{14}^{zz} + A_{23}^{zz} - A_{24}^{zz}},$$

$$\text{SMe}[15, 16] \to -\frac{-A_{13}^{z+} - A_{23}^{z+} + d^{+z}}{A_{13}^{zz} + A_{23}^{zz} - d^{zz} - 2\omega_n}, \quad \text{SMe}[16, 1] \to 0, \quad \text{SMe}[16, 2] \to 0,$$

$$\text{SMe}[16, 3] \to 0, \quad \text{SMe}[16, 4] \to -\frac{2 D^{++}}{A_{13}^{zz} + A_{14}^{zz} + A_{23}^{zz} + A_{24}^{zz} - 2\omega_{e1} - 2\omega_{e2}}, \quad \text{SMe}[16, 5] \to 0,$$

$$\text{SMe}[16, 6] \to -\frac{2 A_{24}^{+-}}{A_{14}^{zz} + A_{23}^{zz} - d^{zz} - D^{zz} - 2\omega_{e2} - 2\omega_n}, \quad \text{SMe}[16, 7] \to -\frac{2 A_{23}^{+-}}{A_{13}^{zz} + A_{24}^{zz} - d^{zz} - D^{zz} - 2\omega_{e2} - 2\omega_n},$$

$$\text{SMe}[16, 8] \to -\frac{-A_{23}^{+z} - A_{24}^{+z} + D^{z+}}{A_{23}^{zz} + A_{24}^{zz} - D^{zz} - 2\omega_{e2}}, \quad \text{SMe}[16, 9] \to -\frac{2 A_{13}^{+-}}{A_{14}^{zz} + A_{23}^{zz} - d^{zz} - D^{zz} - 2\omega_{e1} - 2\omega_n},$$

$$\text{SMe}[16, 10] \to 0, \quad \text{SMe}[16, 11] \to -\frac{2 A_{14}^{+-}}{A_{13}^{zz} + A_{24}^{zz} - d^{zz} - D^{zz} - 2\omega_{e1} - 2\omega_n},$$

$$\text{SMe}[16, 12] \to -\frac{-A_{13}^{+z} - A_{14}^{+z} + D^{+z}}{A_{13}^{zz} + A_{14}^{zz} - D^{zz} - 2\omega_{e1}}, \quad \text{SMe}[16, 13] \to -\frac{2 d^{--}}{A_{13}^{zz} + A_{14}^{zz} + A_{23}^{zz} + A_{24}^{zz} - 4\omega_n},$$

$$\text{SMe}[16, 14] \to -\frac{A_{14}^{z-} + A_{24}^{z-} - d^{z-}}{A_{14}^{zz} + A_{24}^{zz} - d^{zz} - 2\omega_n}, \quad \text{SMe}[16, 15] \to -\frac{A_{13}^{z-} + A_{23}^{z-} - d^{-z}}{A_{13}^{zz} + A_{23}^{zz} - d^{zz} - 2\omega_n}\Big\}\Big\}$$

*In[ ]:=* `SMsol = SM //. solution;`

*In[ ]:=* `SMsol = SMsol⟦1⟧ //. Table[SMe[i, i] → 0, {i, 1, 16}];`

*In[ ]:=* `HamM2 = H0M + 1/2 Commutator[SMsol, VM]`

*Out[ ]=*

[Large expression output — Full expression not available (original memory size: 2.2 MB)]

*In[ ]:=* `HamM2⟦2, 2⟧`
`HamM2⟦3, 3⟧`
`HamM2⟦2, 2⟧ - HamM2⟦3, 3⟧`

*Out[◦]=*

$$\frac{A_{13}^{zz}}{4} - \frac{A_{14}^{zz}}{4} + \frac{A_{23}^{zz}}{4} - \frac{A_{24}^{zz}}{4} - \frac{d^{zz}}{4} + \frac{D^{zz}}{4} - \frac{\omega_{e1}}{2} - \frac{\omega_{e2}}{2} +$$

$$\frac{1}{2}\left( \frac{4\, d^{-+}\, d^{+-}}{A_{13}^{zz} - A_{14}^{zz} + A_{23}^{zz} - A_{24}^{zz}} + \frac{\left(-\frac{A_{13}^{-z}}{2} + \frac{A_{14}^{-z}}{2} - \frac{D^{-z}}{2}\right)\left(-A_{13}^{+z} + A_{14}^{+z} - D^{+z}\right)}{A_{13}^{zz} - A_{14}^{zz} + D^{zz} - 2\,\omega_{e1}} - \right.$$

$$\frac{(A_{13}^{-z} - A_{14}^{-z} + D^{-z})\left(-\frac{A_{13}^{+z}}{2} + \frac{A_{14}^{+z}}{2} - \frac{D^{+z}}{2}\right)}{A_{13}^{zz} - A_{14}^{zz} + D^{zz} - 2\,\omega_{e1}} + \frac{\left(-\frac{A_{23}^{-z}}{2} + \frac{A_{24}^{-z}}{2} - \frac{D^{-z}}{2}\right)\left(-A_{23}^{+z} + A_{24}^{+z} - D^{+z}\right)}{A_{23}^{zz} - A_{24}^{zz} + D^{zz} - 2\,\omega_{e2}} -$$

$$\frac{(A_{23}^{-z} - A_{24}^{-z} + D^{z-})\left(-\frac{A_{23}^{+z}}{2} + \frac{A_{24}^{+z}}{2} - \frac{D^{z+}}{2}\right)}{A_{23}^{zz} - A_{24}^{zz} + D^{zz} - 2\,\omega_{e2}} + \frac{4\, D^{--}\, D^{++}}{A_{13}^{zz} - A_{14}^{zz} + A_{23}^{zz} - A_{24}^{zz} - 2\,\omega_{e1} - 2\,\omega_{e2}} +$$

$$\frac{4\, A_{14}^{-+}\, A_{14}^{+-}}{A_{13}^{zz} - A_{23}^{zz} - d^{zz} + D^{zz} - 2\,\omega_{e1} - 2\,\omega_n} - \frac{4\, A_{13}^{--}\, A_{13}^{++}}{A_{14}^{zz} - A_{23}^{zz} + d^{zz} - D^{zz} + 2\,\omega_{e1} - 2\,\omega_n} -$$

$$\frac{(A_{13}^{z-} + A_{23}^{z-} - d^{-z})\left(-\frac{A_{13}^{z+}}{2} - \frac{A_{23}^{z+}}{2} + \frac{d^{+z}}{2}\right)}{A_{13}^{zz} + A_{23}^{zz} - d^{zz} + 2\,\omega_n} + \frac{\left(-\frac{A_{13}^{z-}}{2} - \frac{A_{23}^{z-}}{2} + \frac{d^{-z}}{2}\right)\left(-A_{13}^{z+} - A_{23}^{z+} + d^{+z}\right)}{A_{13}^{zz} + A_{23}^{zz} - d^{zz} + 2\,\omega_n} -$$

$$\frac{\left(-\frac{A_{14}^{z-}}{2} - \frac{A_{24}^{z-}}{2} - \frac{d^{z-}}{2}\right)\left(-A_{14}^{z+} - A_{24}^{z+} - d^{z+}\right)}{A_{14}^{zz} + A_{24}^{zz} + d^{zz} + 2\,\omega_n} + \frac{(A_{14}^{z-} + A_{24}^{z-} + d^{z-})\left(-\frac{A_{14}^{z+}}{2} - \frac{A_{24}^{z+}}{2} - \frac{d^{z+}}{2}\right)}{A_{14}^{zz} + A_{24}^{zz} + d^{zz} + 2\,\omega_n} +$$

$$\left. \frac{4\, A_{23}^{--}\, A_{23}^{++}}{A_{13}^{zz} - A_{24}^{zz} - d^{zz} + D^{zz} - 2\,\omega_{e2} + 2\,\omega_n} - \frac{4\, A_{24}^{-+}\, A_{24}^{+-}}{A_{14}^{zz} - A_{23}^{zz} + d^{zz} - D^{zz} + 2\,\omega_{e2} + 2\,\omega_n} \right)$$

*Out[◦]=*

$$-\frac{A_{13}^{zz}}{4} + \frac{A_{14}^{zz}}{4} - \frac{A_{23}^{zz}}{4} + \frac{A_{24}^{zz}}{4} - \frac{d^{zz}}{4} + \frac{D^{zz}}{4} - \frac{\omega_{e1}}{2} - \frac{\omega_{e2}}{2} +$$

$$\frac{1}{2}\left( -\frac{4\, d^{-+}\, d^{+-}}{A_{13}^{zz} - A_{14}^{zz} + A_{23}^{zz} - A_{24}^{zz}} - \frac{(A_{13}^{-z} - A_{14}^{-z} - D^{-z})\left(\frac{A_{13}^{+z}}{2} - \frac{A_{14}^{+z}}{2} - \frac{D^{+z}}{2}\right)}{A_{13}^{zz} - A_{14}^{zz} - D^{zz} + 2\,\omega_{e1}} + \right.$$

$$\frac{\left(\frac{A_{13}^{-z}}{2} - \frac{A_{14}^{-z}}{2} - \frac{D^{-z}}{2}\right)\left(-A_{13}^{+z} + A_{14}^{+z} + D^{+z}\right)}{A_{13}^{zz} - A_{14}^{zz} - D^{zz} + 2\,\omega_{e1}} - \frac{(A_{23}^{-z} - A_{24}^{-z} - D^{z-})\left(\frac{A_{23}^{+z}}{2} - \frac{A_{24}^{+z}}{2} - \frac{D^{z+}}{2}\right)}{A_{23}^{zz} - A_{24}^{zz} - D^{zz} + 2\,\omega_{e2}} +$$

$$\frac{\left(\frac{A_{23}^{-z}}{2} - \frac{A_{24}^{-z}}{2} - \frac{D^{z-}}{2}\right)\left(-A_{23}^{+z} + A_{24}^{+z} + D^{z+}\right)}{A_{23}^{zz} - A_{24}^{zz} - D^{zz} + 2\,\omega_{e2}} - \frac{4\, D^{--}\, D^{++}}{A_{13}^{zz} - A_{14}^{zz} + A_{23}^{zz} - A_{24}^{zz} + 2\,\omega_{e1} + 2\,\omega_{e2}} +$$

$$\frac{4\, A_{13}^{-+}\, A_{13}^{+-}}{A_{14}^{zz} - A_{23}^{zz} - d^{zz} + D^{zz} - 2\,\omega_{e1} - 2\,\omega_n} - \frac{4\, A_{14}^{--}\, A_{14}^{++}}{A_{13}^{zz} - A_{24}^{zz} + d^{zz} - D^{zz} + 2\,\omega_{e1} - 2\,\omega_n} -$$

$$\frac{(A_{14}^{z-} + A_{24}^{z-} - d^{z-})\left(-\frac{A_{14}^{z+}}{2} - \frac{A_{24}^{z+}}{2} + \frac{d^{z+}}{2}\right)}{A_{14}^{zz} + A_{24}^{zz} - d^{zz} + 2\,\omega_n} + \frac{\left(-\frac{A_{14}^{z-}}{2} - \frac{A_{24}^{z-}}{2} + \frac{d^{z-}}{2}\right)\left(-A_{14}^{z+} - A_{24}^{z+} + d^{z+}\right)}{A_{14}^{zz} + A_{24}^{zz} - d^{zz} + 2\,\omega_n} -$$

$$\frac{\left(-\frac{A_{13}^{z-}}{2} - \frac{A_{23}^{z-}}{2} - \frac{d^{-z}}{2}\right)\left(-A_{13}^{z+} - A_{23}^{z+} - d^{+z}\right)}{A_{13}^{zz} + A_{23}^{zz} + d^{zz} + 2\,\omega_n} + \frac{(A_{13}^{z-} + A_{23}^{z-} + d^{-z})\left(-\frac{A_{13}^{z+}}{2} - \frac{A_{23}^{z+}}{2} - \frac{d^{+z}}{2}\right)}{A_{13}^{zz} + A_{23}^{zz} + d^{zz} + 2\,\omega_n} +$$

$$\left. \frac{4\, A_{24}^{--}\, A_{24}^{++}}{A_{14}^{zz} - A_{23}^{zz} - d^{zz} + D^{zz} - 2\,\omega_{e2} + 2\,\omega_n} - \frac{4\, A_{23}^{-+}\, A_{23}^{+-}}{A_{13}^{zz} - A_{24}^{zz} + d^{zz} - D^{zz} + 2\,\omega_{e2} + 2\,\omega_n} \right)$$

*Out[ ]=*

$$\frac{A_{13}^{zz}}{2} - \frac{A_{14}^{zz}}{2} + \frac{A_{23}^{zz}}{2} - \frac{A_{24}^{zz}}{2} +$$

$$\frac{1}{2} \left( \frac{4\, d^{-+}\, d^{+-}}{A_{13}^{zz} - A_{14}^{zz} + A_{23}^{zz} - A_{24}^{zz}} + \frac{\left(-\frac{A_{13}^{-z}}{2} + \frac{A_{14}^{-z}}{2} - \frac{D^{-z}}{2}\right)(-A_{13}^{+z} + A_{14}^{+z} - D^{+z})}{A_{13}^{zz} - A_{14}^{zz} + D^{zz} - 2\,\omega_{e1}} - \frac{(A_{13}^{-z} - A_{14}^{-z} + D^{-z})\left(-\frac{A_{13}^{+z}}{2} + \frac{A_{14}^{+z}}{2} - \frac{D^{+z}}{2}\right)}{A_{13}^{zz} - A_{14}^{zz} + D^{zz} - 2\,\omega_{e1}} + \right.$$

$$\frac{\left(-\frac{A_{23}^{-z}}{2} + \frac{A_{24}^{-z}}{2} - \frac{D^{z-}}{2}\right)(-A_{23}^{+z} + A_{24}^{+z} - D^{z+})}{A_{23}^{zz} - A_{24}^{zz} + D^{zz} - 2\,\omega_{e2}} - \frac{(A_{23}^{-z} - A_{24}^{-z} + D^{z-})\left(-\frac{A_{23}^{+z}}{2} + \frac{A_{24}^{+z}}{2} - \frac{D^{z+}}{2}\right)}{A_{23}^{zz} - A_{24}^{zz} + D^{zz} - 2\,\omega_{e2}} +$$

$$\frac{4\, D^{--}\, D^{++}}{A_{13}^{zz} - A_{14}^{zz} + A_{23}^{zz} - A_{24}^{zz} - 2\,\omega_{e1} - 2\,\omega_{e2}} + \frac{4\, A_{14}^{-+}\, A_{14}^{+-}}{A_{13}^{zz} - A_{24}^{zz} - d^{zz} + D^{zz} - 2\,\omega_{e1} - 2\,\omega_{n}} -$$

$$\frac{4\, A_{13}^{--}\, A_{13}^{++}}{A_{14}^{zz} - A_{23}^{zz} + d^{zz} - D^{zz} + 2\,\omega_{e1} - 2\,\omega_{n}} - \frac{(A_{13}^{z-} + A_{23}^{z-} - d^{-z})\left(-\frac{A_{13}^{z+}}{2} - \frac{A_{23}^{z+}}{2} + \frac{d^{+z}}{2}\right)}{A_{13}^{zz} + A_{23}^{zz} - d^{zz} + 2\,\omega_{n}} +$$

$$\frac{\left(-\frac{A_{13}^{z-}}{2} - \frac{A_{23}^{z-}}{2} + \frac{d^{-z}}{2}\right)(-A_{13}^{z+} - A_{23}^{z+} + d^{+z})}{A_{13}^{zz} + A_{23}^{zz} - d^{zz} + 2\,\omega_{n}} -$$

$$\frac{\left(-\frac{A_{14}^{z-}}{2} - \frac{A_{24}^{z-}}{2} - \frac{d^{z-}}{2}\right)(-A_{14}^{z+} - A_{24}^{z+} - d^{z+})}{A_{14}^{zz} + A_{24}^{zz} + d^{zz} + 2\,\omega_{n}} + \frac{(A_{14}^{z-} + A_{24}^{z-} + d^{z-})\left(-\frac{A_{14}^{z+}}{2} - \frac{A_{24}^{z+}}{2} - \frac{d^{z+}}{2}\right)}{A_{14}^{zz} + A_{24}^{zz} + d^{zz} + 2\,\omega_{n}} +$$

$$\left. \frac{4\, A_{23}^{--}\, A_{23}^{++}}{A_{13}^{zz} - A_{24}^{zz} - d^{zz} + D^{zz} - 2\,\omega_{e2} + 2\,\omega_{n}} - \frac{4\, A_{24}^{-+}\, A_{24}^{+-}}{A_{14}^{zz} - A_{23}^{zz} + d^{zz} - D^{zz} + 2\,\omega_{e2} + 2\,\omega_{n}} \right) +$$

$$\frac{1}{2} \left( \frac{4\, d^{-+}\, d^{+-}}{A_{13}^{zz} - A_{14}^{zz} + A_{23}^{zz} - A_{24}^{zz}} + \frac{(A_{13}^{-z} - A_{14}^{-z} - D^{-z})\left(\frac{A_{13}^{+z}}{2} - \frac{A_{14}^{+z}}{2} - \frac{D^{+z}}{2}\right)}{A_{13}^{zz} - A_{14}^{zz} - D^{zz} + 2\,\omega_{e1}} - \right.$$

$$\frac{\left(\frac{A_{13}^{-z}}{2} - \frac{A_{14}^{-z}}{2} - \frac{D^{-z}}{2}\right)(-A_{13}^{+z} + A_{14}^{+z} + D^{+z})}{A_{13}^{zz} - A_{14}^{zz} - D^{zz} + 2\,\omega_{e1}} + \frac{(A_{23}^{-z} - A_{24}^{-z} - D^{z-})\left(\frac{A_{23}^{+z}}{2} - \frac{A_{24}^{+z}}{2} - \frac{D^{z+}}{2}\right)}{A_{23}^{zz} - A_{24}^{zz} - D^{zz} + 2\,\omega_{e2}} -$$

$$\frac{\left(\frac{A_{23}^{-z}}{2} - \frac{A_{24}^{-z}}{2} - \frac{D^{z-}}{2}\right)(-A_{23}^{+z} + A_{24}^{+z} + D^{z+})}{A_{23}^{zz} - A_{24}^{zz} - D^{zz} + 2\,\omega_{e2}} + \frac{4\, D^{--}\, D^{++}}{A_{13}^{zz} - A_{14}^{zz} + A_{23}^{zz} - A_{24}^{zz} + 2\,\omega_{e1} + 2\,\omega_{e2}} -$$

$$\frac{4\, A_{13}^{-+}\, A_{13}^{+-}}{A_{14}^{zz} - A_{23}^{zz} - d^{zz} + D^{zz} - 2\,\omega_{e1} - 2\,\omega_{n}} + \frac{4\, A_{14}^{--}\, A_{14}^{++}}{A_{13}^{zz} - A_{24}^{zz} + d^{zz} - D^{zz} + 2\,\omega_{e1} - 2\,\omega_{n}} +$$

$$\frac{(A_{14}^{z-} + A_{24}^{z-} - d^{z-})\left(-\frac{A_{14}^{z+}}{2} - \frac{A_{24}^{z+}}{2} + \frac{d^{z+}}{2}\right)}{A_{14}^{zz} + A_{24}^{zz} - d^{zz} + 2\,\omega_{n}} - \frac{\left(-\frac{A_{14}^{z-}}{2} - \frac{A_{24}^{z-}}{2} + \frac{d^{z-}}{2}\right)(-A_{14}^{z+} - A_{24}^{z+} + d^{z+})}{A_{14}^{zz} + A_{24}^{zz} - d^{zz} + 2\,\omega_{n}} +$$

$$\frac{\left(-\frac{A_{13}^{z-}}{2} - \frac{A_{23}^{z-}}{2} - \frac{d^{z-}}{2}\right)(-A_{13}^{z+} - A_{23}^{z+} - d^{z+})}{A_{13}^{zz} + A_{23}^{zz} + d^{zz} + 2\,\omega_{n}} - \frac{(A_{13}^{z-} + A_{23}^{z-} + d^{-z})\left(-\frac{A_{13}^{z+}}{2} - \frac{A_{23}^{z+}}{2} - \frac{d^{+z}}{2}\right)}{A_{13}^{zz} + A_{23}^{zz} + d^{zz} + 2\,\omega_{n}} -$$

$$\left. \frac{4\, A_{24}^{--}\, A_{24}^{++}}{A_{14}^{zz} - A_{23}^{zz} - d^{zz} + D^{zz} - 2\,\omega_{e2} + 2\,\omega_{n}} + \frac{4\, A_{23}^{-+}\, A_{23}^{+-}}{A_{13}^{zz} - A_{24}^{zz} + d^{zz} - D^{zz} + 2\,\omega_{e2} + 2\,\omega_{n}} \right)$$

*In[ ]:=* **HamM2[[3, 2]]**

*Out[ ]=*

$$\frac{1}{2}\left(\frac{2 A_{13}^{-+} A_{14}^{+-}}{A_{14}^{zz} - A_{23}^{zz} - d^{zz} + D^{zz} - 2\omega_{e1} - 2\omega_n} + \frac{2 A_{13}^{-+} A_{14}^{+-}}{A_{13}^{zz} - A_{24}^{zz} - d^{zz} + D^{zz} - 2\omega_{e1} - 2\omega_n} - \frac{2 A_{13}^{++} A_{14}^{--}}{A_{14}^{zz} - A_{23}^{zz} + d^{zz} - D^{zz} + 2\omega_{e1} - 2\omega_n} - \frac{2 A_{13}^{++} A_{14}^{--}}{A_{13}^{zz} - A_{24}^{zz} + d^{zz} - D^{zz} + 2\omega_{e1} - 2\omega_n} + \frac{(-A_{13}^{z+} - A_{23}^{z+} + d^{+z})\left(-\frac{A_{14}^{z-}}{2} - \frac{A_{24}^{z-}}{2} + \frac{d^{z-}}{2}\right)}{A_{13}^{zz} + A_{23}^{zz} - d^{zz} + 2\omega_n} - \frac{\left(-\frac{A_{13}^{z+}}{2} - \frac{A_{23}^{z+}}{2} + \frac{d^{+z}}{2}\right)(A_{14}^{z-} + A_{24}^{z-} - d^{z-})}{A_{14}^{zz} + A_{24}^{zz} - d^{zz} + 2\omega_n} - \frac{(-A_{13}^{z+} - A_{23}^{z+} - d^{+z})\left(-\frac{A_{14}^{z-}}{2} - \frac{A_{24}^{z-}}{2} - \frac{d^{z-}}{2}\right)}{A_{13}^{zz} + A_{23}^{zz} + d^{zz} + 2\omega_n} + \frac{\left(-\frac{A_{13}^{z+}}{2} - \frac{A_{23}^{z+}}{2} - \frac{d^{+z}}{2}\right)(A_{14}^{z-} + A_{24}^{z-} + d^{z-})}{A_{14}^{zz} + A_{24}^{zz} + d^{zz} + 2\omega_n} + \frac{2 A_{23}^{++} A_{24}^{--}}{A_{14}^{zz} - A_{23}^{zz} - d^{zz} + D^{zz} - 2\omega_{e2} + 2\omega_n} + \frac{2 A_{23}^{++} A_{24}^{--}}{A_{13}^{zz} - A_{24}^{zz} - d^{zz} + D^{zz} - 2\omega_{e2} + 2\omega_n} - \frac{2 A_{23}^{-+} A_{24}^{+-}}{A_{14}^{zz} - A_{23}^{zz} + d^{zz} - D^{zz} + 2\omega_{e2} + 2\omega_n} - \frac{2 A_{23}^{-+} A_{24}^{+-}}{A_{13}^{zz} - A_{24}^{zz} + d^{zz} - D^{zz} + 2\omega_{e2} + 2\omega_n}\right)$$

*In[ ]:=* `HamM2[[6, 6]] - HamM2[[9, 9]]`

*Out[ ]=*

$\omega_{e1} - \omega_{e2} +$

$\frac{1}{2}\left(-\frac{4\, d^{-+}\, d^{+-}}{A_{13}^{zz} - A_{14}^{zz} - A_{23}^{zz} + A_{24}^{zz}} - \frac{\left(-\frac{A_{13}^{-z}}{2} + \frac{A_{14}^{-z}}{2} - \frac{D^{-z}}{2}\right)\left(-A_{13}^{+z} + A_{14}^{+z} - D^{+z}\right)}{A_{13}^{zz} - A_{14}^{zz} + D^{zz} - 2\,\omega_{e1}} + \frac{(A_{13}^{-z} - A_{14}^{-z} + D^{-z})\left(-\frac{A_{13}^{+z}}{2} + \frac{A_{14}^{+z}}{2} - \frac{D^{+z}}{2}\right)}{A_{13}^{zz} - A_{14}^{zz} + D^{zz} - 2\,\omega_{e1}} - \frac{(A_{23}^{-z} - A_{24}^{-z} - D^{z-})\left(-\frac{A_{23}^{+z}}{2} + \frac{A_{24}^{+z}}{2} + \frac{D^{z+}}{2}\right)}{A_{23}^{zz} - A_{24}^{zz} - D^{zz} - 2\,\omega_{e2}} + \frac{\left(-\frac{A_{23}^{-z}}{2} + \frac{A_{24}^{-z}}{2} + \frac{D^{z-}}{2}\right)(-A_{23}^{+z} + A_{24}^{+z} + D^{z+})}{A_{23}^{zz} - A_{24}^{zz} - D^{zz} - 2\,\omega_{e2}} - \frac{4\, D^{-+}\, D^{+-}}{A_{13}^{zz} - A_{14}^{zz} - A_{23}^{zz} + A_{24}^{zz} - 2\,\omega_{e1} + 2\,\omega_{e2}} + \frac{(A_{14}^{z-} - A_{24}^{z-} - d^{z-})\left(\frac{A_{14}^{z+}}{2} - \frac{A_{24}^{z+}}{2} - \frac{d^{z+}}{2}\right)}{A_{14}^{zz} - A_{24}^{zz} - d^{zz} - 2\,\omega_{n}} - \frac{\left(\frac{A_{14}^{z-}}{2} - \frac{A_{24}^{z-}}{2} - \frac{d^{z-}}{2}\right)(-A_{14}^{z+} + A_{24}^{z+} + d^{z+})}{A_{14}^{zz} - A_{24}^{zz} - d^{zz} - 2\,\omega_{n}} + \frac{\left(\frac{A_{13}^{z-}}{2} - \frac{A_{23}^{z-}}{2} + \frac{d^{-z}}{2}\right)(-A_{13}^{z+} + A_{23}^{z+} - d^{+z})}{A_{13}^{zz} - A_{23}^{zz} + d^{zz} - 2\,\omega_{n}} - \frac{(A_{13}^{z-} - A_{23}^{z-} + d^{-z})\left(\frac{A_{13}^{z+}}{2} - \frac{A_{23}^{z+}}{2} + \frac{d^{+z}}{2}\right)}{A_{13}^{zz} - A_{23}^{zz} + d^{zz} - 2\,\omega_{n}} + \frac{4\, A_{24}^{-+}\, A_{24}^{+-}}{A_{14}^{zz} + A_{23}^{zz} - d^{zz} - D^{zz} - 2\,\omega_{e2} - 2\,\omega_{n}} - \frac{4\, A_{23}^{--}\, A_{23}^{++}}{A_{13}^{zz} + A_{24}^{zz} + d^{zz} + D^{zz} + 2\,\omega_{e2} - 2\,\omega_{n}} - \frac{4\, A_{14}^{--}\, A_{14}^{++}}{A_{13}^{zz} + A_{24}^{zz} + d^{zz} + D^{zz} - 2\,\omega_{e1} + 2\,\omega_{n}} + \frac{4\, A_{13}^{-+}\, A_{13}^{+-}}{A_{14}^{zz} + A_{23}^{zz} - d^{zz} - D^{zz} + 2\,\omega_{e1} + 2\,\omega_{n}}\right) +$

$\frac{1}{2}\left(\frac{4\, d^{-+}\, d^{+-}}{A_{13}^{zz} - A_{14}^{zz} - A_{23}^{zz} + A_{24}^{zz}} - \frac{\left(\frac{A_{13}^{-z}}{2} - \frac{A_{14}^{-z}}{2} + \frac{D^{-z}}{2}\right)(-A_{13}^{+z} + A_{14}^{+z} - D^{+z})}{A_{13}^{zz} - A_{14}^{zz} + D^{zz} + 2\,\omega_{e1}} + \frac{(A_{13}^{-z} - A_{14}^{-z} + D^{-z})\left(\frac{A_{13}^{+z}}{2} - \frac{A_{14}^{+z}}{2} + \frac{D^{+z}}{2}\right)}{A_{13}^{zz} - A_{14}^{zz} + D^{zz} + 2\,\omega_{e1}} + \frac{4\, D^{-+}\, D^{+-}}{A_{13}^{zz} - A_{14}^{zz} - A_{23}^{zz} + A_{24}^{zz} + 2\,\omega_{e1} - 2\,\omega_{e2}} - \frac{(A_{23}^{-z} - A_{24}^{-z} - D^{z-})\left(\frac{A_{23}^{+z}}{2} - \frac{A_{24}^{+z}}{2} - \frac{D^{z+}}{2}\right)}{A_{23}^{zz} - A_{24}^{zz} - D^{zz} + 2\,\omega_{e2}} + \frac{\left(\frac{A_{23}^{-z}}{2} - \frac{A_{24}^{-z}}{2} - \frac{D^{z-}}{2}\right)(-A_{23}^{+z} + A_{24}^{+z} + D^{z+})}{A_{23}^{zz} - A_{24}^{zz} - D^{zz} + 2\,\omega_{e2}} - \frac{4\, A_{13}^{-+}\, A_{13}^{+-}}{A_{14}^{zz} + A_{23}^{zz} - d^{zz} - D^{zz} - 2\,\omega_{e1} - 2\,\omega_{n}} + \frac{4\, A_{14}^{--}\, A_{14}^{++}}{A_{13}^{zz} + A_{24}^{zz} + d^{zz} + D^{zz} + 2\,\omega_{e1} - 2\,\omega_{n}} + \frac{(A_{14}^{z-} - A_{24}^{z-} - d^{z-})\left(-\frac{A_{14}^{z+}}{2} + \frac{A_{24}^{z+}}{2} + \frac{d^{z+}}{2}\right)}{A_{14}^{zz} - A_{24}^{zz} - d^{zz} + 2\,\omega_{n}} - \frac{\left(-\frac{A_{14}^{z-}}{2} + \frac{A_{24}^{z-}}{2} + \frac{d^{z-}}{2}\right)(-A_{14}^{z+} + A_{24}^{z+} + d^{z+})}{A_{14}^{zz} - A_{24}^{zz} - d^{zz} + 2\,\omega_{n}} + \frac{\left(-\frac{A_{13}^{z-}}{2} + \frac{A_{23}^{z-}}{2} - \frac{d^{-z}}{2}\right)(-A_{13}^{z+} + A_{23}^{z+} - d^{+z})}{A_{13}^{zz} - A_{23}^{zz} + d^{zz} + 2\,\omega_{n}} - \frac{(A_{13}^{z-} - A_{23}^{z-} + d^{-z})\left(-\frac{A_{13}^{z+}}{2} + \frac{A_{23}^{z+}}{2} - \frac{d^{+z}}{2}\right)}{A_{13}^{zz} - A_{23}^{zz} + d^{zz} + 2\,\omega_{n}} + \frac{4\, A_{23}^{--}\, A_{23}^{++}}{A_{13}^{zz} + A_{24}^{zz} + d^{zz} + D^{zz} - 2\,\omega_{e2} + 2\,\omega_{n}} - \frac{4\, A_{24}^{-+}\, A_{24}^{+-}}{A_{14}^{zz} + A_{23}^{zz} - d^{zz} - D^{zz} + 2\,\omega_{e2} + 2\,\omega_{n}}\right)$

*In[ ]:=* **HamM2[[5, 3]]**

*Out[ ]=*

$$\frac{1}{2}\left(-\frac{2 A_{13}^{++} d^{-+}}{A_{13}^{zz} - A_{14}^{zz} + A_{23}^{zz} - A_{24}^{zz}} + \frac{(-A_{13}^{+z} + A_{14}^{+z} + D^{+z})\left(\frac{A_{14}^{z+}}{2} - \frac{A_{24}^{z+}}{2} + \frac{d^{z+}}{2}\right)}{A_{13}^{zz} - A_{14}^{zz} - D^{zz} + 2\omega_{e1}} - \frac{(-A_{13}^{+z} - A_{14}^{+z} + D^{+z})\left(-\frac{A_{14}^{z+}}{2} - \frac{A_{24}^{z+}}{2} + \frac{d^{z+}}{2}\right)}{A_{13}^{zz} + A_{14}^{zz} - D^{zz} + 2\omega_{e1}} + \right.$$

$$\frac{2 A_{24}^{++} D^{+-}}{A_{13}^{zz} + A_{14}^{zz} - A_{23}^{zz} - A_{24}^{zz} + 2\omega_{e1} - 2\omega_{e2}} - \frac{2 A_{24}^{-+} D^{++}}{A_{13}^{zz} - A_{14}^{zz} + A_{23}^{zz} - A_{24}^{zz} + 2\omega_{e1} + 2\omega_{e2}} + \frac{2 A_{13}^{+-} d^{++}}{A_{13}^{zz} + A_{14}^{zz} - A_{23}^{zz} - A_{24}^{zz} - 4\omega_n} -$$

$$\frac{\left(\frac{A_{13}^{+z}}{2} - \frac{A_{14}^{+z}}{2} - \frac{D^{+z}}{2}\right)(-A_{14}^{z+} + A_{24}^{z+} - d^{z+})}{A_{14}^{zz} - A_{24}^{zz} + d^{zz} - 2\omega_n} + \frac{2 A_{13}^{+-} d^{++}}{A_{14}^{zz} - A_{23}^{zz} - d^{zz} + D^{zz} - 2\omega_{e1} - 2\omega_n} +$$

$$\frac{2 A_{13}^{++} d^{-+}}{A_{14}^{zz} - A_{23}^{zz} + d^{zz} - D^{zz} + 2\omega_{e1} - 2\omega_n} + \frac{2 A_{24}^{-+} D^{++}}{A_{14}^{zz} - A_{23}^{zz} + d^{zz} - D^{zz} - 2\omega_{e2} - 2\omega_n} +$$

$$\left.\frac{\left(\frac{A_{13}^{+z}}{2} + \frac{A_{14}^{+z}}{2} - \frac{D^{+z}}{2}\right)(-A_{14}^{z+} - A_{24}^{z+} + d^{z+})}{A_{14}^{zz} + A_{24}^{zz} - d^{zz} + 2\omega_n} + \frac{2 A_{24}^{++} D^{+-}}{A_{14}^{zz} - A_{23}^{zz} - d^{zz} + D^{zz} - 2\omega_{e2} + 2\omega_n}\right)$$

*In[ ]:=* **HamM2[[7, 2]]**

*Out[ ]=*

$$\frac{1}{2}\left(\frac{2 d^{+-}\left(\frac{A_{13}^{+z}}{2} - \frac{A_{14}^{+z}}{2} - \frac{D^{+z}}{2}\right)}{A_{13}^{zz} - A_{14}^{zz} + A_{23}^{zz} - A_{24}^{zz}} + \frac{2 d^{+-}\left(-\frac{A_{13}^{+z}}{2} + \frac{A_{14}^{+z}}{2} - \frac{D^{+z}}{2}\right)}{A_{13}^{zz} - A_{14}^{zz} - A_{23}^{zz} + A_{24}^{zz}} + \frac{d^{+-}(-A_{13}^{+z} + A_{14}^{+z} - D^{+z})}{A_{13}^{zz} - A_{14}^{zz} + D^{zz} - 2\omega_{e1}} - \frac{d^{+-}(-A_{13}^{+z} + A_{14}^{+z} + D^{+z})}{A_{13}^{zz} - A_{14}^{zz} - D^{zz} + 2\omega_{e1}} - \right.$$

$$\frac{A_{14}^{+-}(-A_{13}^{z+} + A_{23}^{z+} + d^{+z})}{A_{13}^{zz} - A_{23}^{zz} - d^{zz} - 2\omega_n} - \frac{A_{13}^{++}(A_{14}^{z-} - A_{24}^{z-} + d^{z-})}{A_{14}^{zz} - A_{24}^{zz} + d^{zz} - 2\omega_n} + \frac{2 A_{14}^{+-}\left(\frac{A_{13}^{z+}}{2} - \frac{A_{23}^{z+}}{2} - \frac{d^{+z}}{2}\right)}{A_{13}^{zz} - A_{24}^{zz} - d^{zz} + D^{zz} - 2\omega_{e1} - 2\omega_n} -$$

$$\frac{2 A_{13}^{++}\left(\frac{A_{14}^{z-}}{2} - \frac{A_{24}^{z-}}{2} + \frac{d^{z-}}{2}\right)}{A_{14}^{zz} - A_{23}^{zz} + d^{zz} - D^{zz} + 2\omega_{e1} - 2\omega_n} + \frac{A_{14}^{+-}(-A_{13}^{z+} - A_{23}^{z+} + d^{+z})}{A_{13}^{zz} + A_{23}^{zz} - d^{zz} + 2\omega_n} + \frac{A_{13}^{++}(A_{14}^{z-} + A_{24}^{z-} + d^{z-})}{A_{14}^{zz} + A_{24}^{zz} + d^{zz} + 2\omega_n} -$$

$$\left.\frac{2 A_{13}^{++}\left(-\frac{A_{14}^{z-}}{2} - \frac{A_{24}^{z-}}{2} - \frac{d^{z-}}{2}\right)}{A_{14}^{zz} + A_{23}^{zz} + d^{zz} + D^{zz} - 2\omega_{e1} + 2\omega_n} + \frac{2 A_{14}^{+-}\left(-\frac{A_{13}^{z+}}{2} - \frac{A_{23}^{z+}}{2} + \frac{d^{+z}}{2}\right)}{A_{13}^{zz} + A_{24}^{zz} - d^{zz} - D^{zz} + 2\omega_{e1} + 2\omega_n}\right)$$

*In[ ]:=* **HamM2[[6, 9]]**

*Out[ ]=*

$$\frac{1}{2}\left(-\frac{4 d^{-+} D^{+-}}{A_{13}^{zz} - A_{14}^{zz} - A_{23}^{zz} + A_{24}^{zz}} - \frac{2 d^{-+} D^{+-}}{A_{13}^{zz} - A_{14}^{zz} - A_{23}^{zz} + A_{24}^{zz} + 2\omega_{e1} - 2\omega_{e2}} - \frac{2 d^{-+} D^{+-}}{A_{13}^{zz} - A_{14}^{zz} - A_{23}^{zz} + A_{24}^{zz} - 2\omega_{e1} + 2\omega_{e2}} + \right.$$

$$\frac{2 A_{13}^{+-} A_{24}^{-+}}{A_{14}^{zz} + A_{23}^{zz} - d^{zz} - D^{zz} - 2\omega_{e1} - 2\omega_n} - \frac{2 A_{14}^{++} A_{23}^{--}}{A_{13}^{zz} + A_{24}^{zz} + d^{zz} + D^{zz} + 2\omega_{e1} - 2\omega_n} +$$

$$\frac{2 A_{13}^{+-} A_{24}^{-+}}{A_{14}^{zz} + A_{23}^{zz} - d^{zz} - D^{zz} - 2\omega_{e2} - 2\omega_n} - \frac{2 A_{14}^{++} A_{23}^{--}}{A_{13}^{zz} + A_{24}^{zz} + d^{zz} + D^{zz} + 2\omega_{e2} - 2\omega_n} -$$

$$\frac{2 A_{14}^{++} A_{23}^{--}}{A_{13}^{zz} + A_{24}^{zz} + d^{zz} + D^{zz} - 2\omega_{e1} + 2\omega_n} + \frac{2 A_{13}^{+-} A_{24}^{-+}}{A_{14}^{zz} + A_{23}^{zz} - d^{zz} - D^{zz} + 2\omega_{e1} + 2\omega_n} -$$

$$\left.\frac{2 A_{14}^{++} A_{23}^{--}}{A_{13}^{zz} + A_{24}^{zz} + d^{zz} + D^{zz} - 2\omega_{e2} + 2\omega_n} + \frac{2 A_{13}^{+-} A_{24}^{-+}}{A_{14}^{zz} + A_{23}^{zz} - d^{zz} - D^{zz} + 2\omega_{e2} + 2\omega_n}\right)$$

$\textit{In[ ]:= }$ **HamM2〚6, 10〛**

$\textit{Out[ ]= }$

$$\frac{1}{2}\left(-\frac{A_{23}^{--}\left(-A_{13}^{+z}+A_{14}^{+z}-D^{+z}\right)}{A_{13}^{zz}-A_{14}^{zz}+D^{zz}-2\omega_{e1}}+\frac{A_{23}^{--}\left(-A_{13}^{+z}-A_{14}^{+z}-D^{+z}\right)}{A_{13}^{zz}+A_{14}^{zz}+D^{zz}+2\omega_{e1}}-\right.$$

$$\frac{A_{13}^{+-}\left(A_{23}^{-z}-A_{24}^{-z}-D^{z-}\right)}{A_{23}^{zz}-A_{24}^{zz}-D^{zz}-2\omega_{e2}}-\frac{2\left(\frac{A_{13}^{z-}}{2}-\frac{A_{23}^{z-}}{2}+\frac{d^{-z}}{2}\right)D^{+-}}{A_{13}^{zz}+A_{14}^{zz}-A_{23}^{zz}-A_{24}^{zz}+2\omega_{e1}-2\omega_{e2}}+$$

$$\frac{A_{13}^{+-}\left(A_{23}^{-z}+A_{24}^{-z}-D^{z-}\right)}{A_{23}^{zz}+A_{24}^{zz}-D^{zz}+2\omega_{e2}}-\frac{2\left(-\frac{A_{13}^{z-}}{2}+\frac{A_{23}^{z-}}{2}+\frac{d^{-z}}{2}\right)D^{+-}}{A_{13}^{zz}-A_{14}^{zz}-A_{23}^{zz}+A_{24}^{zz}-2\omega_{e1}+2\omega_{e2}}-\frac{\left(A_{13}^{z-}-A_{23}^{z-}+d^{-z}\right)D^{+-}}{A_{13}^{zz}-A_{23}^{zz}+d^{zz}-2\omega_{n}}-$$

$$\frac{2A_{23}^{--}\left(\frac{A_{13}^{+z}}{2}+\frac{A_{14}^{+z}}{2}+\frac{D^{+z}}{2}\right)}{A_{13}^{zz}+A_{24}^{zz}+d^{zz}+D^{zz}+2\omega_{e2}-2\omega_{n}}+\frac{\left(A_{13}^{z-}-A_{23}^{z-}-d^{-z}\right)D^{+-}}{A_{13}^{zz}-A_{23}^{zz}-d^{zz}+2\omega_{n}}+\frac{2A_{13}^{+-}\left(\frac{A_{23}^{-z}}{2}+\frac{A_{24}^{-z}}{2}-\frac{D^{z-}}{2}\right)}{A_{14}^{zz}+A_{23}^{zz}-d^{zz}-D^{zz}+2\omega_{e1}+2\omega_{n}}-$$

$$\left.\frac{2A_{13}^{+-}\left(-\frac{A_{23}^{-z}}{2}+\frac{A_{24}^{-z}}{2}+\frac{D^{z-}}{2}\right)}{A_{14}^{zz}-A_{23}^{zz}-d^{zz}+D^{zz}+2\omega_{e1}+2\omega_{n}}-\frac{2A_{23}^{--}\left(-\frac{A_{13}^{+z}}{2}+\frac{A_{14}^{+z}}{2}-\frac{D^{+z}}{2}\right)}{A_{13}^{zz}-A_{24}^{zz}-d^{zz}+D^{zz}-2\omega_{e2}+2\omega_{n}}\right)$$

$\textit{In[ ]:= }$ **HamM〚3, 3〛 - HamM〚2, 2〛**

⋯ Part: Part specification HamM〚2, 2〛 is longer than depth of object.

⋯ Part: Part specification HamM〚3, 3〛 is longer than depth of object.

$\textit{In[ ]:= }$ **H0M〚2, 2〛 - H0M〚3, 3〛**
**H0M〚4, 4〛 - H0M〚3, 3〛**

$\textit{Out[ ]= }$

$$\frac{A_{13}^{zz}}{2}-\frac{A_{14}^{zz}}{2}+\frac{A_{23}^{zz}}{2}-\frac{A_{24}^{zz}}{2}$$

$\textit{Out[ ]= }$

$$\frac{A_{13}^{zz}}{2}+\frac{A_{23}^{zz}}{2}+\frac{d^{zz}}{2}+\omega_{n}$$



\clearpage

\end{document}